\documentclass[11pt]{article}
 
\usepackage{amsbsy,amssymb,amsmath,bm}
\usepackage{color,epsfig,rotate}

\usepackage{graphicx}
\usepackage[curve,matrix,arrow]{xy}
\usepackage[all,dvips,color]{xypic}

\newcommand{\Ket}[1]{\left|#1  \right>}
\newcommand{\Bra}[1]{\left<#1  \right|}
\newcommand{\Braket}[1]{\left<#1  \right>}

\newcommand{\ffrac}[2]{\mbox{\footnotesize$\displaystyle\frac{#1}{#2}$}}

\newcommand{\vac}{|0\rangle}

\newcommand{\LQG}{U_{\q} \mathfrak{sl}_2}
\newcommand{\HXXZ}[1]{\mathcal{H}^{\rm XXZ}_{#1}}

\newcommand{\BX}{\mathcal{X}}
\newcommand{\BK}{\mathcal{K}}
\newcommand{\BW}{\mathcal{W}}
\newcommand{\BP}{\mathcal{P}}
\newcommand{\BT}{\mathcal{T}}

\newcommand{\Verma}[1]{\mathcal{V}_{#1}}
\newcommand{\Kac}[1]{\mathcal{K}_{#1}}

\newcommand{\primb}{\varphi}
\newcommand{\primt}{\psi}
\newcommand{\primr}{\rho}

\newcommand{\priml}{\xi}

\newcommand{\res}{quotient}
\newcommand{\rres}{Quotient}

\usepackage{amsmath,amsfonts,amssymb,latexsym,times}
\usepackage{verbatim}
\usepackage{bm}% bold math
\usepackage{epstopdf}
\usepackage[dvipsnames]{xcolor} % Before pstricks: permits OliveGreen, etc.
\usepackage{pstricks,pst-fill,pst-text,pst-node,pstricks-add}
\usepackage{mathrsfs}

\newcommand*{\longhookrightarrow}{\ensuremath{\lhook\joinrel\relbar\joinrel\rightarrow}}

\usepackage{psfrag}

\newcommand{\VK}{\mathcal{K}}
\newcommand{\VX}{\mathcal{X}}
\newcommand{\VP}{\mathcal{S}}

\newcommand{\Hilb}{\mathcal{H}}
\newcommand{\Zxxz}[1]{Z_{\rm XXZ}[#1]}

\newcommand{\chVv}{\Hilb}
\newcommand{\blobalg}{\mathcal{B}(2N,n,y)}
\newcommand{\rep}{\rho}
\newcommand{\Endo}{\mathrm{End}}
\newcommand{\oC}{\mathbb{C}}
\newcommand{\q}{\mathfrak{q}}

\usepackage{rotating}
\usepackage{multirow}

\begin{document}

\title{A physical approach to the classification of  indecomposable Virasoro representations from the blob algebra}

\author{Azat M. Gainutdinov$^{1}$, Jesper Lykke Jacobsen$^{2,3}$, Hubert Saleur$^{1,4}$ and Romain Vasseur$^{1,2}$\\
[2.0mm]
  ${}^1$Institut de Physique Th\'eorique, CEA Saclay,
  91191 Gif Sur Yvette, France \\
  ${}^2$LPTENS, 24 rue Lhomond, 75231 Paris, France \\
  ${}^3$ Universit\'e Pierre et Marie Curie, 4 place Jussieu, 75252 Paris, France \\
  ${}^4$Department of Physics,
  University of Southern California, Los Angeles, CA 90089-0484}

%\date{\today}

%\affiliation{$^1$ Institut de Physique Th\'eorique, CEA Saclay, 91191 Gif Sur Yvette, France \\ $^2$ LPTENS, 24 rue Lhomond, 75231 Paris, France}

\maketitle

\begin{abstract}

In the context of Conformal Field Theory (CFT), many results can be obtained from the representation theory of the Virasoro algebra. While the interest in Logarithmic CFTs has been growing recently, the Virasoro representations corresponding to these quantum field theories remain dauntingly complicated, thus hindering our understanding of various critical phenomena. We extend in this paper the construction of Read and Saleur~\cite{RS2,RS3}, and uncover a deep relationship between the Virasoro algebra and a finite-dimensional algebra characterizing the properties of two-dimensional statistical models, the so-called blob algebra (a proper extension of the Temperley--Lieb algebra). This allows us to explore vast classes of Virasoro representations (projective, tilting, generalized staggered modules, {\it etc.}), and to conjecture a classification of all  possible indecomposable Virasoro modules (with, in particular, $L_0$ Jordan cells of arbitrary rank) that may appear in a consistent physical Logarithmic CFT where Virasoro is the \textit{maximal local} chiral algebra. As  by-products, we solve and analyze algebraically  quantum-group symmetric XXZ spin chains and $\mathfrak{sl}(2|1)$ supersymmetric spin chains with extra spins at the boundary,  together with  the  ``mirror'' spin chain introduced by Martin and Woodcock~\cite{MartinFaithful0}.
\end{abstract}

%\tableofcontents
\section{Introduction}

While the subject of Logarithmic Conformal Field Theory (LCFT) appeared initially a little marginal, its importance has grown steadily over the last twenty years, following the pioneering papers \cite{RozanskySaleur, Gurarie}. A LCFT is a two dimensional conformal field theory whose symmetry algebra -- be it Virasoro, the product $\hbox{Virasoro}\times\overline{\hbox{Virasoro}}$, a current algebra, SUSY extensions,  {\it etc}. -- exhibits reducible but indecomposable modules. It is not hard to see that this leads to logarithmic dependence of (some of) the  correlation functions, as well as indecomposable operator product expansions.

Such unpleasant features are in fact unavoidable in many aspects of condensed matter physics. There, the unitarity of the CFT is not a requirement when one is interested in the description of geometrical, non local features in problems such as $2d$ polymers or percolation (see {\it e.g.}~\cite{VJSperco} for a recent application);  in the description of critical points in $2d$ classical disordered systems, such as the random-bond Ising model; or in the description of some quantum critical points such as occur for non interacting electrons  in a disordered potential in $2+1$ dimensions. Indecomposability also appears  whenever one considers $2d$ supergroup sigma models - such as the ones appearing on the string theory side of the AdS/CFT correspondence -- or supersymmetric sigma models relevant to topological conformal field theory, as fully fledged quantum field theories, that is, beyond their ``minimal'' or ``physical'' sectors. Indecomposability is of course not restricted to two dimensions either, and it is believed to play a role in some four dimensional gauge theories as well~\cite{Nekrasov}.

With a little perspective,  the interest in  LCFTs is a very natural development in algebra, and naturally transposes to the quantum field theory context the extension of the theory of Lie algebras from the semi-simple to the non semi-simple case. As a result, the field of ``non semi-simple CFTs'' is currently witnessing a very fast expansion in mathematics \cite{Huang}.

Despite this growing interest, useful, tangible physical results are hard to come by. Very few LCFTs are fully solved for instance: apart from  the ubiquitous symplectic fermions \cite{Kausch, GaberdielKausch}, the list includes -- and probably restricts to -- the $GL(1|1)$ and $SL(2|1)$ WZW models, and the $SL(2)$ WZW model at level $k=-{1\over 2}$ (see {\it e.g.}~\cite{CreutzigRidout}). In particular, bulk four point functions in the $c=0$ case \cite{SaleurDerrida} remain sadly unavailable to this day.

One of the difficulties of LCFTs is the absence of methods to understand and control the amount of indecomposability one might expect to encounter in the general case. While models such as WZW theories on supergroups can be (partly) tackled because they admit semi-classical limits where intuition from supergeometry can be used, the world of, say, LCFTs at $c=0$ appears overwhelmingly hard to tame abstractly. This difficulty can in fact be given a clear mathematical formulation:  for instance, it is known that the theory of representations of the Virasoro algebra is \textit{wild} which means roughly that it is as complicated as can be \cite{wild,Germoni, wild1}.  Among the rather modest questions whose answer is not known is, for instance, the question of how large 
are the $L_0$ Jordan cells appearing in a given (chiral or non-chiral) CFT, such as the long searched for ``CFT for percolation''. 

As is often the case however, physics provides powerful constraints which can be used to restrict the wilderness of the algebraic problem. For instance, there is a large body of evidence that only a certain kind  of modules appear in at least the simplest incarnations of supergroup WZW models, and this information can be put to very powerful use in concrete examples \cite{GotzQuellaSchomerus, QuellaSchomerus}. Rather than getting  into abstract considerations, it thus seems that important progress will be obtained by first studying in detail concrete models.

While supergroup WZW models can be tackled quite explicitly by using their semi-classical limit,  the theories at $c=0$ that we expect to describe, say, percolation or the self avoiding walk problem do not offer such natural handles. In their case, it has turned out in fact to be most useful to have a closer look at  lattice regularizations.

Lattice models indeed can be formulated in terms of associative lattice algebras such as the well known Temperley--Lieb algebra, which admits by now many generalizations, including the so called blob  or one-boundary Temperley--Lieb algebra, the Birman--Wenzl--Murakami algebra, the Jones annular algebra, {\it etc}. The non-semi simplicity of the conformal field theoretic description of the continuum limit of the lattice models is prefigured naturally in the non semi-simplicity of these lattice algebras. A remarkable fact -- first observed from characters identities many years back in some restricted cases~\cite{PasquierSaleur} -- is that the lattice algebras seem to exhibit, even in finite size, the exact same pattern of indecomposability as the continuum limit theory~\cite{RS3,GV}. While there are proposed mechanisms explaining this coincidence -- in the simplest case, the common presence of an underlying ``Lusztig'' quantum group symmetry \cite{Gainut} -- the fact is not entirely understood at the moment. 

Nevertheless, a wealth of information has been obtained recently about indecomposable modules of the Virasoro algebra using this technique \cite{PRZ,RS2,RS3,Jorgen1}, in agreement with  results in abstract Virasoro representation theory~\cite{MathieuRidout1,KytolaRidout} of so-called staggered modules. The idea can be naturally extended to gain information about fusion \cite{RS3,RP,Jorgen2,GV}, and even to measure the logarithmic couplings \cite{DJS,VJS}, whose value plays a key role in LCFTs. The extension to the case of $\hbox{Virasoro}\times\overline{\hbox{Virasoro}}$ and thus bulk logarithmic theories is an important but challenging direction -- for recent progress here, see  \cite{GRS1,GRS2,GRS3,VGJS}.

The analysis in \cite{RS2,RS3} focussed mostly on lattice models with open boundary conditions, which are described algebraically in terms of the Temperley--Lieb algebra, and deeply related in the continuum limit to Virasoro algebra modules with lowest conformal weights $h_{1,s}$ (see eq.~\eqref{conf-weight}), while a ``dilute''  version gives access to conformal weights $h_{r,1}$. It is of utmost importance, in order to better understand all physically relevant indecomposable Virasoro modules, to generalize these ideas to conformal weights $h_{r,s}$ with both indices tunable at will. It was suggested in \cite{PRZ} that  this should involve allowing for some extra degrees of freedom at the boundary, and in \cite{JS} that this idea corresponds, algebraically, to considering a one parameter extension of the TL algebra known as the blob or one-boundary Temperley--Lieb algebra. 

The purpose of this paper is to study in details the correspondence between this algebra, the lattice models where it naturally plays a role,  and the Virasoro algebra appearing in the continuum limit. While this looks like a rather technical endeavor, we believe that we are reaping a lot of rewards in the end, by obtaining what we believe (but do not prove) to be the  first complete classification of Virasoro indecomposable modules possibly appearing in physical LCFTs where the Virasoro is the \textit{maximal local} chiral algebra. 

Obviously, this conjecture requires a careful definition of `physical' LCFTs, this shall be discussed in details in Sec.~\ref{sec:class-ind-Vir}.
In a nutshell, we call {\sl physical} LCFTs  those conformal theories which are  fixed points of  interacting, non unitary, quantum field theories with well defined local actions\footnote{The superprojective sigma models at topological angle $\theta=\pi$ are among possible examples of such non unitary theories.}.  Assuming that such physical LCFTs describing critical phenomena in percolation or self-avoiding walks do exist, we can expect that they must also admit some lattice regularizations involving local Hamiltonians and local degrees of freedom. Such lattice regularizations are typically  quantum spin-chains with local interactions (susy spin-chains~\cite{ReadSaleur01}, XXZ models~\cite{RS3}, {\it etc}). It is worth noting that in this context of local theories, the well known statistical loop models (see {\it e.g.}~\cite{JS}) that  are formulated using non local Boltzman weights, do not, in general, lead to  physical field theories  -- it is known that their continuum limit does not coincide with any local field theory. Though the loop models can still be described using \textit{local} field theories, they must be considered as some subsectors of these local field theories which we believe can be obtained as scaling limits of quantum spin-chains. Note that even though we define physical LCFTs as scaling limits of local quantum spin chains, thus excluding loop models, the latters can of course be thought of as interesting and physically-reasonable
from the point of view of statistical mechanics. We will actually come back to loop models in this paper as the blob algebra provides new indecomposable modules for those theories too.

%\begin{equation}
%\parbox[t]{90ex}{\textit{``text''}}
%\end{equation}

%is restored only after adding to these subsectors more degrees of freedom. We believe this will eventually results in the continuum limit of the quantum spin-chains.

%, that is, that their properties can be studied by considering models defined on large, but finite, lattices, and exploring the thermodynamic and scaling limits. The point is now that the representation theory of algebras occurring for such finite lattice models -- such as the Temperley Lieb algebra or the blob algebra -- is well under control. 

 % physically relevant indecomposable Virasoro modules by simply defining them as scaling limit of spin-chains modules.
  
% Of course loop models can be described using local field theories, but these must contain more degrees of freedom than those in the loop models, the latter appearing then only as a sort of subsector. The additional degrees of freedom do restore self-duality, as one can observe  for instance when going from the usual Temperley Lieb loop models to the XXZ spin chain \cite{RS3}.

\bigskip

\begin{figure}
\begin{center}
\includegraphics[scale=0.82]{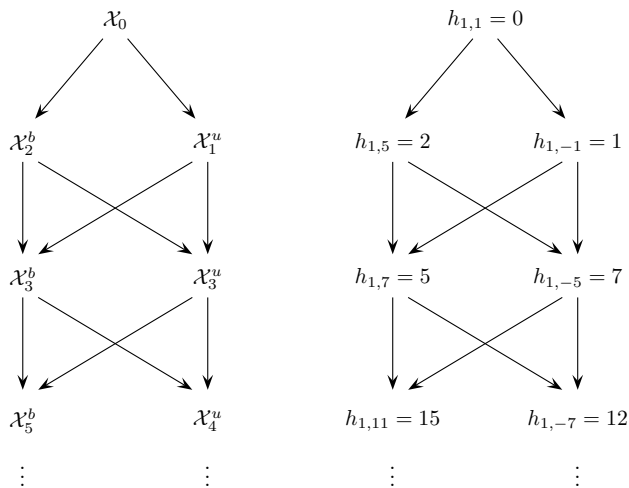}
\end{center}
  \caption{Structure of the standard module $\BW_{0}$ in terms of simple modules over the blob algebra $\mathcal{B}(2N,n,y)$ (or $\mathcal{B}^{b}(2N,n,y)$) for $r=1$ and $\q=\mathrm{e}^{i\pi/3}$, and the corresponding $c=0$ Verma module in the continuum limit. Simple Virasoro modules are denoted by their conformal weights.}
  \label{figStd_perco_1}
\end{figure}

\begin{figure}
\begin{center}
\includegraphics[scale=0.82]{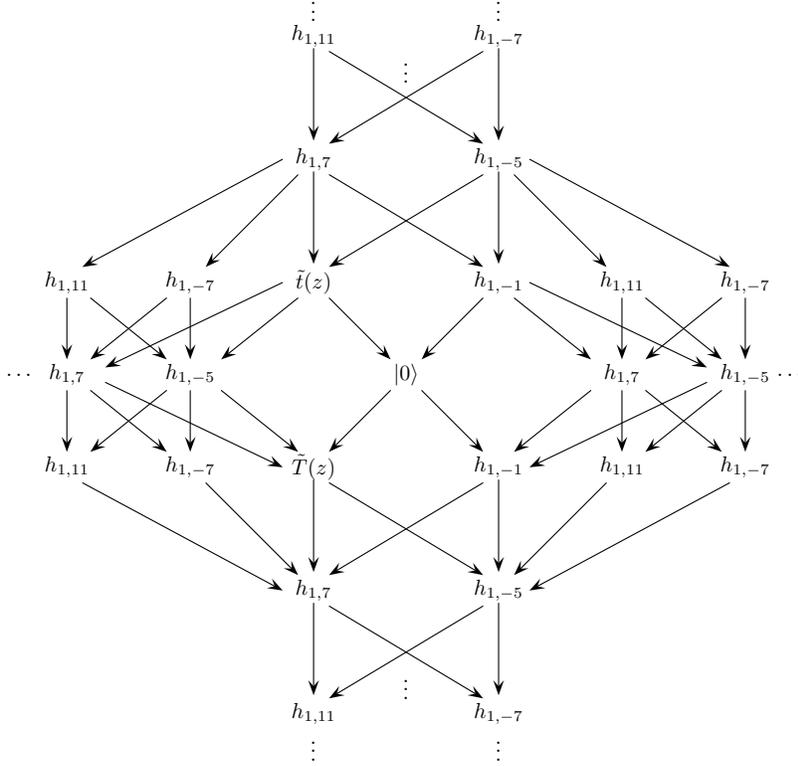}
\end{center}
  \caption{Structure in terms of simple subquotients of the ``vacuum'' indecomposable Virasoro module
  % from Fig.~\ref{fig:tilt-lim} 
  arising in the scaling limit of a faithful spin-chain representation
  of the blob algebra at $\q=\mathrm{e}^{i\pi/3}$ and $r=1$. The corresponding central charge is $c=0$. Our ``vacuum'' state $\vac$ corresponds to an operator
  with conformal weight $h_{1,1}=0$, its descendant on the level $2$ is  a primary field marked by $\tilde{T}(z)$ which has a logarithmic partner $\tilde{t}(z)$ with conformal weight $h_{1,5}=2$;
  the other fields are simply represented by their conformal weight.}
  \label{figVirIndec}
\end{figure}

Our paper is written from the point of view of \textit{modules}. Indeed, it is unclear  for now whether there is a  complete, general relation between  the blob (an associative algebra) and the Virasoro (a Lie algebra) at an abstract level, but what we can do is to compare in details the structure of modules for the two algebras. The following discussion is thus split into two main parts. In the first one, we discuss the correspondence  in the simplest case, by considering, on the Virasoro side, Verma modules, and on the blob side standard modules -- together with their various types of submodules. An example of such correspondence is given at central charge $c=0$ in Fig.~\ref{figStd_perco_1} (the notations for the blob algebra will be detailed in the following.) In the second part, we discuss the correspondence for a more general class of modules, in particular those  that can appear in physical models, and whose structure is constrained, in particular by self-duality. These
  modules, called \textit{tilting}, are different and more complicated than those considered in the first part, and their analysis requires some technical tools, which are discussed and summarized in a separate technical section. An example of such a module is provided in Fig.~\ref{figVirIndec} where  fields are simply represented by their conformal weight and the arrows increasing the conformal weight correspond to negative Virasoro modes while the arrows decreasing the weight represent the action of positive modes. We also note that  such a ``tilting'' module contains Jordan cells of arbitrary rank, for example, there is a rank-$3$ Jordan cell in $L_0$ for the conformal weight $h_{1,7}=5$, a rank-$4$ Jordan cell for the conformal weight $h_{1,11}=15$ and so on.
 
In more detail, our paper is organized as follows. Section~2 contains the basics: we review the blob algebra and its variants, some general aspects of its representation theory,  how it appears in lattice models, and what it has to do with the Virasoro algebra. In section~3, we discuss the relationship between the blob and   the Virasoro algebra at the level of well-known modules such as Verma modules. We demonstrate in particular  a detailed correspondence  -- summarized in table~1 --  between standard modules, quotients of standards and simple modules for the blob algebra on the one hand, and Verma, Kac and simple modules for the Virasoro algebra on the other hand. 

While in the case of unitary conformal field theories, irreducible Virasoro (or some appropriate chiral algebra) modules are all that is needed, many kinds of more complicated, reducible but indecomposable modules, may appear in a LCFT.  In fact, one of the purposes of the present analysis - and more generally of the associative algebraic approach to LCFTs - is to gain some general understanding of these modules. While the task is very difficult on the Virasoro side, it is much easier on the blob algebra side, thanks in part to mathematical theorems about  so called cellular and other types of algebras studied recently in the mathematical literature \cite{GL1}. In order to benefit from these results, the knowledge of a few definitions (such as projectiveness, self-duality, and tilting modules) is necessary. For convenience, we have gathered all such technical definitions in an algebraic reminder in section 4. We also describe the general structure of projective modules for the blob algebra in this section. 

We are then able to study in section~5 the correspondence between the blob algebra modules appearing in the XXZ and supergroup boundary spin-chains on the one hand, and Virasoro staggered modules appearing in the continuum limit on the other. As a by-product, a large class of conformally invariant boundary conditions  is introduced as the continuum limit of this family of boundary spin-chains. 

Based on these examples and warm-ups, we take one step further in sections~6 and~7, which are arguably  the main sections of this paper. In section~6, we discuss the  ``mirror''  spin-chain  first introduced in~\cite{MartinFaithful0}, which has the property of providing a {\sl faithful} representation of the blob algebra. We analyze the corresponding Hilbert space decomposition in terms of  tilting modules, which  requires the use of mathematical concepts introduced in section~4. The continuum limit of these modules is then studied in section~7. In this last section, we provide a tentative classification of new families of indecomposable {\sl self-dual} Virasoro modules, where in particular the Virasoro generator $L_0$ can have Jordan cells of arbitrary rank.  \textit{We conjecture in fact that for a LCFT with Virasoro as the maximal local chiral algebra, all possible physical indecomposable modules should be obtained as the scaling limit of lattice tilting modules of the blob algebra or (sub)quotients thereof}. This potentially tames\footnote{Somehow in  the same way the analysis of the semi-classical limit does for the WZW cases~\cite{SchomerusSaleur,SaleurSchomerus}.} the wilderness issue~\cite{wild,wild1}.
%We also provide a qualitative description of the continuum limit of mirror spin-chains for generic bulk fugacity and specific values of the boundary fugacity. These cases correspond to generic central charges, and we find in the continuum limit indecomposable Virasoro modules with $L_0$ Jordan cells of at most rank~$2$. We also compute the corresponding logarithmic couplings. At the end of  

 Although the blob algebra plays a central role in our classification of physically relevant Virasoro modules, it also enters into the definition of non-local lattice models -- such as boundary loop models~\cite{JS} -- that do not in general lead to physical field theories in the sense defined above. Even though we do not consider these loop models as proper lattice regularizations of physical LCFTs, they can obviously be considered as interesting in their own right, and relevant Virasoro modules in that context can also be described by the blob algebra. In section~7, we  present a large family of indecomposable blob modules which can potentially describe boundary loop models. All these modules are realized in the continuum limit as {\sl non} self-dual (in contrast to spin-chain modules) staggered Virasoro modules and their proper generalizations when three or more Verma modules are glued with each other. 
% This nice correspondence between the representation theory of the blob algebra on the one hand, and known results about indecomposable representations for Virasoro at different central charges on the other hand, motivates our tentative classification of indecomposable Virasoro modules given at the end of section~6.
 A few conclusions are gathered in section~8. Finally, three appendices contain rather technical and supplementary material.

\subsection*{Notations}
For convenience, we collect here some notations that we shall use throughout this paper:
\begin{itemize}
\item $\mathrm{TL}(2N,n)$: Temperley-Lieb algebra on $2N$ sites with loop fugacity $n$,
\item $\mathcal{B}(2N,n,y)$: blob algebra on $2N$ sites, with fugacity $y$ for boundary loops,
\item $\BW^{b/u}_{j}$: Standard modules over the blob algebra corresponding to $j$ through lines in the blobbed ($b$) or unblobbed ($u$) sector,
\item $\mathcal{B}^{b}(2N,n,y) = b \mathcal{B}(2N,n,y) b$: the JS blob algebra (see section~\ref{subsecSecJSblob}),
\item $\hat{\BW}^{b/u}_{j} = b \BW^{b/u}_{j} $: Standard modules over $\mathcal{B}^{b}$ (see section~\ref{subsecSecJSblob}),
\item $\Verma{h}$: Verma modules over the Virasoro algebra with conformal weight $h$,
\item $\Verma{r,s}$: Verma modules with the weight $h_{r,s}$ where $r$ and $s$ are the Kac labels,
\item $\BK^{b/u}_{j}$: quotients of standard modules over the blob algebra (see section~\ref{subSecSinglyCritical}),
\item $\VK_{r,s}$: Kac modules -- quotients of Verma modules over the Virasoro algebra (see section~\ref{subSecSinglyCritical}),
\item $P^{b}_k$ and $P^{u}_k$: Jones-Wenzl projectors (see eq.~\eqref{eqJonesWenzl}),
\item $\mathcal{B}^{\rm res}(2N,n,y) = \mathcal{B}(2N,n,y) / P^{u}_{r} $: the ``\res'' blob algebra 
%(defined for $r(y)$ integer)  
(see  section~\ref{sec:quot-blob}). %and studied intensively throughout section~\ref{secboundaryXXZrestrictedStagg},
%\item $\BK^{b}_{j} \equiv \BW^{b}_{j}/\BW^{u}_{j+r}$ and $\BK^{u}_{j}\equiv \BW^{u}_{j}/\BW^{u}_{r-j}$: Quotient standard modules of $\mathcal{B}^{\rm res}$
\item $\BX^{b/u}_{j}$: Simple modules over the blob algebra,
\item $\BP^{b/u}_{j}$: Projective modules over the blob algebra,
\item $\mathcal{T}^{b/u}_{j}$: Tilting modules over the blob algebra (used in section~\ref{sectionFaithfulMirrorSpinChain}),
\item $U_{\q} \mathfrak{sl}_2$: Quantum group $\q$-deformation of SU$(2)$,
\item $(j)$: Standard (Weyl) modules of $U_{\q} \mathfrak{sl}_2$,
\item $(j)_0$: Simple modules over $U_{\q} \mathfrak{sl}_2$,
\item $T_j$: Tilting modules over $U_{\q} \mathfrak{sl}_2$,
\item $\VP_{r,s}$: Staggered modules over the Virasoro algebra.
\end{itemize}

Before starting, we  emphasize that, while this paper deals with rather mathematical concepts, its style and emphasis are definitely non rigorous, though the analysis on the  lattice part is rather formal and mathematically precise. More rigorous work in the same spirit can be found in~\cite{GRS1,GRS2,GRS3}. Note also that our analysis uses extensively the thorough numerical conjectures of~\cite{JS} and the algebraic results of~\cite{Martin}.

\section{Preliminaries}
\label{section1}

In this section, we define the blob algebra and its generic representation theory.
We motivate the introduction of this algebra by discussing boundary conditions 
in loop models and we discuss how a slightly deformed version of the blob algebra, that
we shall call ``JS blob algebra'', is in fact more natural from a physical point of view~\cite{JS}. 
We will also introduce some notations that we will use throughout this paper.

\subsection{The blob algebra and its standard modules}
\label{subsecblobalgeRep}

\paragraph{blob algebra.} In this paper, we shall study models based on the blob algebra, also called one-boundary Temperley-Lieb algebra 
$\mathcal{B}(2N,n,y)$. This algebra -- which now plays an important role in the study of affine Hecke algebras~\cite{GL}, and admits various generalizations as well \cite{DubailTwoBd} -- seems to have been introduced 
first in the mathematical physics literature in~\cite{MartinSaleur}. It is a two parameters algebra, conveniently defined using decorated Temperley Lieb diagrams. We will assume in the following  that the reader is familiar with the Temperley-Lieb algebra (see~\cite{RidoutSaintAubin} for a recent review), its representation
theory when $n$ is generic and its interpretation as an algebra of diagrams. To define the blob algebra, consider now  $L=2N$ ($N \in \mathbb{N}/2$) strands and all the words 
written with the $2N-1$ generators $e_i$ ($1 \leq i \leq 2N-1$) and an extra generator $b$, subject to the relations
\begin{subequations} \label{TLdef}
\begin{eqnarray}
\left[ e_i , e_j \right] &=&0 \ (\left|i-j \right| \geq 2 )\\
\left[ e_i , b \right] &=&0 \ ( i \geq 2 )\\
e_i ^2 &=& n e_i\\
e_i e_{i \pm 1} e_i &=& e_i \\
b ^2 &=& b\\
e_1 b e_1 &=& y e_1.
\end{eqnarray}
\end{subequations}
This algebra is of course nothing but the usual Temperley-Lieb algebra with an extra boundary 
operator $b$, which decorates lines with a ``blob'', and gives to blobbed loops  a different weight $y$~\cite{MartinSaleur}.

We also define $\q=\mathrm{e}^{i \gamma}$ and $\gamma = \frac{\pi}{x+1}$ with $x \in \mathbb{R}$, and 
 parametrize the weights as
\begin{equation}
\displaystyle{ n=\q+\q^{-1} = \left[ 2 \right]_{\q}}
\end{equation}
and
\begin{equation}
\label{eqDefr}
\displaystyle y=\frac{\sin \left( (r+1) \gamma \right)}{\sin \left( r \gamma \right)} = \frac{ \left[ r+1 \right]_{\q}}{ \left[ r \right]_{\q}}.
\end{equation}
where $r$ is a real number and  we   use the $\q$-numbers
\begin{equation}
\displaystyle \left[ x\right]_{\q} = \dfrac{\q^{x}-\q^{-x}}{\q-\q^{-1}}.
\end{equation}
In the notations of Martin and Woodcock~\cite{Martin} (a reference we will use frequently in what follows), $y_{-}=y$, $m_{-}=-r$  and $y_{+}= \left[ 2 \right]_{\q} - y_{-}$. We can also introduce $m_{+}$ so that  $y_+=\frac{\sin \left( (m_+ -1) \gamma \right)}{\sin \left( m_+ \gamma \right)}$ with $m_+ =x+1 - m_-$.
We restrict ourselves to the case $r \geq 0$, as negative $r$'s are simply related to the case $ r \geq 0$ by a switch between $y_-$ and $y_+$.
Note also that we see from eq.~\eqref{eqDefr} that one can choose $r$ to be in what we shall call  the ``fundamental domain'' $0<r < x+1$. 

The operators $e_i$ and $b$ are best understood diagrammatically. Introducing the notation
$$
\begin{pspicture}(0.,0.)(1.0,1.0)
	\psline[](0.,0.)(0.,0.6)
	\psline[](0.5,0.)(0.5,0.6)
	\rput{0}(-0.7,0.25){$e_i=$}
\end{pspicture}\hdots \ \ \
\begin{pspicture}(0.,0.)(1.0,1.0)
	\psellipticarc{-}(0.25,0.)(0.25,-0.25){180}{0}
	\psellipticarc{-}(0.25,0.6)(0.25,0.25){180}{0}
	\rput{0}(0.,-0.3){$^i$}
	\rput{0}(0.5,-0.3){$^{i+1}$}
\end{pspicture} \hdots \ \ \
\begin{pspicture}(0.,0.)(1.0,1.0)
	\psline[](0.,0.)(0.,0.6)
	\psline[](0.5,0.)(0.5,0.6)
\end{pspicture},
$$
and 
$$
\begin{pspicture}(0.,0.)(1.0,1.0)
 \psdots[dotstyle=*,linecolor=black,dotscale= 1.5 1.5](0.0,0.3)
	\psline[](0.,0.)(0.,0.6)
	\psline[](0.5,0.)(0.5,0.6)
	\rput{0}(-0.7,0.25){$b=$}
\end{pspicture}  \hdots \ \ \
 \begin{pspicture}(0.,0.)(1.0,1.0)
	\psline[](0.,0.)(0.,0.6)
	\psline[](0.5,0.)(0.5,0.6)
\end{pspicture},
$$
the equations~\eqref{TLdef} can now be interpreted geometrically. The
composition law then corresponds to stacking the diagrams,
where it is assumed that every closed loop carries a weight $n$, while closed
loops carrying a blob symbol $\bullet$ (or more) get a different weight $y$.
Within this geometrical setup, the algebra $\mathcal{B}(2N,n,y)$ itself 
can be thought of as an algebra of diagrams.

\paragraph{Generically irreducible representations.} 

The blob algebra has the important property \cite{GL1} of being ``cellular'', which roughly means,
 that when considered as an algebra of diagrams,
the states can be cut in half, forming ``reduced states'' in the physicist language of the transfer 
matrix. These reduced states may be used as a basis to construct the so-called ``cell'' or \textit{standard modules}
of the algebra, which are in fact  the {\it irreducible} representations for $r$ generic (non integer).

The standard modules for the blob algebra are well known and can be easily constructed geometrically. 
They are parametrized by the number of through lines $0\leq 2j\leq 2N$ propagating through the system, {\it i.e.}, extending from the bottom to the top in the diagrammatic representation. Within a 
standard module, through lines cannot be contracted by the TL generators $e_i$.
In addition, the standard modules carry a label ($u$ or $b$) whose
meaning we now explain. We introduce an ``unblob'' or ``antiblob'' operator $u = 1-b$, which is an idempotent just like $b$ itself.
It is represented by a symbol `$\square$' instead of `$\bullet$'.
Moreover, the blob operator $b$ and the unblob operator $u$ are orthogonal projectors.
When there are through lines in the system ($j>0$), only the leftmost line is exposed to the boundary where $b$ and $u$ act.
We can choose a basis so that the leftmost through line is either blobbed by the symbol `$\bullet$' or unblobbed, with symbol `$\square$', and it has to stay that way
 under the action of the transfer matrix or spin-chain hamiltonian (see below). 
 We then denote $\BW_{0}$ the standard module with no through lines, and $\BW^{b}_{j}$ ({\it resp.} $\BW^{u}_{j}$)
the standard module with $2j$ through lines in the blobbed ({\it resp.}, unblobbed) sector. 

\begin{figure}
\begin{center}
\includegraphics[width=10.0cm]{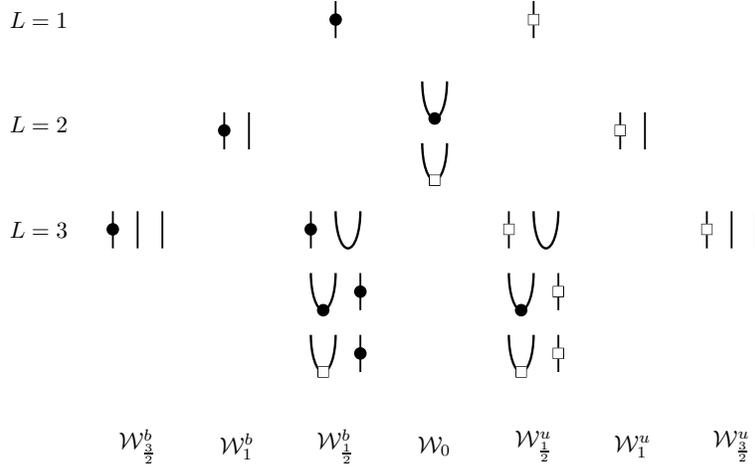}
\end{center}
  \caption{Pascal triangle construction of the standard modules of the blob algebra. The diagram is here truncated to $L=3$.
  The symbols $\bullet$ and $\square$ represent the blob $b$ and antiblob $1-b$ operators, respectively. 
   In the left (resp.\ right) part of the diagram, a step one row down and to the left (resp.\ right) corresponds to adding a through line on the right of the diagram; and a step one row down and to the right (resp.\ left) corresponds to bending the rightmost through line to the right of the diagram.}
  \label{figPascalStd}
\end{figure}

Following Ref.~\cite{MartinSaleur}, the standard modules can be constructed iteratively
using a Pascal triangle construction (see Fig.~\ref{figPascalStd}).
The action of the algebra on states can be obtained geometrically by stacking the diagrams 
on top of one another, just as for the Temperley-Lieb algebra with the convention 
that the TL generators $e_i$ cannot contract two through lines. 
This convention implies that each standard module is stable under the action of the algebra.
The blobbed and unblobbed standard modules
have the same dimension
\begin{equation} \label{eqDimStdblob}
\displaystyle d_j \equiv \mathrm{dim} \BW^{b/u}_{j} = \left( \begin{array}{c} 2N  \\ N - j  \end{array} \right) \,.
\end{equation}
Indeed, the Pascal construction of the state is compatible with the standard recursion relation
${2N-1 \choose N-j} + {2N-1 \choose N-j-1} = {2N \choose N-j}$.
The dimension of $\mathcal{B}(2N,n,y)$ is therefore
 \begin{equation}
\displaystyle \mathrm{dim} \mathcal{B}(2N,n,y) = \sum_{j=-N}^{N} \left( \begin{array}{c} 2N  \\ N - j  \end{array} \right)^2 = \left( \begin{array}{c} 4N  \\ 2N \end{array} \right).
\end{equation}
These standard modules are irreducible for any $\q$ and when $r$ is not an integer (generic). Finally note that throughout all this paper, we shall mostly restrict to the case $N$ integer, so that $L$ is even and $j$ is integer.
In any case $2j$ has the same parity as $L$.

\paragraph{Scalar product.}
 We also mention the existence of a natural symmetric bilinear form for the blob algebra, that we will loosely call scalar product.
In a given standard module, the scalar product of two basis states $\Ket{v_1}$ and $\Ket{v_2}$, denoted $\Braket{v_1 , v_2}$, is defined geometrically
by taking the mirror image of $\Ket{v_1}$ by a horizontal reflection, and gluing it on top of $\Ket{v_2}$. If a pair of through lines is contracted in the process,
we define the scalar product to be $0$, otherwise, its value is simply given by the weights of the loops in the resulting diagram. Let us illustrate this with a few
examples 
\begin{subequations} 
\begin{eqnarray}
\psset{xunit=2.5mm,yunit=2.5mm}
\Braket{ \
\begin{pspicture}(0,0)(5,1)
%\psgrid[subgriddiv=1,griddots=10,gridlabels=10pt](0,0)(5,1)
 \psdots[dotstyle=*,linecolor=black,dotscale= 1.0 1.0](1.5,-0.42)
 \psellipticarc[linecolor=black,linewidth=1.0pt]{-}(1.5,1.0)(1.5,1.42){180}{360}
 \psellipticarc[linecolor=black,linewidth=1.0pt]{-}(1.5,1.0)(0.5,0.71){180}{360}
 \psline[linecolor=black](4,-0.42)(4,1)
 \psdots[dotstyle=*,linecolor=black,dotscale= 1.0 1.0](4,0.3)
 \psline[linecolor=black](5,-0.42)(5,1)
 
\end{pspicture} \ , 
\begin{pspicture}(0,0)(5,1)
%\psgrid[subgriddiv=1,griddots=10,gridlabels=10pt](0,0)(5,1)
 \psdots[dotstyle=*,linecolor=black,dotscale= 1.0 1.0](1.5,-0.42)
 \psellipticarc[linecolor=black,linewidth=1.0pt]{-}(1.5,1.0)(1.5,1.42){180}{360}
 \psellipticarc[linecolor=black,linewidth=1.0pt]{-}(1.5,1.0)(0.5,0.71){180}{360}
 \psline[linecolor=black](4,-0.42)(4,1)
 \psdots[dotstyle=*,linecolor=black,dotscale= 1.0 1.0](4,0.3)
 \psline[linecolor=black](5,-0.42)(5,1)
 
\end{pspicture} \ } &=&  \
\psset{xunit=2.5mm,yunit=2.5mm}
\begin{pspicture}(0,0)(5,2.5)
%\psgrid[subgriddiv=1,griddots=10,gridlabels=10pt](0,0)(5,1)

 \psdots[dotstyle=*,linecolor=black,dotscale= 1.0 1.0](1.5,2.42)
  \psellipticarc[linecolor=black,linewidth=1.0pt]{-}(1.5,1.0)(1.5,1.42){0}{180}
 \psellipticarc[linecolor=black,linewidth=1.0pt]{-}(1.5,1.0)(0.5,0.71){0}{180}
 \psline[linecolor=black](4,0.58)(4,2)
 \psdots[dotstyle=*,linecolor=black,dotscale= 1.0 1.0](4,1.3)
 \psline[linecolor=black](5,0.58)(5,2)
 
  \psdots[dotstyle=*,linecolor=black,dotscale= 1.0 1.0](1.5,-0.42)
 \psellipticarc[linecolor=black,linewidth=1.0pt]{-}(1.5,1.0)(1.5,1.42){180}{360}
 \psellipticarc[linecolor=black,linewidth=1.0pt]{-}(1.5,1.0)(0.5,0.71){180}{360}
 \psline[linecolor=black](4,-0.42)(4,1)
 \psdots[dotstyle=*,linecolor=black,dotscale= 1.0 1.0](4,0.3)
 \psline[linecolor=black](5,-0.42)(5,1)
 
\end{pspicture} \
=
 n y, \\ 
\psset{xunit=2.5mm,yunit=2.5mm}
\Braket{ \
\begin{pspicture}(0,0)(5,1)
%\psgrid[subgriddiv=1,griddots=10,gridlabels=10pt](0,0)(5,1)
 \psdots[dotstyle=*,linecolor=black,dotscale= 1.0 1.0](1.5,-0.42)
 \psellipticarc[linecolor=black,linewidth=1.0pt]{-}(1.5,1.0)(1.5,1.42){180}{360}
 \psellipticarc[linecolor=black,linewidth=1.0pt]{-}(1.5,1.0)(0.5,0.71){180}{360}
 \psline[linecolor=black](4,-0.42)(4,1)
 \psdots[dotstyle=*,linecolor=black,dotscale= 1.0 1.0](4,0.3)
 \psline[linecolor=black](5,-0.42)(5,1)
 
\end{pspicture} \ , 
\begin{pspicture}(0,0)(5,1)
%\psgrid[subgriddiv=1,griddots=10,gridlabels=10pt](0,0)(5,1)
 \psline[linecolor=black](0,-0.42)(0,1)
 \psdots[dotstyle=*,linecolor=black,dotscale= 1.0 1.0](0,0.3)
 
 \psdots[dotstyle=*,linecolor=black,dotscale= 1.0 1.0](2.5,-0.42)
 \psellipticarc[linecolor=black,linewidth=1.0pt]{-}(2.5,1.0)(1.5,1.42){180}{360}
 \psellipticarc[linecolor=black,linewidth=1.0pt]{-}(2.5,1.0)(0.5,0.71){180}{360}
 \psline[linecolor=black](5,-0.42)(5,1)
 
\end{pspicture} \ }  &=&  \
\psset{xunit=2.5mm,yunit=2.5mm}
\begin{pspicture}(0,0)(5,2.5)
%\psgrid[subgriddiv=1,griddots=10,gridlabels=10pt](0,0)(5,1)

 \psellipticarc[linecolor=black,linewidth=1.0pt]{-}(1.5,1.0)(1.5,1.42){0}{180}
 \psellipticarc[linecolor=black,linewidth=1.0pt]{-}(1.5,1.0)(0.5,0.71){0}{180}
 \psline[linecolor=black](4,0.58)(4,2)
 \psdots[dotstyle=*,linecolor=black,dotscale= 1.0 1.0](4,1.3)
 \psline[linecolor=black](5,0.58)(5,2)
 
 \psline[linecolor=black](0,-0.42)(0,1)
 \psdots[dotstyle=*,linecolor=black,dotscale= 1.0 1.0](0,0.3)
 
 \psdots[dotstyle=*,linecolor=black,dotscale= 1.0 1.0](2.5,-0.42)
 \psellipticarc[linecolor=black,linewidth=1.0pt]{-}(2.5,1.0)(1.5,1.42){180}{360}
 \psellipticarc[linecolor=black,linewidth=1.0pt]{-}(2.5,1.0)(0.5,0.71){180}{360}
 \psline[linecolor=black](5,-0.42)(5,1)
 
\end{pspicture} \
=1, \\ 
\psset{xunit=2.5mm,yunit=2.5mm}
\Braket{ \
\begin{pspicture}(0,0)(5,1)
%\psgrid[subgriddiv=1,griddots=10,gridlabels=10pt](0,0)(5,1)
 \psline[linecolor=black](0,-0.42)(0,1)
 \psdots[dotstyle=*,linecolor=black,dotscale= 1.0 1.0](0,0.3)
  \psline[linecolor=black](1,-0.42)(1,1)
 \psellipticarc[linecolor=black,linewidth=1.0pt]{-}(2.5,1.0)(0.5,1.42){180}{360}
  \psellipticarc[linecolor=black,linewidth=1.0pt]{-}(4.5,1.0)(0.5,1.42){180}{360}

\end{pspicture} \ , 
\begin{pspicture}(0,0)(5,1)
%\psgrid[subgriddiv=1,griddots=10,gridlabels=10pt](0,0)(5,1)
 \psdots[dotstyle=*,linecolor=black,dotscale= 1.0 1.0](1.5,-0.42)
 \psellipticarc[linecolor=black,linewidth=1.0pt]{-}(1.5,1.0)(1.5,1.42){180}{360}
 \psellipticarc[linecolor=black,linewidth=1.0pt]{-}(1.5,1.0)(0.5,0.71){180}{360}
 \psline[linecolor=black](4,-0.42)(4,1)
 \psdots[dotstyle=*,linecolor=black,dotscale= 1.0 1.0](4,0.3)
 \psline[linecolor=black](5,-0.42)(5,1)
 
\end{pspicture} \ } &=&  \
\psset{xunit=2.5mm,yunit=2.5mm}
\begin{pspicture}(0,0)(5,2.5)
%\psgrid[subgriddiv=1,griddots=10,gridlabels=10pt](0,0)(5,1)

 \psline[linecolor=black](0,0.64)(0,2)
 \psdots[dotstyle=*,linecolor=black,dotscale= 1.0 1.0](0,1.3)
  \psline[linecolor=black](1,0.64)(1,2)
 \psellipticarc[linecolor=black,linewidth=1.0pt]{-}(2.5,1.0)(0.5,1.42){0}{180}
  \psellipticarc[linecolor=black,linewidth=1.0pt]{-}(4.5,1.0)(0.5,1.42){0}{180}

  \psdots[dotstyle=*,linecolor=black,dotscale= 1.0 1.0](1.5,-0.42)
 \psellipticarc[linecolor=black,linewidth=1.0pt]{-}(1.5,1.0)(1.5,1.42){180}{360}
 \psellipticarc[linecolor=black,linewidth=1.0pt]{-}(1.5,1.0)(0.5,0.71){180}{360}
 \psline[linecolor=black](4,-0.42)(4,1)
 \psdots[dotstyle=*,linecolor=black,dotscale= 1.0 1.0](4,0.3)
 \psline[linecolor=black](5,-0.42)(5,1)
 
\end{pspicture} \
= 0.
\end{eqnarray}
\end{subequations}
An important point is that the generators of the blob algebra are self-adjoint for this scalar product, that is, $b^\dag=b$ and $e_i^\dag=e_i$.

\subsection{The JS blob algebra and its standard modules}
\label{subsecSecJSblob}

\paragraph{The JS algebra}

For reasons that will become clear later, it will be useful for us to consider 
a smaller algebra, that we will call \textit{the JS blob algebra} \cite{JS,JScomb}, obtained via
\begin{equation}\label{def-JS}
\mathcal{B}^{b}(2N,n,y) \equiv b \mathcal{B}(2N,n,y) b,
\end{equation}
 Note that by this definition, any element of the JS algebra $\mathcal{B}^{b}(2N,n,y)$ is an element of $\mathcal{B}(2N,n,y)$ sandwiched between two $b$'s.  In terms of diagrams, the 
 action of the generators of $\mathcal{B}^{b}(2N,n,y)$
guarantees that every loop touching the left boundary 
 carries a blob symbol. 
 
A  crucial observation 
is that when $y \neq 0$, this algebra is isomorphic to another blob 
algebra on a smaller number of sites and with different parameters.
More explicitly, we have an isomorphism of associative algebras
\begin{equation}\label{JS-alg-iso}
\mathcal{B}^{b}(2N,n,y) \simeq \mathcal{B}(2N-1,n,y^{-1}).
\end{equation}
 The proof of this statement is elementary, 
denoting $b$ and $b e_i b$, for $1 \leq i \leq 2N-1$, the generators of $\mathcal{B}^{b}(2N,n,y)$,
one can readily check that the identity in the JS algebra is  $1'\equiv b$ and  the generators 
\begin{equation}\label{JSblob-iso}
b' \equiv y^{-1} b e_1 b, \quad \text{and} \quad e_i' \equiv  b e_{i+1} b
\end{equation}
  satisfy  all the relations of the blob algebra $\mathcal{B}(2N-1,n,y^{-1})$. 

In the  particular case $y=0$,  the element $b'$ from~\eqref{JSblob-iso} is not defined. Introducing then  $b' = b e_1 b$ and $e_i' \equiv  b e_{i+1} b$, the algebra $ \mathcal{B}^{b}(2N,n,y=0)$  is then  isomorphic to
 yet another variant of the  blob algebra generated by $e'_i$, with $1\leq i\leq 2N-2$, and $b'$  with the relations~\eqref{TLdef} except the two last ones which  read now
 \begin{equation}\label{b-square-zero}
  b'^2=0,\qquad
  e'_1 b' e'_1 = e'_1.
 \end{equation}
 Notice that, since $b'^2=0$,   there is no possible isomorphism of this algebra  with a blob algebra. Representation theory in this case is more peculiar and will be discussed separately below in this section and in Sec.~\ref{case-2} and~\ref{sec:proj-gen}.

 \paragraph{Standard modules over the JS algebra}
 
The standard modules over the JS blob algebra $\mathcal{B}^{b}(2N,n,y)$ are readily 
obtained from those of the original blob algebra by acting with $b$, {\it i.e.}, they are images of $b$ under its action in $\BW^{b/u}_{j}$.
We will use a hat to denote the representations of $\mathcal{B}^{b}(2N,n,y)$ ({\it e.g.} $\hat{\BW}^{b/u}_{j}=b \BW^{b/u}_{j}$).

Recalling the isomorphism~\eqref{JS-alg-iso} and~\eqref{JSblob-iso} of associative algebras, the correspondence with the standard modules over $\mathcal{B}(2N-1,n,y^{-1})$ is the following. 
We find that the JS blobbed modules labeled with $j$ are {\it unblobbed} standard modules 
with $2(j-\frac{1}{2})=2j-1$ through lines for $\mathcal{B}(2N-1,n,y^{-1})$, and that JS unblobbed modules 
with $2j$ through lines are {\it blobbed} standard modules 
with $2(j+\frac{1}{2})=2j+1$ through lines.
In other words, we have the isomorphisms $\hat{\BW}^{b}_{j} \simeq \BW^{u}_{j-\frac{1}{2}}$ and $\hat{\BW}^{u}_{j} \simeq \BW^{b}_{j+\frac{1}{2}}$,
with a positive integer~$j$ not exceeding $N$ (note that $\hat{\BW}^{u}_{N} = 0$), and where modules in the right-hand side of these equations are over $\mathcal{B}(2N-1,n,y^{-1})$.
The rigorous proof of this statement is rather technical so we shall not go into these formal details here.
We note that the module with no through lines should be considered as unblobbed \cite{JScomb} and hence $\hat{\BW}_{0} \simeq \BW^{b}_{\frac{1}{2}}$.
Note also that using the dimensions of the standard modules for $\mathcal{B}(2N-1,n,y^{-1})$, 
we immediately find the dimensions of the JS standard modules
\begin{subequations} \label{eq_dimb}
\begin{eqnarray}
\hat{d}^b_j \equiv \mathrm{dim} \hat{\BW}^{b}_{j} & = & \binom{2N - 1}{N - j}, \\
\hat{d}^u_j \equiv \mathrm{dim} \hat{\BW}^{u}_{j} & = & \binom{2N - 1}{N - j -1}. 
\end{eqnarray}
\end{subequations}
These dimensions can also be obtained using combinatorics~\cite{JScomb}. 

%Although these observations provide a nice algebraical framework
%to explain some of the results obtained in Ref.~\cite{JS}, one might
%wonder why this algebra $\mathcal{B}^{b}(2N,n,y)$ should be important as, after all,
%it is isomorphic to the blob algebra itself. 
%The answer to this question
%will arise when we turn to the representation theory of these algebras.
In the case  of non-zero $y$, we will use  the algebra isomorphism $\mathcal{B}^{b}(2N,n,y) \simeq \mathcal{B}(2N-1,n,y^{-1})$
to study the representation theory of $\mathcal{B}^{b}(2N,n,y)$ using the results of~\cite{Martin}. The case 
 $y=0$ requires more work because it is not covered in~\cite{Martin}\footnote{Note however that this case is part of the results in~\cite{GL} that uses the representation theory of the affine TL algebra along with the so-called braid translator. We follow here a more direct route.}. In the case $y=0$,  we find that the   two non isomorphic modules $\BW^u_{j}$ and $ \BW^b_{j+1}$ 
over the blob algebra $\mathcal{B}(2N,n,0)$  become isomorphic 
for the JS blob algebra:
\begin{equation}\label{eqIsoCaseGeom}
\hat{\BW}^u_{j} \equiv b \BW^u_{j} \simeq \hat{\BW}^b_{j+1} \equiv b \BW^b_{j+1},\quad {\rm when} \ y=0.
\end{equation}
To show this, let us introduce the linear map $\phi: b \BW^b_{j+1} \longrightarrow b \BW^u_{j}$, whose action 
on the basis states of $\hat{\BW}^b_{j+1} \equiv b \BW^b_{j+1}$ is given by
$$
\begin{pspicture}(0,0)(9,2)
\psset{xunit=10mm,yunit=10mm}
%\psgrid[subgriddiv=1,griddots=10,gridlabels=10pt](0,0)(9,2)

 \rput[Bc](0.5,1){$\phi$}
 \rput[Bc](1.0,1){$:$}

 \rput[Bc](1.5,1){$\dots$}
 \psline[linecolor=black](2,0.5)(2,1.5)
 \psdots[dotstyle=*,linecolor=black,dotscale= 1.5 1.5](2,1)
 \rput[Bc](2.5,1){$\dots$}
 \psline[linecolor=black](3,0.5)(3,1.5)
 \rput[Bc](3.5,1){$\dots$}
 \psline[linecolor=black](4,0.5)(4,1.5) 
  \rput[Bc](4.5,1){$\dots$} 

  \rput[Bc](5.25,1){$\longmapsto$}  
  
\rput[Bc](6,1){$\dots$}
 \psline[linecolor=black](6.5,0.5)(6.5,1.5)
 \rput[Bc](7,1){$\dots$}
 \psline[linecolor=black](7.5,0.5)(7.5,1.5)
 \rput[Bc](8,1){$\dots$}
 \psline[linecolor=black](8.5,0.5)(8.5,1.5) 
 \psdots[dotstyle=square,linecolor=black,dotscale= 1.5 1.5](8.5,1) 
  \rput[Bc](9,1){$\dots$} 
  
  \psellipticarc[linecolor=black](7,0.5)(0.5,0.35){180}{360}
 \psdots[dotstyle=*,linecolor=black,dotscale= 1.5 1.5](7,0.15)
     
\end{pspicture}
$$
where we have represented only the first three leftmost through lines of the state in $\hat{\BW}^b_{j+1}$.
In simple words, $\phi$ joins the first two leftmost through lines of the state in $\hat{\BW}^b_{j+1}$,
and puts an antiblob symbol on the third one if the latter exists. The resulting state belongs to $\hat{\BW}^u_{j} \equiv  b \BW^u_{j}$
indeed, with the convention $\hat{\BW}^u_{0} \equiv b \BW_0 $. It is straightforward to check
that this is a one to one mapping so that $\phi$ is an isomorphism between vector spaces.
Moreover, when $y=0$, $\phi$ is an intertwining operator of $\mathcal{B}^{b}(2N,n,y=0)$.
The main point in this last claim is that 
$\psset{xunit=2mm,yunit=2mm}
\begin{pspicture}(0,0)(1.5,1)
 \psdots[dotstyle=*,linecolor=black,dotscale= 1.0 1.0](0.5,-0.42)
 \psellipticarc[linecolor=black,linewidth=1.0pt]{-}(0.5,1.0)(0.5,1.42){180}{360}
\end{pspicture}$
behaves exactly like two through lines when $y=0$, in particular, it is annihilated
by the corresponding Temperley-Lieb generator
$
\psset{xunit=3mm,yunit=3mm} \begin{pspicture}(0,0)(1,1)
  \psellipticarc[linecolor=black,linewidth=1.0pt]{-}(0.5,1)(0.355,0.355){180}{360}
  \psellipticarc[linecolor=black,linewidth=1.0pt]{-}(0.5,0)(0.355,0.355){0}{180}
\end{pspicture}$. The representations $\hat{\BW}^u_{j} $ and $\hat{\BW}^b_{j+1}$ are
therefore equivalent when the weight of the blobbed loops vanishes $y=0$. We will denote this representation for the JS algebra as $\BW_{j+\frac{1}{2}}$, without the superscript
\begin{equation}\label{BWj-def}
\BW_{j+\frac{1}{2}} \equiv \hat{\BW}^u_{j} \simeq \hat{\BW}^b_{j+1},\qquad 0\leq j\leq N-1.
\end{equation}
 The isomorphisms~\eqref{eqIsoCaseGeom}  will turn out to be  important in what follows\footnote{
 Note that such isomorphisms can be checked numerically, at least for generic $\q$ when the modules $\BW_{j}$ are simple. In particular, we found that the transfer
matrix  (see below) has exactly the same spectrum in the modules $\hat{\BW}_{0}$ and $\hat{\BW}^{b}_{1}$, {\it etc}.}.

\subsection{The blob algebra in statistical mechanics models}
\label{subsecblobalgStatmech}

\paragraph{Transfer matrix and loop models}

We will study models defined by their \textit{transfer matrix}
\begin{equation}\label{eq_Transfer}
\displaystyle T = \left(1 + \lambda b \right) \prod_{i=1}^{N-1} \left( 1 +  e_{2i} \right) \prod_{i=1}^{N} \left( 1 + e_{2i-1} \right) ,
\end{equation}
or by the $1+1$-dimensional \textit{quantum hamiltonian}
\begin{equation}\label{eq_Hamiltoniang}
\displaystyle H = - \alpha b  - \sum_{i=1}^{2N-1} e_i.
\end{equation}
The transfer matrix acting on standard modules should be thought of in terms of a geometrical loop model.
Let us represent its action in terms of plaquettes
$$
\begin{pspicture}(0,0)(9,2.25)
\psset{xunit=10mm,yunit=10mm}
%\psgrid[subgriddiv=1,griddots=10,gridlabels=10pt](0,0)(9,2.25)

 \rput[Bc](0.5,1){$T$}
 \rput[Bc](1.5,1){$=$}
\psline[linecolor=black](2.25,0.75)(3,1.5)
\psline[linecolor=black](3,1.5)(3.75,0.75)
\psline[linecolor=black](2.25,0.75)(3,0)
\psline[linecolor=black](3,0)(3.75,0.75)

\psline[linecolor=black](3.75,0.75)(4.5,1.5)
\psline[linecolor=black](4.5,1.5)(5.25,0.75)
\psline[linecolor=black](3.75,0.75)(4.5,0)
\psline[linecolor=black](4.5,0)(5.25,0.75)

\psline[linecolor=black](5.25,0.75)(6,1.5)
\psline[linecolor=black](6,1.5)(6.75,0.75)
\psline[linecolor=black](5.25,0.75)(6,0)
\psline[linecolor=black](6,0)(6.75,0.75)

\psline[linecolor=black](6.75,0.75)(7.5,1.5)
\psline[linecolor=black](7.5,1.5)(8.25,0.75)
\psline[linecolor=black](6.75,0.75)(7.5,0)
\psline[linecolor=black](7.5,0)(8.25,0.75)

\psline[linecolor=black](3,1.5)(3.75,2.25)
\psline[linecolor=black](3.75,2.25)(4.5,1.5)
\psline[linecolor=black](4.5,1.5)(5.25,2.25)
\psline[linecolor=black](5.25,2.25)(6,1.5)
\psline[linecolor=black](6,1.5)(6.75,2.25)
\psline[linecolor=black](6.75,2.25)(7.5,1.5)

\psline[linecolor=black](7.5,1.5)(8.25,2.25)
\psline[linecolor=black](8.25,2.25)(8.25,0.75)

\psellipticarc[linecolor=blue](7.5,1.5)(0.53033,0.53033){315}{45}

\psline[linecolor=black](2.25,2.25)(3,1.5)
\psline[linecolor=black](2.25,2.25)(2.25,0.75)

\end{pspicture}
$$
where in this example $N=4$ ($L=8$). The rightmost half plaquette corresponds 
to a free boundary condition whereas the first half plaquette may or may not
carry a blob symbol.
The bulk plaquettes carry the usual action of the Temperley-Lieb generators 
$$
\begin{pspicture}(0,0)(10,2)
\psset{xunit=10mm,yunit=10mm}
%\psgrid[subgriddiv=1,griddots=10,gridlabels=10pt](0,0)(10,2)

 \rput[Bc](0.5,1){$1+e_i$}
 \rput[Bc](1.5,1){$=$}
\psline[linecolor=black](2.25,1)(3,1.75)
\psline[linecolor=black](3,1.75)(3.75,1)
\psline[linecolor=black](2.25,1)(3,0.25)
\psline[linecolor=black](3,0.25)(3.75,1)

 \rput[Bc](4.5,1){$=$}

\psline[linecolor=black](5.25,1)(6,1.75)
\psline[linecolor=black](6,1.75)(6.75,1)
\psline[linecolor=black](5.25,1)(6,0.25)
\psline[linecolor=black](6,0.25)(6.75,1)

\psellipticarc[linecolor=blue](5.25,1)(0.53033,0.53033){315}{45}
\psellipticarc[linecolor=blue](6.75,1)(0.53033,0.53033){135}{225}

 \rput[Bc](7.5,1){$+$}

\psline[linecolor=black](8.25,1)(9,1.75)
\psline[linecolor=black](9,1.75)(9.75,1)
\psline[linecolor=black](8.25,1)(9,0.25)
\psline[linecolor=black](9,0.25)(9.75,1)

\psellipticarc[linecolor=blue](9,0.25)(0.53033,0.53033){45}{135}
\psellipticarc[linecolor=blue](9,1.75)(0.53033,0.53033){225}{315}

\end{pspicture}
$$
whilst the first one represents 
$$
\begin{pspicture}(0,0)(10,2)
\psset{xunit=10mm,yunit=10mm}
%\psgrid[subgriddiv=1,griddots=10,gridlabels=10pt](0,0)(10,2)

 \rput[Bc](0.5,1){$1+\lambda b$}
 \rput[Bc](1.5,1){$=$}
\psline[linecolor=black](3,1.75)(3.75,1)
\psline[linecolor=black](3,0.25)(3.75,1)
\psline[linecolor=black](3,1.75)(3,0.25)

 \rput[Bc](4.5,1){$=$}

\psline[linecolor=black](6,1.75)(6.75,1)
\psline[linecolor=black](6,0.25)(6.75,1)
\psline[linecolor=black](6,1.75)(6,0.25)

\psellipticarc[linecolor=blue](6.75,1)(0.53033,0.53033){135}{225}

 \rput[Bc](7.5,1){$+$}
 \rput[Bc](8.25,1){$\lambda$}

\psline[linecolor=black](9,1.75)(9.75,1)
\psline[linecolor=black](9,0.25)(9.75,1)
\psline[linecolor=black](9,1.75)(9,0.25)

\psellipticarc[linecolor=blue](9.75,1)(0.53033,0.53033){135}{225}
 \psdots[dotstyle=*,linecolor=blue,dotscale= 1.5 1.5](9.22,1)

\end{pspicture}
$$
The action of this transfer matrix builds up loop configurations, 
where the edges touching the boundary may or may not be blobbed,
depending on the parameter $\lambda$. In addition to these
$\lambda$ factors, the Boltzmann weight of 
a configuration is computed by attributing a weight $y$ to closed loops carrying
at least one blob symbol, and a weight $n$ to the other loops.  
The blob algebra is completely symmetric between blob $b$ and antiblob
$1-b$ operators, and it is  convenient to introduce the diagrammatic representation
$$
\begin{pspicture}(0,0)(4,2)
\psset{xunit=10mm,yunit=10mm}
%\psgrid[subgriddiv=1,griddots=10,gridlabels=10pt](0,0)(4,2)

 \rput[Bc](1,1){$1- b$}
 \rput[Bc](2.05,1){$=$}

\psline[linecolor=black](3,1.75)(3.75,1)
\psline[linecolor=black](3,0.25)(3.75,1)
\psline[linecolor=black](3,1.75)(3,0.25)

\psellipticarc[linecolor=blue](3.75,1)(0.53033,0.53033){135}{225}
 \psdots[dotstyle=square,linecolor=blue,dotscale= 1.5 1.5](3.22,1)

\end{pspicture}
$$
where the box represents the antiblob operator, and each loop carrying 
at least one box will be weighted by $n-y$. Note also that configurations
with at least one loop that is blobbed and unblobbed at the same time
should be excluded from the partition function, since $b(1-b)=0$.

\begin{figure} 
\begin{center}
\includegraphics[width=14.0cm]{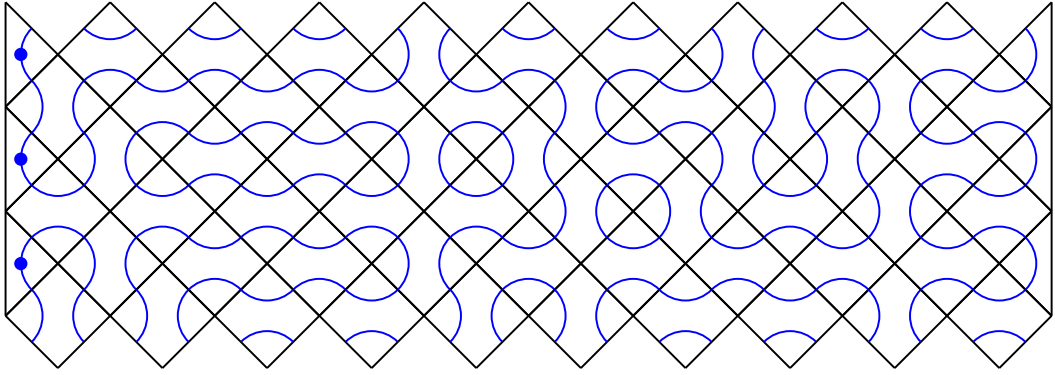}
\caption{Example of loop configuration on $L=20$ sites, after 3 iterations of the transfer matrix $T$ in the limit $\lambda = \infty$.
  Every loop touching the left boundary is blobbed and hence carries a weight $y$. The other loops that do not carry a blob
  symbol get a weight $n$.}
  \label{FigConfig}
\end{center}
\end{figure}

Thanks to the extensive numerical study of~\cite{JS} and the Bethe ansatz results of~\cite{Nichols1,Nichols2},
the phase diagram of this statistical model in terms of the boundary coupling $\lambda$ is now well understood, and can be summarized by the diagram below:
$$
\begin{pspicture}(0,0)(9,2)
\psset{xunit=10mm,yunit=10mm}
%\psgrid[subgriddiv=1,griddots=10,gridlabels=10pt](0,0)(9,2)

 \rput[Bc](0,1.5){$\lambda=-1$}
 \rput[Bc](4.5,1.5){$\lambda=0$}
 \rput[Bc](9,1.5){$\lambda=\infty$}
   
 \psline[linecolor=black,arrowsize=5pt]{->}(4.5,1)(7.25,1)
 \psline[linecolor=black](9,1)(0,1) 
 \psline[linecolor=black,arrowsize=5pt]{->}(4.5,1)(2.25,1)
  
 \psdots[dotstyle=*,linecolor=red,dotscale= 1.0 1.0](9,1)
 \psdots[dotstyle=*,linecolor=red,dotscale= 1.0 1.0](0,1)
 \psdots[dotstyle=*,linecolor=red,dotscale= 1.0 1.0](4.5,1)
       
\end{pspicture}
$$
Here, the fixed point $\lambda=0$ corresponds to free boundary conditions, and is described algebraically 
by the Temperley-Lieb algebra. The perturbation by the boundary coupling $\lambda$ is relevant in the renormalization
group sense, so that the system flows to one of the two conformally invariant boundary conditions where every loop
touching the left boundary carries a blob or unblob symbol. These two fixed points are related by a simple switch
between the weights of blobbed and unblobbed loops, and in the following we will focus only on the fixed point $\lambda=\infty$.

The limit $\lambda=\infty$ in the transfer matrix $\lambda^{-1} T$,  corresponds to blobbing
every loop touching the boundary; 
the transfer matrix  reads then schematically
$$
\begin{pspicture}(0,0)(9,2.25)
\psset{xunit=10mm,yunit=10mm}
%\psgrid[subgriddiv=1,griddots=10,gridlabels=10pt](0,0)(9,2.25)

\psellipticarc[linecolor=blue](3,1.5)(0.53033,0.53033){135}{225}
 \psdots[dotstyle=*,linecolor=blue,dotscale= 1.5 1.5](2.47,1.5)

 \rput[Bc](0,1){$\lim_{\lambda \rightarrow \infty} \lambda^{-1}T$}
 \rput[Bc](1.5,1){$=$}
\psline[linecolor=black](2.25,0.75)(3,1.5)
\psline[linecolor=black](3,1.5)(3.75,0.75)
\psline[linecolor=black](2.25,0.75)(3,0)
\psline[linecolor=black](3,0)(3.75,0.75)

\psline[linecolor=black](3.75,0.75)(4.5,1.5)
\psline[linecolor=black](4.5,1.5)(5.25,0.75)
\psline[linecolor=black](3.75,0.75)(4.5,0)
\psline[linecolor=black](4.5,0)(5.25,0.75)

\psline[linecolor=black](5.25,0.75)(6,1.5)
\psline[linecolor=black](6,1.5)(6.75,0.75)
\psline[linecolor=black](5.25,0.75)(6,0)
\psline[linecolor=black](6,0)(6.75,0.75)

\psline[linecolor=black](6.75,0.75)(7.5,1.5)
\psline[linecolor=black](7.5,1.5)(8.25,0.75)
\psline[linecolor=black](6.75,0.75)(7.5,0)
\psline[linecolor=black](7.5,0)(8.25,0.75)

\psline[linecolor=black](3,1.5)(3.75,2.25)
\psline[linecolor=black](3.75,2.25)(4.5,1.5)
\psline[linecolor=black](4.5,1.5)(5.25,2.25)
\psline[linecolor=black](5.25,2.25)(6,1.5)
\psline[linecolor=black](6,1.5)(6.75,2.25)
\psline[linecolor=black](6.75,2.25)(7.5,1.5)

\psline[linecolor=black](7.5,1.5)(8.25,2.25)
\psline[linecolor=black](8.25,2.25)(8.25,0.75)

\psellipticarc[linecolor=blue](7.5,1.5)(0.53033,0.53033){315}{45}

\psline[linecolor=black](2.25,2.25)(3,1.5)
\psline[linecolor=black](2.25,2.25)(2.25,0.75)

\end{pspicture}
$$
and an example of loop configuration in this case is shown in Fig.~\ref{FigConfig}. The limit $\lim_{\lambda \rightarrow \infty}$
 has the advantage of reducing
the dimension of the Hilbert space while keeping the correct universality class in the continuum limit: the relevant algebra indeed is now the JS blob algebra $\mathcal{B}^{b}(2N,n,y)$.
In addition of being more
pleasant physically and more convenient numerically, we will
show in the following that this limit leads also to the correct 
algebraic object to consider for studying the correspondence with
the Virasoro algebra.

\subsection{Connections with the Virasoro algebra}\label{subsecSpec}

\paragraph{The scaling limit}

At the renormalization group fixed point $\lambda=\infty$, we expect statistical mechanics systems with transfer matrix (\ref{eq_Transfer}) to exhibit  conformal invariance at low energy, and their 
scaling limit to be described by a conformal field theory. 

In agreement with the renormalization group flow described previously,  one finds  (numerically as well as analytically~\cite{JS, Nichols1, Nichols2}) that the low energy properties 
are indeed independent of the parameter $\lambda > 0$ (or $\alpha>0$ in the Hamiltonian point of view). The central charge can be extracted from the scaling of the ground state eigenvalue 
$\lambda_0(L)$~\cite{CardyFiniteSize}, and it  takes the well known form
\begin{equation}\label{eq-c}
\displaystyle c = 1 - \frac{6}{x (x+1)},
\end{equation}
where we recall that $x$ parametrizes the loop weight $n = 2 \cos \left( \frac{\pi}{x+1} \right)$.

The subdominant eigenvalues $\lambda_{\phi}(L)$ are related to the conformal dimension $h_{\phi}$ (critical
exponent) of the operators in the underlying boundary conformal field theory
\begin{equation}
\displaystyle \lambda_{\phi}(L) \simeq \lambda_0(L) \ \mathrm{e}^{-\frac{\pi}{L} h_{\phi}}.
\end{equation}

Which exponents are obtained, and with which multiplicity, depends on the sector of the transfer matrix, and the particular representation of the blob algebra one considers: so far, we have implicitly assumed that we were dealing with a loop model, but in fact many other representations are possible, and relevant physically as well as formally. 

Restricting for now to loop models, one can compute the low-energy spectrum
in the scaling limit and identify all the exponents $h_i$ occurring in the spectrum of the transfer matrix (or the  hamiltonian) acting  in the  standard modules $\BW^{b/u}_{j}$ over the blob algebra. It is convenient to gather the results  into a generating function
(character) $\mathrm{Tr}_{\BW^{b/u}_{j}} q^{L_0-c/24} \equiv \sum_{i} q^{h_{i}-c/24}$, where the sum is taken over all low-energy states.
In the sector with no through lines, one finds~\cite{JS} 
\begin{equation}
\displaystyle Z_0 = \mathrm{Tr}_{\BW_0} q^{L_0-c/24} = \frac{q^{-c/24}}{P(q)} q^{h_{r,r}},
\end{equation}
with
\begin{equation} 
P(q) = q^{-1/24} \eta (q) = \prod_{n=1}^{\infty} (1-q^n)
\end{equation}
and
\begin{equation}\label{conf-weight}
\displaystyle h_{r,s} = \frac{ \left[(x+1)r - xs \right]^2 - 1}{4 x (x+1)}.
\end{equation}
On the other hand, the generating functions~\cite{JS} for the blobbed and unblobbed sectors read
\begin{subequations}\label{spec-bu}
\begin{eqnarray}
Z^{b}_j & = & \mathrm{Tr}_{\BW^b_j} q^{L_0-c/24} = \frac{q^{-c/24}}{P(q)} q^{h_{r,r+2j}}, \label{spec-b}\\
Z^{u}_j & = & \mathrm{Tr}_{\BW^u_j} q^{L_0-c/24} = \frac{q^{-c/24}}{P(q)} q^{h_{r,r-2j}}.\label{spec-u}
\end{eqnarray}
\end{subequations}
%As usual in CFT, the parameter $q$ may be seen as a formal parameter in the generating function of the spectrum,
%or as $q=\mathrm{e}^{-M \pi/L}$ in the computation of the universal part of the partition function of the statistical
%model defined by the transfer matrix~\eqref{eq_Transfer} on a strip $M \times L$, with $M \gg L$.

Some  important points  are worth mentioning before going further:

\begin{itemize}

\item In all these formulas involving critical exponents, we have $r \in \left( 0, x+1 \right) $. In particular, we have to keep in mind that $r$ is defined modulo $x+1$. However,
the correspondence~\eqref{spec-bu} with the conformal exponents is obviously only correct in the fundamental domain  $r \in \left( 0, x+1 \right) $. Recall nevertheless the symmetry of the conformal weights: $h_{r,s}=h_{r+x,s+x+1}$, for any real numbers $r$ and $s$. Using this symmetry we actually cover the full extended Kac table, when $r,s\in\mathbb{Z}$.

\item  Algebraically, there is a symmetry between the  blobbed and unblobbed sectors  -- realized, in practice, by swapping the loop weights 
$y_{-} = y  \rightarrow \left[2\right]_{\q} - y_{-} $ and $y_{+} = \left[2\right]_{\q} - y_{-}  \rightarrow y_{-} $ while switching the modules $\BW^b_j$ and $\BW^u_j$. 
This is because of the symmetric role of the operators $b$ and $1-b$ in the definition of the blob algebra.
This symmetry is however {\sl broken} by the choice of the transfer matrix, and the critical exponents do depend on the sign of the $b$ coupling, which we chose to be $\lambda > 0$ here. To illustrate this point, consider 
a  simple example  with $\q=\mathrm{e}^{i \pi /3}$. For $r=1$ ($y_- = 1$ and $y_+ = 0$) and $j=2$ in the blobbed sector, we find a  critical exponent $h_{1,5}=2$. For
$r=2$ ($y_- = 0$ and $y_+ = 1$) and still $j=2$, we find in the unblobbed sector a completely different exponent  $h_{2,-2}=\frac{33}{8}$.
Therefore, unblobbed and blobbed exponents {\sl are not} switched when swapping the blobbed and unblobbed loop weights.

\end{itemize}

We comment finally on the spectrum of the standard modules over the JS blob algebra. It was argued in~\cite{JS}
that, in agreement with the renormalization group flow,  the full spectrum of critical exponents obtained for $\mathcal{B}^{b}(2N,n,y)$
was the same as for $\mathcal{B}(2N,n,y)$.
 This can also be checked directly using 
our formulas~\eqref{spec-b} and~\eqref{spec-u} for the critical exponents. We first note that the change 
$y \rightarrow y^{-1}$ in the isomorphism $\mathcal{B}^{b}(2N,n,y) = b \mathcal{B}(2N,n,y) b \simeq \mathcal{B}(2N-1,n,y^{-1})$
corresponds to the change of the Kac label $r \rightarrow x-r$. Using the isomorphism of the standard modules, for $y\ne0$,
one finds that the critical exponent corresponding to $\hat{\BW}^{b}_{j}$ reads
\begin{equation*}
h_{x-r,x-r-2(j-\frac{1}{2})} = h_{x-r,x+1-r-2j} = h_{-r,-r-2j} = h_{r,r+2j}
\end{equation*}
as it is expected. An analogous argument shows that the critical exponent corresponding 
to $\hat{\BW}^{u}_{j}$ is indeed $h_{r,r-2j}$ as expected.

For the case $y=0$,  the critical exponent corresponding to ${\BW}_{j+\frac{1}{2}}$ is $h_{x,x-2j}$. Note that we have an identity $h_{x,x-2j} = h_{x,x+2(j+1)}$ compatible with the isomorphism~\eqref{BWj-def}.
The case $y=0$ will be discussed in more detail below.

\paragraph{Standard versus Verma modules}
Consider now the fully generic case, where  $\q$ is not a root of unity and $r$ is non integer. In this case, the standard modules are irreducible,
and the central charge of the corresponding CFT $c(\q)$ is also generic. It is then interesting to observe that the generating function 
\begin{equation} \label{eqcharirrVerma}
Z^{b/u}_j = \frac{q^{-c/24}}{P(q)} q^{h_{r,r \pm 2j}}
\end{equation}
coincides with  the character of the Verma module 
$\mathcal{V}_{h_{r,r \pm 2j}}$, which is also irreducible when $\q$ and $r$
are generic. 
Recall~\cite{YellowBook} that a Virasoro  Verma module $\mathcal{V}_h$ is  simply defined by letting the lowering part of the Virasoro algebra $\mathfrak{vir}^-$
to act freely on a highest weight state $\phi_h$. 
The character of the Verma module $\mathrm{Tr}_{\mathcal{V}_h} q^{L_0-c/24}$ is then given by the right-hand side of eq.~\eqref{eqcharirrVerma}
when $h=h_{r,r \pm 2j}$.

We thus observe an intriguing connection between two a priori quite different objects: standard modules for an associative algebra, and Verma modules for an (infinite dimensional) Lie algebra. 
Similar connections have been observed in other contexts. Their physical origin is that one presumably can, in the scaling limit, build expressions in terms of the lattice generators whose action, on a properly chosen subset of states (the scaling states) provides a representation of the Virasoro algebra. Except in very favorable (free of interactions) cases~\cite{KadanoffCeva,KooSaleur,GRS3} however, such a construction has never been fully completed. We will give below a sketch of how this might go for the blob algebra. 
Note also that non isomorphic standard modules over the blob algebra with a fixed value of $0<r<x+1$ give rise to non isomorphic Verma modules in the scaling limit\footnote{Indeed, recalling the isomorphism of the Verma modules  $\mathcal{V}_{h_{r,s}}$ and  $\mathcal{V}_{h_{x-r,x+1-s}}$, {\it i.e.}, for those having the same conformal weight $h_{r,s}=h_{x-r,x+1-s}$, see eq.~\eqref{conf-weight}, we  note that these Verma modules can be obtained from different standard modules $\BW_j^{b}$ and $\BW_{j-\frac{1}{2}}^u$, with $s=r+2j$. However, these are over different  blob algebras defined by different $y$-parameters $[r+1]/r$ and $[x-r+1]/[x-r]$, respectively, and defined for different parities of the total number of sites (one should fix the parity of the number of sites before taking the limit).
% -- we will take the limit for even number $L$ of sites. 
}. 

%We have thus for any central charge $c<1$ and the conformal weight of the form $h_{r,r+n}$, with $r\in\mathbb{R}$ and $n\in\mathbb{Z}$, 
%Indeed recalling the isomorphism of  Verma modules  $\mathcal{V}_{h_{r,s}} \simeq \mathcal{V}_{h_{x-r,x+1-s}}$ --- {\it i.e.}, for those having the same conformal weight $h_{r,s}=h_{x-r,x+1-s}$, see eq.~\eqref{conf-weight}, we  note that these Verma modules can be obtained from different standard modules $\BW_j^{b}$ and $\BW_{j-\frac{1}{2}}^u$, with $s=r+2j$, but these are over different  blob algebras defined by the $y$-parameters $[r+1]/r$ and $[x-r+1]/[x-r]$, respectively, and  on the total number of sites of different parities. 
%%correspond in the scaling limit to isomorphic Verma modules $\mathcal{V}_{h_{r,s}}$ and  $\mathcal{V}_{h_{x-r,x+1-s}}$, respectively. 
%This means that before taking the limit we should fix parity of the number of sites in the lattice -- we will take the limit for even number $L$ of sites. 

It is tempting and fruitful however to remain at a more global level, and wonder whether the relationship  between standard and Verma modules in the generic case carries over to degenerate cases. Of course, the correspondence between the generating functions for standard modules and the characters of Verma modules holds for non-generic values of $c$, or integer $x$, as well.  On the Virasoro side, it is well known that in these cases the Verma modules become reducible, and exhibit a more or less intricate structure of sub-modules.  The same is true for the blob algebra. The decomposition of Verma/standard modules  can be summarized by some simple data: the set of simple (irreducible) sub-modules, together with arrows indicating maps by elements of the algebra.  The key question is then, does the connection between the two algebras still hold at this more detailed level? This is the point we shall address in  the next section.

\paragraph{From the JS algebra to Virasoro generators}\label{sec:JS-Vir}

The relationship between central charge and critical exponents on the one hand,  and the scaling of the transfer matrix (Hamiltonian) eigenvalues corresponds formally to the fact that the Hamiltonian  behaves like
\begin{equation}
\displaystyle H = E_S + E_B L + \frac{\pi v_F}{L} \left( L_0 - \frac{c}{24}\right) + \mathcal{O}(L^{-2}),
\end{equation}
where $v_F = \frac{\pi \sin \gamma}{\gamma}$ is the Fermi velocity~\cite{PasquierSaleur}, $E_S$ and $E_B$ are non-universal
quantities, and $L_0$ is the usual Virasoro zero mode. We recall that $n = \q + \q^{-1}$ and $\q=\mathrm{e}^{i \gamma}$, with $\gamma = \frac{\pi}{x+1}$.

A little more precisely, the scaling limit is obtained by considering a finite number $M$
of low-energy eigenvectors of the Hamiltonian, computing the matrix elements of the desired operators
between these $M$ scaling states in the limit of large size $L \rightarrow \infty$, and then taking the limit
$M \rightarrow \infty$. This double limit procedure is necessary to obtain a quantum field theory describing
the low-energy physics of a given lattice system. In particular, we expect a certain combination of the 
local Hamiltonian densities to reproduce in the scaling limit, on this basis of low-energy scaling states,  the action of the Virasoro
generators $L_n$ . This was discussed in the context of the TL algebra in Ref.~\cite{KooSaleur},
following the pioneering work of Kadanoff and Ceva~\cite{KadanoffCeva}.

In our example of the Hamiltonian $H = -\alpha b - \sum_i e_i$, it is rather easy to come up with a conjecture
for the expression of Virasoro generators on the lattice. Recall first that one can choose safely any $\alpha>0$
while remaining in the same universality class. Second, using the isomorphism~\eqref{JS-alg-iso}
between the JS blob algebra and another blob algebra, one can rewrite the Hamiltonian as $H= - \sum_i b e_i b$,
defined on $L+1$ sites up to a switch $y \rightarrow y^{-1}$. In fact, since we have seen
that the JS version of the blob algebra $b \mathcal{B} b$ is the physically more correct object to consider,
we will work with
$H= - \sum_i b e_i b$ as our fundamental Hamiltonian defined on $L$ sites, acting on a
Hilbert space $\mathcal{H}$ where all states are in the image of the operator $b$
so that $b \mathcal{H} = \mathcal{H}$.

Actually, when acting on $\mathcal{H}$ we only need one factor of $b$ in the expression of the Hamiltonian, so that 
the ``blobbed'' Hamiltonian $H$ is simply related to the usual TL Hamiltonian $H_{\rm TL} = - \sum {e_i} $
by $H=b H_{\rm TL}$. This is because any state $\Ket{v} \in \mathcal{H}$ satisfy $b \Ket{v} = \Ket{v}$.
Meanwhile, the $b$ factor in $H=b H_{\rm TL}$ ensures that the image of $H$ is indeed in $\mathcal{H}$.
In the continuum limit, we expect the Hamiltonian to tend, up to a scaling factor, to the Virasoro
generator $L_0$. Therefore, we expect the following lattice regularization for this generator $ L^{(2N)}_0 \propto b \sum e_i$.
To obtain a lattice version of the other Virasoro generators $L^{(2N)}_n$, following~\cite{KooSaleur}, it is
useful to think in terms of an {\it anisotropic} version of our lattice model
\begin{equation}
\displaystyle T =  b \prod_{i=1}^{N-1} \left( 1 + \epsilon_y e_{2i} \right) \prod_{i=1}^{N} \left( \epsilon_x + e_{2i-1} \right) ,
\end{equation}
where
\begin{equation}
\epsilon_y = \frac{1}{\epsilon_x} = \frac{\sin u}{ \sin \left( \gamma - u\right)}.
\end{equation}
Here, $u$ parametrizes the anisotropy (spectral parameter) and $\gamma = \frac{\pi}{x+1}$.
An important point is that the derivation of the lattice generators $L^{(2N)}_n$ in~\cite{KooSaleur} relies only on a definition of the stress energy tensor $T(z)$ in terms of the straining of the lattice (we refer the reader to~\cite{KooSaleur} for more details),
which is a bulk property that we do not expect to be modified by the blob operator $b$.
This means that the lattice expression of the Virasoro generators in the blob case should take the same form 
as in~\cite{KooSaleur}, only with an additional $b$ multiplicative factor. 

%The stress energy tensor $T(z)$ is then defined in terms of the anisotropy angle $\theta$,
%meaning that the time direction has been stretched out by a factor $\tan (\theta/2)$ with respect to the space-time isotropic point
%$u = \gamma/2$. Note that $\theta$ is a bulk property that we do not expect to be modified by the blob operator $b$.
%Using the results of Ref.~\cite{KooSaleur}, we thus expect
%\begin{equation}
%\displaystyle \theta = \frac{\pi \left(\gamma - u \right)}{\gamma}.
%\end{equation}
%Therefore, the expansion of $T(z)$ in terms of $u$ is the same as in the TL case expect for the overall $b$ factor.

We thus conjecture that the Virasoro generators on the lattice in terms of the elements of $\mathcal{B}^b=b \mathcal{B} b$ take the form\footnote{ It is worth pointing out that all the operators in this expression belong to the JS blob algebra $b\mathcal{B} b$. However, the $b$ factors on the right are not strictly necessary since the Hilbert space we are working with satisfies $b \mathcal{H} = \mathcal{H}$, so we might as well use $b e_i$ instead of $b e_i b$. }
\begin{equation}
\displaystyle L^{(L=2N)}_n = \frac{L}{\pi} \left[ - \frac{1}{v_F} \sum_{i=1}^{L-1} (b e_i b - e_{\infty}) \cos \left( \frac{n i \pi}{L} \right) +  \frac{1}{v_F^2}  \sum_{i=1}^{L-2} \left[ b e_i b, b e_{i+1} b \right] \sin \left( \frac{n i \pi}{L} \right)\right] + \frac{c}{24} \delta_{n,0} b,
\label{eq_KooSaleur}
\end{equation}
where $e_{\infty}$ is the groundstate energy density $e_{\infty}=\lim_{L \rightarrow \infty } E_{0}(L)/L$ for $r=1$. 
Using Bethe ansatz methods, one can show~\cite{PottsBethe} that  
\begin{equation}
\label{eq_einf}
\displaystyle e_{\infty} = \sin^2 \gamma \int_{-\infty}^{+\infty} \frac{{\rm d} x}{\cosh \pi x} \ \frac{1}{\cosh 2 \gamma x - 2 \cos \gamma}.
\end{equation}

The $L^{(2N)}_n$ operators are expected to tend to the Virasoro modes in the double limit sense
explained above, that is, their matrix elements between scaling (low-energy) states
should converge to Virasoro operators.
In the case $r=1$,  this lattice regularization of the Virasoro algebra 
 proved  very useful to measure numerically Virasoro matrix elements in~\cite{KooSaleur},
and in particular, it was used extensively to determine
the so-called indecomposability parameters in the context of logarithmic CFTs~\cite{VJS}.
We leave the verification of this formula in other cases for future work.

\section{From the representation theory of the (JS) blob algebra to Virasoro}
\label{section3}

We now wish to study what happens to the similarities between the  standard modules over the (JS) blob algebra and the Virasoro Verma modules 
in degenerate cases. To make a long story short, we will see that everything keeps working quite beautifully. Only in the case  $y=0$ will there be a small subtlety
related to the distinction between what we called JS blob algebra and the blob 
algebra, and we shall show that only
the JS version completely matches the known results of Virasoro representation theory.
The main results of this section are gathered in Tab.~\ref{tabblobVir}.

\subsection{Singly critical cases and Kac modules}
\label{subSecSinglyCritical}

Let us turn to a slightly more involved case which is usually referred to as ``singly critical''
for the blob algebra representation theory~\cite{Martin}. Keeping $\q$ generic, let us recall the parametrization~\eqref{eqDefr} of the blob parameter $y$ and consider the case
$r \in \mathbb{N}^{*}$ integer, for which the standard modules are in general not irreducible anymore. Then for positive\footnote{The case $r$ negative is obviously obtained by switching the blob $b$
and unblobbed $u$ labels everywhere.} $r$, it is well known~\cite{MartinSaleur} that one has the following injective homomorphisms (embeddings)
\begin{subequations}
\begin{eqnarray}
\BW^{u}_{j+r} & \longhookrightarrow & \BW^{b}_{j}, \qquad  j=0, \ffrac{1}{2},1, \dots, N-r, \\
\BW^{u}_{r-j} & \longhookrightarrow & \BW^{u}_{j}, \qquad j=\ffrac{1}{2},1, \dots, \ffrac{r-1}{2}.
\end{eqnarray}
\end{subequations}
The other standard modules $ \BW^{u}_{j}$ with $j=\frac{r}{2}, \dots, N$ remain irreducible.
Because of these embeddings, we can define the quotients
$\BK^{b}_{j} \equiv \BW^{b}_{j}/\BW^{u}_{j+r}$, and similarly $\BK^{u}_{j}\equiv \BW^{u}_{j}/\BW^{u}_{r-j}$ for $j=\frac{1}{2},1,\dots, \frac{r-1}{2}$,
which are irreducible as long as $\q$ remains generic. We shall refer to these modules as  ``quotient'' modules over the blob algebra.

These quotient representations $\BK^{b}_{j}$ and $\BK^{u}_{j}$ have dimensions
\begin{subequations} 
\label{eqCharKacSingle}
\begin{eqnarray}
 \mathrm{dim} (\BK^{b}_{j}) &=& \binom{2N}{N + j} - \binom{2N}{N + j + r}, \\
 \mathrm{dim} (\BK^{u}_{j}) &=& \binom{2N}{N + j} - \binom{2N}{N - j + r},
\end{eqnarray}
\end{subequations}
and their generating functions of scaling states are given by the Virasoro characters
\begin{subequations} 
\label{eq_charBres}
\begin{eqnarray}
\mathrm{Tr}_{\BK^{b}_{j}} q^{L_0-c/24} = \frac{q^{-c/24}}{P(q)} \left( q^{h_{r,r+2j}} - q^{h_{r,-r-2j}} \right) \equiv K_{r,r+2j} ,\\
\mathrm{Tr}_{\BK^{u}_{j}} q^{L_0-c/24} = \frac{q^{-c/24}}{P(q)} \left( q^{h_{r,r-2j}} - q^{h_{r,-r+2j}} \right) \equiv K_{r,r-2j} ,
\end{eqnarray}
\end{subequations}
using~\eqref{spec-b} and~\eqref{spec-u}.

The crucial observation is that the corresponding Verma module $\mathcal{V}_{h_{r,r \pm 2j}}$ becomes also reducible
when $r$ is integer, and has a proper submodule at level $r(r\pm 2j)$ isomorphic to $\mathcal{V}_{h_{r,-r \mp 2j}}$ which is precisely 
the Verma module associated with $\BW^{u}_{r \pm j}$. We represent the structure of this Verma module by the diagram
\begin{equation}
\mathcal{V}_{h_{r,r \pm 2j}}=
\begin{array}{ccc}
h_{r,r \pm 2j} &&\\
&\hskip-.2cm\searrow&\\
&&\hskip-.3cm h_{r,-r \mp 2j},
\end{array} 
\end{equation}
where the arrow represents the action of $\mathfrak{vir}$, of $L_n$'s with $n<0$ here, and we denote irreducible subquotients by the corresponding conformal weights.  The quotient $\mathcal{V}_{h_{r,r \pm 2j}}/\mathcal{V}_{h_{r,-r \mp 2j}}$
is called a \textit{Kac module}\footnote{We note that there exist several inequivalent definitions of Kac modules in the literature. One of them uses quotients of Verma modules, the convention chosen here, and another uses quotients (and submodules) of Feigin--Fuchs modules, see e.g.~\cite{JR}.} and we denote it by $\VK_{r,s}\equiv\mathcal{V}_{h_{r,s}}/\mathcal{V}_{h_{r,-s}}$. This module is irreducible when the central charge $c$ is generic. Its character 
\begin{equation}
\mathrm{Tr}_{\Kac{r,r\pm2j}} q^{L_0-c/24} = q^{-c/24} \frac{q^{h_{r,r \pm 2j}}-q^{h_{r,-r \mp 2j}}}{P(q)},
\end{equation}
corresponds exactly to the generating functions~\eqref{eq_charBres}.

An explicit construction of these quotient modules in a basis was given in~\cite{Martin} and~\cite{JS} (see also~\cite{SaleurBauer, TheseJerome}). Following~\cite{JS}, they can be described in terms of $r-1$ ``ghost'' strings added to the left of the usual Temperley Lieb standard modules with $2j$ through lines. Using the so-called $\q$-symmetrizer operator~\cite{JS} (closely related to the Jones-Wenzl projectors that we will introduce in the next paragraph) and the usual TL generators, it is possible to define representations that coincide with these quotients.
We remark that this construction is related to the $(r,s)$ integrable boundary conditions of Ref.~\cite{PRZ},
that uses similar projectors, although the models there correspond rather to a {\it fusion}
of blob modules that we shall not discuss here.

\paragraph{The quotient blob algebra}\label{sec:quot-blob}

To be more precise, it is also useful to define a quotient of the blob algebra that will have the quotient modules $\BK^{b/u}_{j}$ as standard (cell) modules.
We follow here Ref.~\cite{Martin}, 
and introduce a generalization of the Jones-Wenzl projectors to the blob algebra, that is, we  construct  projectors $P^{u}_k$ on the unblobbed sector with number of through lines  $j\geq k$, defined recursively by
\begin{subequations} \label{eqJonesWenzl}
\begin{eqnarray}
P^{u}_1 & = & 1-b, \\
%P^{u}_{k+1} & = &  P^{u}_{k} - \frac{\left[ k-1+m_+ \right]_{\q} }{\left[ k+m_+ \right]_{\q}} P^{u}_{k} e_k P^{u}_{k} ,
P^{u}_{k+1} & = &  P^{u}_{k} - \frac{\left[ k-1-r \right]_{\q} }{\left[ k-r \right]_{\q}} P^{u}_{k} e_k P^{u}_{k}.
\end{eqnarray}
\end{subequations}
We then introduce the quotient (only for $r$  positive integer) algebra\footnote{Note that for $r<0$, we could introduce similar projectors $P^b_{-r}$ onto blobbed sectors with  number of through lines  $j\geq -r$.}
\begin{equation}\label{eqDefRestrictedblob}
\mathcal{B}^{\rm res}(2N,n,y) \equiv \mathcal{B}(2N,n,y) / P^{u}_{r} , 
\end{equation}
which is nothing but the
blob algebra where the ideal spanned by the projector $P^{u}_{r}$ has been quotiented out. We shall call this algebra \textit{the \res~blob algebra}, and we will
see that the representation theory of this algebra is  simpler than that of $\mathcal{B}(2N,n,y)$. In particular, the projective  modules (see next section for some reminders on this important concept) over $\mathcal{B}^{\rm res}(2N,n,y)$
can be at worse of  diamond shape whereas projective covers for $\mathcal{B}(2N,n,y)$ are much more complicated. 

Taking the quotient~\eqref{eqDefRestrictedblob} by the two-sided ideal generated by $P^{u}_{r}$ means that we remove all modules (and subquotients) isomorphic to $\BW^u_{j\geq r}$.
Then note that $\mathcal{B}^{\rm res}(2N,n,n)$ is isomorphic
to the Temperley--Lieb algebra $\mathrm{TL}(2N,n)$, as can be readily shown ($r=1$ in this case) by setting to zero all the diagrams that contain the antiblob symbol. The quotient blob algebra $\mathcal{B}^{\rm res}(2N,n,y)$ is still a cellular algebra and its standard modules are given precisely 
by $\BK^{b}_{j}$ (and $\BK^{u}_{j}$, for $j=1,$ \dots$, \frac{r-1}{2}$ ). 
In particular, one can show using the RSOS picture that 
the projector $P^{u}_{r}$ acts as $0$ in the ghost string representation~\cite{TheseJerome} (for a definition of RSOS models, see {\it e.g.}~\cite{MartinBook}).

\subsection{Doubly critical cases: blob algebra representation theory at $\q=\mathrm{e}^{i\pi/3}$ and $r \in \mathbb{N}^{*}$}

Finally, we turn to the most interesting case, with $\q$ root of unity or $x$ integer (which is 
necessary to describe most of the physically relevant examples), and $r \geq 0$ integer in the fundamental
domain $r=1,\dots,x$, so the rational conformal weights lie within the (extended) Kac table.
The representation theory of the blob algebra is much richer is this case. For simplicity, we shall focus
on the example of boundary Percolation ($\q=\mathrm{e}^{i\pi/3}$, $c=0$).
The generalization to other roots of unity is straightforward. In this section, we will follow closely~\cite{Martin}\footnote{Note that when $x$ is integer, the parameter $m_-=-r \mod (x+1)$ is defined modulo $x+1$. In order to be consistent with~\cite{Martin}, we fix its value as $m_-= x+1-r$, see section~7 in~\cite{Martin}.}.

\subsubsection{Case $r=1$}
\label{sec:st-mod-r1}

\begin{figure}
\begin{center}
\begin{pspicture}(0,0)(10,2)
\psset{xunit=10mm,yunit=10mm}
%\psgrid[subgriddiv=1,griddots=10,gridlabels=10pt](0,0)(10,2)

\psline[linecolor=black,linewidth=0.5pt,arrowsize=5pt]{-}(0,1)(10.0,1.0) 

 \rput[Bc](0,1.5){$-4$}
 \rput[Bc](1,1.5){$-3$}
 \rput[Bc](2,1.5){$-2$}
 \rput[Bc](3,1.5){$-1$}   
 \rput[Bc](4,1.5){$0$}   
 \rput[Bc](5,1.5){$1$}   
 \rput[Bc](6,1.5){$2$}   
 \rput[Bc](7,1.5){$3$}   
 \rput[Bc](8,1.5){$4$}   
 \rput[Bc](9,1.5){$5$}   
 \rput[Bc](10,1.5){$6$}   
 
 \rput[Bc](0,2){$\BW^{b}_{2}$}
 \rput[Bc](1,2){$\BW^{b}_{\frac{3}{2}}$}
 \rput[Bc](2,2){$\BW^{b}_{1}$}
 \rput[Bc](3,2){$\BW^{b}_{\frac{1}{2}}$}   
 \rput[Bc](4,2){$\BW_{0}$}   
 \rput[Bc](5,2){$\BW^{u}_{\frac{1}{2}}$}   
 \rput[Bc](6,2){$\BW^{u}_{1}$}   
 \rput[Bc](7,2){$\BW^{u}_{\frac{3}{2}}$}   
 \rput[Bc](8,2){$\BW^{u}_{2}$}   
 \rput[Bc](9,2){$\BW^{u}_{\frac{5}{2}}$}   
 \rput[Bc](10,2){$\BW^{u}_{3}$}    

 \rput[Bc](10.5,1.0){$\dots$}   
 \rput[Bc](-0.5,1.0){$\dots$}  
  
\psline[linecolor=black,linewidth=0.5pt,arrowsize=5pt]{-}(0,0.5)(0,1.0) 
\psline[linecolor=black,linewidth=0.5pt,arrowsize=5pt]{-}(1,0.5)(1,1.0) 
\psline[linecolor=red,linewidth=1.5pt,arrowsize=5pt]{-}(2,1)(2,0) 
\psline[linecolor=black,linewidth=0.5pt,arrowsize=5pt]{-}(3,0.5)(3,1.0) 
\psline[linecolor=black,linewidth=0.5pt,arrowsize=5pt]{-}(4,0.5)(4,1.0) 
\psline[linecolor=red,linewidth=1.5pt,arrowsize=5pt]{-}(5,1)(5,0) 
\psline[linecolor=black,linewidth=0.5pt,arrowsize=5pt]{-}(6,0.5)(6,1.0) 
\psline[linecolor=black,linewidth=0.5pt,arrowsize=5pt]{-}(7,0.5)(7,1.0) 
\psline[linecolor=red,linewidth=1.5pt,arrowsize=5pt]{-}(8,1)(8,0) 
\psline[linecolor=black,linewidth=0.5pt,arrowsize=5pt]{-}(9,0.5)(9,1.0) 
\psline[linecolor=black,linewidth=0.5pt,arrowsize=5pt]{-}(10,0.5)(10,1.0) 

\end{pspicture}
\end{center}
  \caption{Weight diagram for $\q=\mathrm{e}^{i \pi/3}$ and $r=1$. The fundamental alcove is $\left(-2,1 \right)$,
   the structure of the standard modules is given by reflexion through the walls
  (see text and Ref.~\cite{Martin} for more details).}
  \label{figpercoWeyl}
\end{figure}
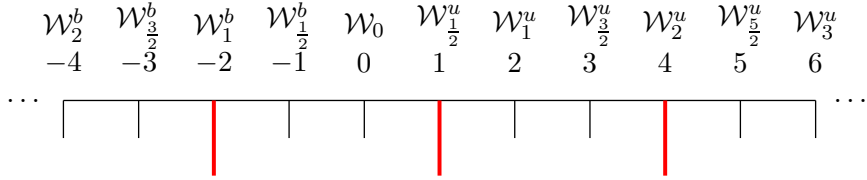

\begin{figure}
\begin{center}
\begin{pspicture}(0,0)(10,3)
\psset{xunit=10mm,yunit=10mm}
%\psgrid[subgriddiv=1,griddots=10,gridlabels=10pt](0,0)(10,3)

 \rput[Bc](-0.5,3){$\BX^{u}_{\frac{1}{2}}$}
\psline[linecolor=black,linewidth=0.5pt,arrowsize=5pt]{->}(0,3)(1.5,3) 
 \rput[Bc](2,3){$\BX^{b}_{\frac{5}{2}}$}
\psline[linecolor=black,linewidth=0.5pt,arrowsize=5pt]{->}(2.5,3)(4,3) 
 \rput[Bc](4.5,3){$\BX^{u}_{\frac{7}{2}}$}
\psline[linecolor=black,linewidth=0.5pt,arrowsize=5pt]{->}(5,3)(6.5,3) 
 \rput[Bc](7,3){$\BX^{b}_{\frac{11}{2}}$}
\psline[linecolor=black,linewidth=0.5pt,arrowsize=5pt]{->}(7.5,3)(9,3) 
 \rput[Bc](9.5,3){$\dots$}
 
  \rput[Bc](-0.5,1){$h_{1,3}$}
\psline[linecolor=black,linewidth=0.5pt,arrowsize=5pt]{->}(0,1)(1.5,1) 
 \rput[Bc](2,1){$h_{1,6}$}
\psline[linecolor=black,linewidth=0.5pt,arrowsize=5pt]{->}(2.5,1)(4,1) 
 \rput[Bc](4.5,1){$h_{1,9}$}
\psline[linecolor=black,linewidth=0.5pt,arrowsize=5pt]{->}(5,1)(6.5,1) 
 \rput[Bc](7,1){$h_{1,12}$}
\psline[linecolor=black,linewidth=0.5pt,arrowsize=5pt]{->}(7.5,1)(9,1) 
 \rput[Bc](9.5,1){$\dots$}
 
\end{pspicture}
\end{center}
  \caption{Subquotient structure of $\BW^{u}_{\frac{1}{2}}$ of $\mathcal{B}(2N,n,y)$ (or $\mathcal{B}^{b}(2N,n,y)$) for $r=1$ and $\q=\mathrm{e}^{i\pi/3}$.
  The continuum limit gives the correct structure for the associated Verma module (using the symmetries of the Kac table).}
  \label{figStd_perco_2}
\end{figure}

The structure of the standard modules of $\mathcal{B}(2N,n,y)$ is described in~\cite{Martin}, we will work them out explicitly here.
Let us begin with the case $r=1$ which corresponds to giving the weight $y=n=\left[ 2 \right]_{\q}$ to blobbed loops.
We shall see that everything goes over smoothly to Virasoro here, but the situation will be more intricate later.

The case $r=1$ may seem trivial as it seems to reduce to the ordinary Temperley--Lieb case. However, one must
 remember that the blob algebra is in fact  bigger, and reduces to the TL algebra only after taking
a quotient by the ideal generated by the Jones--Wenzl projector $P^u_1$.   
The structure of the standard modules can be obtained from a weight diagram~\cite{Martin}, similar to those
encountered in Lie algebra theory.
The weight diagram in Fig.~\ref{figpercoWeyl} is constructed with walls at $r (\mathrm{mod} \ 3)$, and we call the fundamental alcove the domain $\left(-2,1 \right)$ containing the origin.
There is a correspondence between the integers $p$ on the weight diagram and the standard modules as $\left| p\right|$ is the number $2j$ of through lines in the 
blobbed ({\it resp.}, unblobbed) sector if $p<0$ ({\it resp.}, $p>0$). The structure of the standard modules in the fundamental alcove is given by reflexions through the walls  which lie outside the interval defined by  the weight and the origin  (see~\cite{Martin}). These (allowable) reflections correspond to embeddings of the standard modules. To see how reflection works, let us work out a simple example of structure. Using Fig.~\ref{figpercoWeyl}, one can see that 
the node $p=0$ (that corresponds to the module $\BW_{0}$) gives rise, after reflections with respect to the  walls at $p=-2$ and $p=1$, to two nodes at $p=-4$ ($\BW^b_{2}$) and $p=2$ ($\BW^u_{1}$). This means that $\BW_{0}$ has two maximal proper submodules isomorphic to $\BW^b_{2}$ and $\BW^u_{1}$. The latter are themselves reducible, 
and their structure can be worked out from the weight diagram in a similar way: the allowable reflections acting on $p=-4$ give two nodes at $p=\pm6$ ($\BW^{b/u}_{3}$) and the same for $p=2$.  
Finally, the structure of the vacuum sector ($p=0$) is given in Fig.~\ref{figStd_perco_1}.
We denote $\BX^{b/u}_{j}$ the simple module (subquotient) over the blob algebra. Once again, the arrows represent the action of the algebra.
Of course, when $N$ is finite, at some point the braided ladder structure will stop. The structure of $\BW^{b}_{\frac{1}{2}}$ ($p=-1$) has similar  braided ladder structure. The other modules, corresponding to weights outside the fundamental alcove (and not on walls), are submodules of those in the fundamental alcove. Their structure can be readily deduced following the rule: only arrows emanating from the corresponding simple node (subquotient) matter, that is to say, we discard
all the nodes that are not descendants of it.

The dimension of the simple modules over the blob algebra can be computed by elementary calculations using the embedding diagrams for the standard modules; for example, using the  ladder (or ``braided'') structure of $\BW_0$ we find by alternating subtractions
\begin{align}
{\rm dim} \BX_0 &= {\rm dim} \BW_0 - {\rm dim} \BW^b_2 - {\rm dim} \BW^u_1 + {\rm dim} \BW^b_3 + {\rm dim} \BW^u_3 -\dots \notag \\
   &  = \sum_{p=0}^{\infty} \left[2 \binom{2N}{N - 3 p} - \binom{2N}{N - 3 p -1} - \binom{2N}{N - 3 p - 2} \right] - \binom{2N}{N} \notag\\
   &= 1.
\end{align}
The ladder structure of $\BW_0$ is of course
reminiscent of the Virasoro algebra representation theory.
The corresponding Verma module of the identity operator is also
shown in Fig.~\ref{figStd_perco_1}. We see again in this example
that the structure of the module given by the blob algebra representation
theory is in perfect agreement with what is expected on the 
Virasoro side. 
In particular, we recover in the scaling limit the characters given by the Rocha-Caridi formula~\cite{RochaCaridi,YellowBook},
with for example the generating function of levels for the simple module $\BX_0$    
\begin{align}
\chi_{1,1} &= \frac{1}{P(q)} \left(1-q-q^{2}+q^{5}+q^{7} - \dots \right) \notag \\
   &  = \sum_{n \in \mathbb{Z}}  \frac{q^{(12 n-1)^2/24}-q^{(12 n+7)^2/24}}{\eta(q)} = 1,
\end{align}
where we have used the Euler pentagonal identity.
Note that for $1\leq r\leq x-1$ the lattice formulation of a restriction to a given simple blob module, with $j\leq\frac{x-r}{2}$ for the blobbed sector and with $j\leq\frac{r-1}{2}$ in the unblobbed one, already exists and 
can be obtained by considering different boundary conditions in RSOS models~\cite{SaleurBauer,TheseJerome}. These boundary conditions correspond to the Virasoro content of the usual minimal models.  
The structure of $\BW^{b}_{\frac{1}{2}}$ ($p=-1$) gives the same pattern in the scaling limit\footnote{Actually, this module gives in the limit the Verma module $\Verma{h_{1,2}}$ isomorphic to $\Verma{h_{1,1}}$. We thus find that two different modules  $\BW^{b}_{\frac{1}{2}}$ and  $\BW_{0}$ give in the limit the same Verma module. Note however that these standard modules are actually over  different blob algebras defined for different (odd and even, respectively) number of sites. 
%, see also the discussion about the general correspondence between standard and Verma  modules in section~\ref{subsecSpec}. 
% Before taking the limit one should fix parity of $L$
 %the number of sites -- we will take the limit for even number $L$ of sites. 
}.

 We now describe the structure of standard modules with weights on a wall. Using the allowable reflections we see that the standard modules have a chain structure.
For example, the case $p = 1 (\mathrm{mod} \ 3)$ corresponds to a representation on a wall\footnote{This
module will only appear in odd length systems, as it has $j$ half integer.} and the structure of $\BW^{u}_{\frac{1}{2}}$ ($p=1$) is given in Fig.~\ref{figStd_perco_2}. 
The dimension of the simple modules can be obtained from a single subtraction in this case.
Once again, the results are completely consistent with what is expected from Virasoro representation theory.
The generating function of the simple module $\BX^{u}_{\frac{1}{2}}$ reads
\begin{equation}
\displaystyle \mathrm{Tr}_{\BX^{u}_{\frac{1}{2}}} q^{L_0-c/24}  = \frac{q^{h_{1,0}}-q^{h_{1,6}}}{P(q)}= \frac{q^{h_{1,3}}-q^{h_{1,-3}}}{P(q)} \equiv \chi_{1,3},
\end{equation}
which agrees with the general  formula for Virasoro characters of irreducible representations at the walls:
\begin{equation}
\displaystyle \chi_{1,3+3k} = \frac{q^{h_{1,3k+3}}-q^{h_{1,-3k-3}}}{P(q)}.
\end{equation}
We use here the standard notation $\chi_{r,s}$ for the characters of  irreducible Virasoro representations of weight $h_{r,s}$.

Following the same line of thought, it is easy to obtain the structure of all the standard modules $\BW^{b/u}_j$,
and we find that their corresponding diagrams are identical (in the limit) to the corresponding Verma modules 
$\Verma{1,1\pm2j}$ over the Virasoro algebra at $c=0$.   The important point is that {\sl non-isomorphic}  standard modules
% (over the same algebra of course) 
 give rise to {\sl non-isomorphic}  Verma modules in the scaling limit for a fixed parity of $L$.
We also see on this example $r=1$ that all blob simple modules have a scaling limit described
by Virasoro simple modules. 
Therefore, not only does the representation theory of the blob algebra share most features of  that of the Virasoro algebra, but
the character identities we derived also show that the (scaling limit of the) spectrum of the transfer matrix 
acting on simple (irreducible) blob modules leads exactly to characters of irreducible
Virasoro modules. This  suggests  that the connection
between the two algebras is actually deeper, an aspect we now study further.

 Using the technique we have just described,  
we actually found for any $y \neq 0$ a perfect match between blob algebra results and Virasoro. We also refer the reader to App.~C where the general structure of the standard modules is given along with the dimensions of a few simple modules. Using the algebras isomorphism
 $ \mathcal{B}^{b}(2N,n,y) \simeq \mathcal{B}(2N-1,n,y^{-1})$ discussed in Sec.~\ref{subsecSecJSblob}, we  can of  course rederive the results
%, {\it i.e.} blob $\leftrightarrow \mathfrak{vir}$, 
for the JS blob algebra, which will read the same, after putting hats on top on every module.
%We therefore find 
%a correspondence between the representation theories of the JS algebra and the Virasoro.

\subsubsection{Case $r=2$ (or $y=0$)}
\label{case-2}

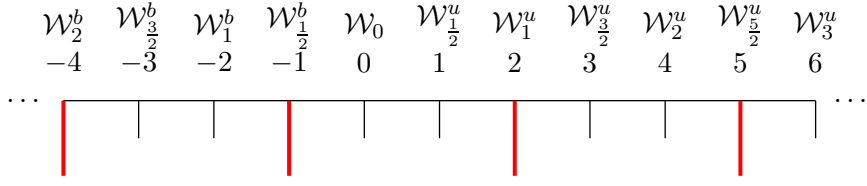
\begin{figure}
\begin{center}
\begin{pspicture}(0,0)(10,2)
\psset{xunit=10mm,yunit=10mm}
%\psgrid[subgriddiv=1,griddots=10,gridlabels=10pt](0,0)(10,2)

\psline[linecolor=black,linewidth=0.5pt,arrowsize=5pt]{-}(0,1)(10.0,1.0) 

 \rput[Bc](0,1.5){$-4$}
 \rput[Bc](1,1.5){$-3$}
 \rput[Bc](2,1.5){$-2$}
 \rput[Bc](3,1.5){$-1$}   
 \rput[Bc](4,1.5){$0$}   
 \rput[Bc](5,1.5){$1$}   
 \rput[Bc](6,1.5){$2$}   
 \rput[Bc](7,1.5){$3$}   
 \rput[Bc](8,1.5){$4$}   
 \rput[Bc](9,1.5){$5$}   
 \rput[Bc](10,1.5){$6$}   
 
  \rput[Bc](0,2){$\BW^{b}_{2}$}
 \rput[Bc](1,2){$\BW^{b}_{\frac{3}{2}}$}
 \rput[Bc](2,2){$\BW^{b}_{1}$}
 \rput[Bc](3,2){$\BW^{b}_{\frac{1}{2}}$}   
 \rput[Bc](4,2){$\BW_{0}$}   
 \rput[Bc](5,2){$\BW^{u}_{\frac{1}{2}}$}   
 \rput[Bc](6,2){$\BW^{u}_{1}$}   
 \rput[Bc](7,2){$\BW^{u}_{\frac{3}{2}}$}   
 \rput[Bc](8,2){$\BW^{u}_{2}$}   
 \rput[Bc](9,2){$\BW^{u}_{\frac{5}{2}}$}   
 \rput[Bc](10,2){$\BW^{u}_{3}$}   

 \rput[Bc](10.5,1.0){$\dots$}   
 \rput[Bc](-0.5,1.0){$\dots$}  
  
\psline[linecolor=red,linewidth=1.5pt,arrowsize=5pt]{-}(0,0)(0,1.0) 
\psline[linecolor=black,linewidth=0.5pt,arrowsize=5pt]{-}(1,0.5)(1,1.0) 
\psline[linecolor=black,linewidth=0.5pt,arrowsize=5pt]{-}(2,0.5)(2,1.0) 
\psline[linecolor=red,linewidth=1.5pt,arrowsize=5pt]{-}(3,0)(3,1.0) 
\psline[linecolor=black,linewidth=0.5pt,arrowsize=5pt]{-}(4,0.5)(4,1.0) 
\psline[linecolor=black,linewidth=0.5pt,arrowsize=5pt]{-}(5,0.5)(5,1.0) 
\psline[linecolor=red,linewidth=1.5pt,arrowsize=5pt]{-}(6,0)(6,1.0) 
\psline[linecolor=black,linewidth=0.5pt,arrowsize=5pt]{-}(7,0.5)(7,1.0) 
\psline[linecolor=black,linewidth=0.5pt,arrowsize=5pt]{-}(8,0.5)(8,1.0) 
\psline[linecolor=red,linewidth=1.5pt,arrowsize=5pt]{-}(9,0)(9,1.0) 
\psline[linecolor=black,linewidth=0.5pt,arrowsize=5pt]{-}(10,0.5)(10,1.0) 

\end{pspicture}
\end{center}
  \caption{Weight diagram for $\q=\mathrm{e}^{i \pi/3}$ and $r=2$.}
  \label{figpercoWeyl2}
\end{figure}

\begin{figure}
\begin{center}
\begin{pspicture}(0,0)(10,4)
\psset{xunit=10mm,yunit=10mm}
%\psgrid[subgriddiv=1,griddots=10,gridlabels=10pt](0,0)(10,3)

 \rput[Bc](-0.5,4){$\BX^{b}_{\frac{1}{2}}$}
\psline[linecolor=black,linewidth=0.5pt,arrowsize=5pt]{->}(0,4)(1.5,4) 
 \rput[Bc](2,4){$\BX^{u}_{\frac{5}{2}}$}
\psline[linecolor=black,linewidth=0.5pt,arrowsize=5pt]{->}(2.5,4)(4,4) 
 \rput[Bc](4.5,4){$\BX^{b}_{\frac{7}{2}}$}
\psline[linecolor=black,linewidth=0.5pt,arrowsize=5pt]{->}(5,4)(6.5,4) 
 \rput[Bc](7,4){$\BX^{u}_{\frac{11}{2}}$}
\psline[linecolor=black,linewidth=0.5pt,arrowsize=5pt]{->}(7.5,4)(9,4) 
 \rput[Bc](9.5,4){$\dots$}

  \rput[Bc](-0.5,2){$\hat{\BX^{b}_{\frac{1}{2}}}$}
\psline[linecolor=black,linewidth=0.5pt,arrowsize=5pt]{->}(0,2)(1.5,2) 
 \rput[Bc](2.75,2){$\hat{\BX^{u}_{\frac{5}{2}}} \simeq \hat{\BX^{b}_{\frac{7}{2}}}$}
\psline[linecolor=black,linewidth=0.5pt,arrowsize=5pt]{->}(4,2)(5.5,2) 
 \rput[Bc](6.75,2){$\hat{\BX^{u}_{\frac{11}{2}}} \simeq \hat{\BX^{b}_{\frac{13}{2}}}$}
\psline[linecolor=black,linewidth=0.5pt,arrowsize=5pt]{->}(8,2)(9.5,2) 
 \rput[Bc](10,2){$\dots$}
  
  \rput[Bc](-0.5,0){$h_{2,3}$}
\psline[linecolor=black,linewidth=0.5pt,arrowsize=5pt]{->}(0,0)(1.5,0) 
 \rput[Bc](2.75,0){$h_{2,-3}=h_{2,9}$}
\psline[linecolor=black,linewidth=0.5pt,arrowsize=5pt]{->}(4,0)(5.5,0) 
 \rput[Bc](6.75,0){$h_{2,-9}=h_{2,15}$}
\psline[linecolor=black,linewidth=0.5pt,arrowsize=5pt]{->}(8,0)(9.5,0) 
 \rput[Bc](10,0){$\dots$}
 
\end{pspicture}
\end{center}
  \caption{Subquotient structure of $\BW^{b}_{\frac{1}{2}}$ for $\mathcal{B}(2N,n,y)$ with $r=2$ and $\q=\mathrm{e}^{i\pi/3}$.
  The structure of the same module for the algebra $\mathcal{B}^{b}(2N,n,y)$ is slightly different as some modules in the previous picture are isomorphic.
  We also show the corresponding Verma module.}
  \label{figStd_perco_3}
\end{figure}

\begin{figure}
\begin{center}
\includegraphics[scale=0.82]{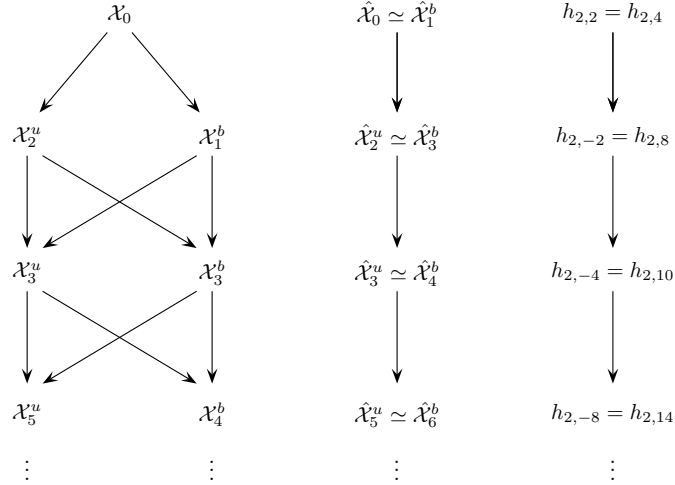}
\end{center}
  \caption{Structure of the standard module $\BW_{0}$ for $\mathcal{B}(2N,n,y)$ with $r=2$ and $\q=\mathrm{e}^{i\pi/3}$ (left).
  When ``restricted'' to the algebra $\mathcal{B}^b(2N,n,y)$ (that is, after acting with the operator $b$), the module $\hat{\BW}_{0}= b \BW_{0}$ has a single chain structure (middle).
  The resulting structure for the Verma module is given by a chain of embeddings (right). }
  \label{figStd_perco_4}
\end{figure}

The case $r=2$ is slightly more complicated, and the difference between the two algebras $\mathcal{B}$ and $\mathcal{B}^b$ will become relevant.
The weight diagram is given in Fig.~\ref{figpercoWeyl2}. Note that this case corresponds to $y=0$ and $\left[ 2 \right]_{\q} - y = \left[ 2 \right]_{\q} $,
so it corresponds to a switch in the weights of the blobbed and unblobbed loops with respect to the case $r=1$. 
The representation theory is consistent with that observation,
since this case is related to the $r=1$ case by the transformation $b \rightarrow u$ and $u \rightarrow b$ everywhere.
Note that the Virasoro representation theory for $r=2$ is less degenerate: 
for example there are no ladder/braid modules in this case (because $x=2 \left| \right. r = 2$).
Having $y=0$ also corresponds to the case where the JS blob algebra $\mathcal{B}^b$ has a less degenerate 
representation theory than that of the blob algebra as discussed in Sec.~\ref{section1}.
We shall see that this turns out to be of crucial importance when 
the results are compared with Virasoro.
%whereas there are definitely some standard modules with a braided structure for the blob algebra. 

Let us start with the case $p=-1$ which corresponds to a representation sitting on a wall in the weight diagram.
The chain structure for the blob algebra is described in Fig.~\ref{figStd_perco_3}.
If we try to go and take the continuum limit,
different simple nodes in the module $\BW^{b}_{\frac{1}{2}}$ correspond to the same conformal exponents ({\it e.g.} $\BW^{u}_{\frac{5}{2}}$ and  $\BW^{b}_{\frac{7}{2}}$ 
both have $h_{2,-3}=h_{2,9}$).  The corresponding Verma module also has a chain structure but without these pairs
of isomorphic nodes.
In fact, this happens because the JS blob algebra  $\mathcal{B}^b$, which is physically the proper algebra to consider, has a distinguished representation theory at the point $y=0$, as
 discussed in Sec.~\ref{section1}. Using isomorphisms described below eq.~\eqref{eqIsoCaseGeom}, embeddings like $\BW^{b}_{\frac{7}{2}} \longhookrightarrow \BW^{u}_{\frac{5}{2}}$ for the blob algebra
actually become isomorphisms when the corresponding modules over the JS blob algebra are considered:
for example, $\hat{\BW}^{u}_{\frac{5}{2}} \simeq \hat{\BW}^{b}_{\frac{7}{2}}$. 
The structure of standard modules over the JS algebra  $\mathcal{B}^b$ can then be easily  obtained 
using  these isomorphisms. 
The  answer is given in Fig.~\ref{figStd_perco_3} along with the associated Verma module. 
The generating function of the associated simple modules reads
\begin{equation}
\displaystyle \mathrm{Tr}_{\hat{\BX}^{b}_{\frac{1}{2}+3 k}} q^{L_0-c/24}  = \frac{q^{6 k^2}-q^{6 (k+1)^2}}{\eta(q)} =\frac{q^{h_{2,3+6k}}-q^{h_{2,-3-6k}}}{P(q)} \equiv \chi_{2,3+6k},
\end{equation}
which is the character of the irreducible Virasoro module with conformal weight $h_{2,3+6 k}$ indeed.

The case $p=0$ is given in Fig.~\ref{figStd_perco_4}. We find a braided structure for the full algebra $\mathcal{B}$, but once again 
 using the results of Sec.~\ref{section1}, we see that 
the embeddings for the blob algebra modules $\BW^{b}_{1}  \longhookrightarrow \BW_{0}$, $\BW^{b}_{3} \longhookrightarrow \BW^u_{2}$,  {\it etc.},
become isomorphisms for the JS blob algebra modules $\hat{\BW}_{0} \simeq \hat{\BW}^{b}_{1}$, $\hat{\BW}^u_{2} \simeq \hat{\BW}^{b}_{3}$, {\it etc.}, see eq.~\eqref{eqIsoCaseGeom}.
Note that in this case we now have a full chain of embeddings for JS algebra standard modules as
\begin{equation}
\hat{\BW}_{0} \hookleftarrow\hat{\BW}^u_{2}  \hookleftarrow\hat{\BW}^b_{4}  \hookleftarrow\dots
\end{equation}
which means that each term in this chain has the subquotient structure of the chain type and not the braid one.
The standard module $\hat{\BW}_{0} = b \BW_{0}$ thus has a chain structure as shown in Fig.~\ref{figStd_perco_4}.
Using now the structure for the standard modules over the JS algebra, we can deduce the generating functions of the associated conformal weights in the  scaling  limit. These functions of the simple JS algebra modules coincide with
known results from Virasoro representation theory again~\cite{Characters}
\begin{subequations}
\begin{eqnarray}
 \mathrm{Tr}_{\hat{\BX}_{0}} q^{L_0-c/24} & = & \frac{q^{h_{2,2}}-q^{h_{2,-2}}}{P(q)} = \chi_{2,2}, \\
   \mathrm{Tr}_{\hat{\BX}^{b}_{3k+1}} q^{L_0-c/24}  & = & \frac{q^{(12k-2)^2/24}-q^{(12k+2)^2/24}}{\eta(q)}=\chi_{2,2+6k}, \\
 \mathrm{Tr}_{\hat{\BX}^{b}_{3k+2}} q^{L_0-c/24} & = & \frac{q^{(12k+2)^2/24}-q^{(12k+10)^2/24}}{\eta(q)} = \chi_{2,4+6k}, \quad k>0.
\end{eqnarray}
\end{subequations}
We also remark that the structure of the standard modules over the JS algebra  $\mathcal{B}^b(2N,n,y=0)$ 
can be deduced using the results of~\cite{GL} (see Thm.~6.1 and Cor.~6.15), where the representation theory of the periodic/affine TL algebra~\cite{GL2} is combined with the so-called braid translator  (a homomorphism from the affine TL to the blob algebra).
We also refer to App.~C where some general results on the dimensions of simple modules over the JS blob algebra are presented.

Note finally that our results for the JS blob algebra can be checked using the scalar product introduced in Sec.~\ref{subsecblobalgeRep}
and the concept of Gram matrix. The Gram matrix $G=\Braket{., .}$ in a given standard module in defined by the (symmetric) matrix of scalar products $G_{i,j}=\Braket{v_i,v_j}$
between the basis elements of the module. The study of this matrix turns out to be very convenient to understand the representation theory of the blob algebra,
just like the study of the Virasoro bilinear form (for which $L_{n}^\dag = L_{-n}$) and the Kac determinant are crucial in Virasoro representation theory.
In the case of the JS blob algebra for $y=0$, the symmetric bilinear form in the representation $\hat{\BW}^u_{j} \simeq \hat{\BW}^b_{j+1}$
can be computed in the module $\hat{\BW}^b_{j+1} \equiv b \BW^b_{j+1}$  using the rules given in section~\ref{subsecblobalgeRep}.
Using this symmetric, generically non degenerate, bilinear form,
one can compute the Gram matrix in any JS blob standard modules to check our results. For example, one can easily show that the dimension
of the  simple quotient $\hat{\BX}_j$ in a standard module $\hat{\BW}_j$ is given by the rank of the Gram matrix $\mathrm{dim} \ \hat{\BX}_j = \mathrm{dim} \ \hat{\BW}_j - \mathrm{Ker} \ G = \mathrm{rank}  \ G $.   Thanks to this relation, we computed the dimension of all the simple subquotients in Fig.~\ref{figStd_perco_4} for various sizes $L$, and checked
that the results are consistent with a single chain structure, that is, we found $\mathrm{dim} \ \hat{\BX}_0 = \mathrm{dim} \ \hat{\BW}_0 - \mathrm{dim} \ \hat{\BW}_3^b $,  
 $\mathrm{dim} \ \hat{\BX}_3^b = \mathrm{dim} \ \hat{\BW}_3^b - \mathrm{dim} \ \hat{\BW}_4^b $, {\it etc}. 
 This provides an additional check that the structure of the modules for the JS blob algebra is definitely different from the structure given by the blob algebra in the case $y=0$.

\subsection{The \res~blob algebra standard modules at $\q$ a root of unity and Kac modules}
\label{subsecRestric-stand}

We now turn to the representation theory of the quotient algebra $\mathcal{B}^{\rm res}(2N,n,y) \equiv \mathcal{B}(2N,n,y) / P^{u}_{r} $, with $r$ integer and $\q$ root of unity. We shall focus on the examples with $r=1,2$ and $\q=\mathrm{e}^{i\pi/3}$. We also recall that $r \in \left( 0, x+1 \right) $ by definition, so $r=1,2$ are the only allowed integer values for $\q=\mathrm{e}^{i\pi/3}$.
The standard modules are given by the quotient $\BK^{b}_{j} \equiv \BW^{b}_{j}/\BW^{u}_{j+r}$.
The operator content is given by the Kac character, with a single subtraction
\begin{equation}
K_{r,r+2j}  =   \frac{q^{h_{r,r+2j}} - q^{h_{r,-r-2j}}}{P(q)} = \frac{q^{(r-4j)^2/24} - q^{(4j+5r)^2/24}}{\eta(q)}.
\end{equation}
The structure of the representations $\BK^{b}_{j}$ can be easily recovered from the results  for the full blob algebra
of the previous subsection by taking the quotient by $\BW^{u}_{j+r}$. 
To begin with, let us return to $r=1$ which reduces to the ordinary Temperley-Lieb algebra. 
For $j$ integer, we obtain (see Fig.~\ref{figStd_perco_1})
\begin{equation}
\BK^{b}_{j}\equiv \BW^{b}_{j}/\BW^{u}_{j+1}=~~~~~\left\{\begin{array}{cl}
\begin{array}{ccc}
\BX^{b}_{j}&&\\
&\hskip-.2cm\searrow&\\
&&\hskip-.3cm\BX^{b}_{j+2}
\end{array} &\hbox{$j\equiv0 \ ($mod \ $ 3)$}\nonumber\\

\BX^{b}_{j} &\hbox{$j\equiv1 \ ($mod \ $ 3)$}\nonumber\\%\label{diag:quot-stand-1}\\

\begin{array}{ccc}
\BX^{b}_{j}&&\\
&\hskip-.2cm\searrow&\\
&&\hskip-.3cm\BX^{b}_{j+1}
\end{array} &\hbox{$j\equiv2 \ ($mod \ $ 3)$}\nonumber\\
\end{array}\right.
\end{equation}
Of course, we recover here some well-known results for the Temperley-Lieb algebra~\cite{Martin1,Martin2,RS3}.
The simple modules over the blob algebra are expected to go over to Virasoro simple modules in the scaling limit,
and the standard modules $\BK^{b}_{j}$ over the \res~blob algebra correspond to Kac modules $\Kac{r,r+2j}$ introduced in section~\ref{subSecSinglyCritical}.
The structure of the standard $\BK^{b}_{j}$ indeed matches what is expected from Virasoro representation theory.
For example, for $j=0$, taking the quotient of the Verma module $\Verma{h_{1,1}}$ by the singular vector at the level~$1$, we obtain the Virasoro Kac module $\Kac{1,1}$ with the diagram $h_{1,1}\to h_{1,5}$, with $h_{1,5}$ critical exponent for the subquotient $\BX^b_2$ (see eq.~\eqref{spec-b}).
%In terms of characters, we have
The match for all other cases with $j>0$, is recovered by simply using the Kac character identities (see {\it e.g.}~\cite{Characters})
\begin{subequations}
\begin{eqnarray}
    K_{1,1+6p}&=&\chi_{1,1+6p}+\chi_{1,5+6p},\\
    K_{1,3+6p}&=&\chi_{1,3+6p},\\
    K_{1,5+6p}&=&\chi_{1,5+6p}+\chi_{1,7+6p},
\end{eqnarray}
\end{subequations}
where $p \in \mathbb{N}$ and $\chi_{r,s}$ is the character of the Virasoro simple module labeled by the conformal weight $h_{r,s}$.
Of course, all these results could have been obtained from the Temperley-Lieb algebra only. To obtain new results,
let us consider the case $r=2$; we still focus on $j$ integer, the case with $j$ half-integer being very similar.
Since $y=0$ in that case, we consider the JS blob algebra rather than the blob algebra; see the discussion in section~\ref{case-2}.
The representation theory of the JS blob algebra carried out in Section~\ref{case-2} (see Fig.~\ref{figStd_perco_4}) gives then the structure of the quotient modules $\hat{\BK}^{b}_{j}\equiv \hat{\BW}^{b}_{j}/\hat{\BW}^{u}_{j+2}$. Taking a quotient by $\hat{\BW}^{u}_{j+2}$ using the  diagram in Fig.~\ref{figStd_perco_4}, we obtain
\begin{equation}
\hat{\BK}^{b}_{j}\equiv \hat{\BW}^{b}_{j}/\hat{\BW}^{u}_{j+2}=~~~~~\left\{\begin{array}{cl}
\hat{\BX}_{0} &\hbox{$j=0$}\nonumber\\
\begin{array}{ccc}
\hat{\BX}^{b}_{j}&&\\
&\hskip-.2cm\searrow&\\
&&\hskip-.3cm\hat{\BX}^{b}_{j+1}
\end{array} &\hbox{$j\equiv0 \ ($mod \ $ 3)$, $j>0$}\nonumber\\
\begin{array}{ccc}
\hat{\BX}^{b}_{j}&&\\
&\hskip-.2cm\searrow&\\
&&\hskip-.3cm\hat{\BX}^{b}_{j+2}
\end{array} &\hbox{$j\equiv1 \ ($mod \ $ 3)$}\nonumber\\
\hat{\BX}^{b}_{j} &\hbox{$j\equiv2 \ ($mod \ $ 3)$}\nonumber\\
\end{array}\right.
\end{equation}
Let us compare these results with the expected properties of Virasoro Kac modules $\Kac{2,2+2j}$.
We find that the quotient modules $\hat{\BK}^{b}_{j}\equiv \hat{\BW}^{b}_{j}/\hat{\BW}^{u}_{j+2}$ have 
exactly the same subquotient structure as the Kac modules over the Virasoro algebra with $c=0$. Indeed, the Kac module $\VK_{2,2}$ is irreducible, and the standard module $\hat{\BK}^{b}_{0}$ of the critical exponent $h_{2,2}$ is also irreducible\footnote{Note that the standard module $\BK_0$ for the  \res~blob algebra has a proper submodule $\BX^b_1$
whereas  $\hat{\BK}_{0}\equiv \hat{\BW}^{b}_{0}/\hat{\BW}^{u}_{2}$ is irreducible. This is yet another example
where we see that one should consider the JS blob algebra in order to match results on the Virasoro 
side for $y=0$, in agreement with the physical arguments in section 2.}. For $j=1$,  taking the quotient of the Verma module $\Verma{h_{2,4}}$ by the singular vector at level~$8$, we obtain the Virasoro Kac module $\Kac{2,4}$ with diagram $h_{2,4}\to h_{2,8}$, where $h_{2,8}$ is the critical exponent corresponding to the subquotient $\hat{\BX}^b_3$ as expected, and similarly for other $j$.
Indeed, in terms of characters, we have
\begin{subequations}
\begin{eqnarray}
    K_{2,2} &=& \chi_{2,2}, \\
    K_{2,2+6p}&=&\chi_{2,2+6p}+\chi_{2,4+6p}, \qquad p>0,\\
    K_{2,4+6p}&=&\chi_{2,4+6p}+\chi_{2,8+6p},\qquad p\geq0,\\
    K_{2,6+6p}&=&\chi_{2,6+6p}, \qquad\qquad\qquad\; p\geq 0.
\end{eqnarray}
\end{subequations}

\subsection{Dictionary of the correspondence between the (JS) blob algebra and Virasoro}

\begin{table}
\renewcommand{\arraystretch}{1.5}
\setlength{\tabcolsep}{1cm}
\begin{center}
\begin{tabular}{|c|c|c|}
  \hline
  JS blob algebra $\mathcal{B}^b(2N,n,y) = b \mathcal{B}(2N,n,y) b$ & Virasoro algebra \\
  \hline
  \hline
    $n = \q+\q^{-1} = 2 \cos \frac{\pi}{x+1}$ & central charge $c=1-\frac{6}{x(x+1)}$\\
  \hline
   $y = \frac{\left[r+1 \right]_{\q}}{\left[r \right]_{\q}}$ & row of the Kac table $r$\\
  \hline
  Standard module $\hat{\BW}^{b/u}_j = b \BW^{b/u}_j $ & Verma module $\mathcal{V}_{h_{r,r\pm2j}}$ \\
   \hline
     \rres~module $ \hat{\BK}^{b/u}_j =  \hat{\BW}^{b/u}_j /\hat{\BW}^{u}_{r \pm j}  $ ($r$ integer)& Kac module $\mathcal{V}_{h_{r,r\pm2j}}/\mathcal{V}_{h_{r,-r\mp2j}}$ \\
   \hline
     Simple module $\hat{\BX}^{b/u}_j $ & Simple module $h_{r,r\pm2j}$ \\
   \hline   
     $L^{(2N)}_{n}$ from eq.~\eqref{eq_KooSaleur} & Virasoro generators $L_n$ \\
        \hline   
    Inner product with $b^\dag = b$, $e_i^\dag = e_i$ & Virasoro bilinear form $L_{n}^\dag = L_{-n}$ \\
   \hline 
\end{tabular}
\end{center}
  \caption{Correspondence between the JS blob algebra $\mathcal{B}^b(2N,n,y) = b \mathcal{B}(2N,n,y) b$ and Virasoro.}
  \label{tabblobVir}
\end{table}

In conclusion, we have uncovered a correspondence (for which a more precise formulation would require the language of category theory, see~\cite{GRS1,GRS2})
between the JS blob algebra $\mathcal{B}^b(2N,n,y) = b \mathcal{B}(2N,n,y) b$ 
and (the universal enveloping algebra of) Virasoro, which is summarized in Tab.~\ref{tabblobVir}. Of course, one could try to go deeper, and investigate in particular 
whether the conjectured expressions for the Virasoro generators in terms of  JS blob algebra generators, {\it etc.}, are correct. While this can be carried out numerically for some examples, a full,
analytical check seems out of reach for now. But it seems very reasonable to conjecture   that
 the standard, their quotients, and simple modules of the
blob algebra introduced in the foregoing sections  provide  lattice regularizations of 
their well-know Virasoro modules counterparts, and that the action of the Virasoro modes
in these modules can be obtained from  eq.~\eqref{eq_KooSaleur}. This will turn out to have very powerful consequences.

\section{An algebraic reminder : projective modules and self-duality}\label{section4}
{\it Note: We gather in this section the definitions of several mathematical concepts that are needed in order to fully appreciate the rest of this paper, such as projectiveness, self-duality, and tilting modules.
The readers not interested in technical details 
might want to skip this rather technical section at a first reading and go directly to the more physical sections~\ref{secboundaryXXZrestrictedStagg} and~\ref{sectionFaithfulMirrorSpinChain}.}

While for unitary models, Verma modules and  irreducible modules are all the physicist must ever worry about, the non-unitary case is, alas, much more complicated\footnote{Of course, some non-unitary models, such as the non-unitary minimal models, are quite simple, rational, CFTs that require only the understanding of irreducible Virasoro modules. However, in general, non-unitarity opens the door to much more complicated, irrational CFTs described by indecomposable Virasoro modules.}. The few known examples of  logarithmic CFTs indeed involve various types of indecomposable modules such as the now well known staggered modules \cite{MathieuRidout1,KytolaRidout}, and it is not clear what kinds of modules to expect in more complicated LCFTs. There are, in fact, results indicating that, from an algebraic point of view, the  possible indecomposable  Virasoro modules are essentially impossible to classify -- that is, in technical terms, that the representation theory is wild \cite{Germoni}. Nevertheless, experience with WZW models \cite{SchomerusSaleur,SaleurSchomerus} has shown that only a small -- and actually tamable -- subset of this wilderness of modules may play a role in physical theories. 
Our goal is to explore and use the correspondence we have uncovered  between the blob and the Virasoro algebras to gain further understanding of the Virasoro indecomposable modules that may appear in a given LCFT.  To see what this may mean in our case, and to be able to borrow powerful results from the related mathematics, we need some simple technical concepts that we review here.

\subsection{Projective modules}\label{sec:proj-gen}

In logarithmic CFTs, projective modules attracted some attention because  these modules have many properties that are physically appealing. For instance, in superalgebras or quantum groups, the tensor product of any module with a projective module is a sum of projective modules, {\it etc}. We  note that in the earlier physics literature, projective modules have been defined a bit loosely as `maximal indecomposable modules' or as indecomposable modules that cannot appear as submodules of still bigger indecomposable modules. This is actually true only in the case of projective modules that are also self-dual or injective. The definition of a self-dual module will be given in the next section. Meanwhile, by definition, an injective hull of a simple module is the maximal indecomposable that can contain this simple module as a submodule. Then, any injective module is a direct sum of injective hulls and if a submodule in a bigger module is injective then it is a direct summand. In 
 contrast, projective modules are defined as direct sums of projective covers, where the projective cover of a simple module is the unique (for finite-dimensional algebras) indecomposable module of maximal dimension that can cover the simple module, {\it i.e.}, the projective cover contains it as a top subquotient. Then, if a subquotient of a bigger module is projective then it is a direct summand\footnote{Note that the distinction between the notions \textit{subquotient} and \textit{submodule} is crucial here.}. Therefore, the projectiveness property does not necessarily imply injectiveness and {\it vice-versa}. In previously studied examples of spin-chains or LCFTs, usually projectives were also injective modules and thus whenever such a projective module appeared, it was necessarily a direct summand.

However, it was then realized~\cite{RS2,RS3} that projective is not necessarily the appropriate concept. Projective modules appear as direct summands of free modules like the regular representation of an algebra, but for the spin chains -- and thus presumably the LCFTs -- the direct summands are, more generally, \textit{tilting} modules which are defined and discussed in details below in Sec.~\ref{tilt-mod}. In many cases, the tiltings are also indecomposable projective, but this is not necessarily the case. In general, there are tilting modules which are not projective, and projective modules which are not tilting. A detailed discussion with examples appears in~\cite{RS3} for the TL action in $U_{\q} \mathfrak{sl}_2$ spin chains, for $\q=i$ and $\q=e^{i\pi/3}$, and also below in our context of the blob algebra. 

Nevertheless, before discussing  spin-chains and the reasons why tilting modules are more important objects, we give some results about projective modules for the blob algebra. Then, using the correspondence summarized  in Tab.~\ref{tabblobVir} between the JS blob algebra and  Virasoro, we can conjecture the structure of projective modules over the Virasoro algebra; see a discussion below in this section. 

In the theory of cellular algebras~\cite{GL1}, there is a general theorem that allows one  to obtain the subquotient structure of  projective covers knowing the subquotient structure of the standard (cell) modules. The essential part of this theorem can be expressed as a reciprocity property of projectives.  Let $[\BW:\BX]$ and  $[\BP:\BW]$ denote the number of appearance of $\BX$ in a diagram for a standard module $\BW$ and the number of appearance of $\BW$ in a diagram for the projective cover $\BP$, respectively. Then, the reciprocity property reads
\begin{equation}\label{recip}
[\BP:\BW'] = [\BW':\BX],
\end{equation}
{\it i.e.}, the projective cover
 $\BP$ that covers $\BX$ is composed of the standard modules $\BW'$ that have the irreducible module
$\BX$ as a subquotient.

\medskip
Having an indecomposable (and reducible) module $M$, we call {\it socle} its maximum semisimple submodule -- in terms of nodes and arrows in the subquotient diagram for $M$, the socle is the direct sum of all nodes having only ingoing arrows. Similarly, the \textit{top} of a module $M$ is the maximal subquotient with respect to the property that a quotient of $M$ is a semisimple module, {\it i.e.}, it is the subquotient of $M$ having only outgoing arrows.

\paragraph{Generic projectives at $y=0$}
We could thus use the cellularity of the JS blob algebra to describe its projective modules. We will explain in more detail how this is done below, but for now we discuss  the special case of the JS algebra when $y=0$, corresponding, up to an isomorphism given in Sec.~\ref{subsecSecJSblob}, to an algebra with the relation $b'^2~=~0$; see eq.~\eqref{b-square-zero}. We discussed the peculiarity of the standard modules above in Sec.~\ref{subsecSecJSblob}. 
Let us work out the representation theory  for this case (with generic $\q=e^{i\pi/(x+1)}$ and non-rational number $r=x$) a bit more.
The spaces  $\BW^{b}_{j}$ of blobbed diagrams are still well defined modules over the JS algebra in this case,
but with $b'$ acting trivially on the (blobbed by definition) leftmost string as $b'^2=0$. In terms of diagrams,
this means that we adopt the new rules
\begin{equation}\label{new-rules}
\begin{pspicture}(0,0)(9,2)
\psset{xunit=10mm,yunit=10mm}
%\psgrid[subgriddiv=1,griddots=10,gridlabels=10pt](0,0)(9,2)

 \psline[linecolor=black](1,0)(1,2)
 \psdots[dotstyle=square,linecolor=black,dotscale= 1.5 1.5](1,0.5)
 \psdots[dotstyle=*,linecolor=black,dotscale= 1.5 1.5](1,1.5) 
 \rput[Bc](2,1){$=$}
 \psline[linecolor=black](3,0)(3,2)
 \psdots[dotstyle=*,linecolor=black,dotscale= 1.5 1.5](3,1)

  \psline[linecolor=black](6,0)(6,2)
 \psdots[dotstyle=*,linecolor=black,dotscale= 1.5 1.5](6,0.5)
 \psdots[dotstyle=*,linecolor=black,dotscale= 1.5 1.5](6,1.5) 
 \rput[Bc](7,1){$=$}
 \rput[Bc](8,1){$0$}
\end{pspicture}
\end{equation}
Note that these modules are simple because $\q$ is generic and $r$ is irrational. When we attempt to define the action of the JS 
algebra on the space of unblobbed diagrams $\BW^{u}_{j}$,  we find that acting with the  operator $b'$ on the leftmost
string (unblobbed by definition) we end up with the blobbed module $\BW^{b}_{j}$ as $b'(1-b')=b'$ in this case; see the left figure in eq.~\eqref{new-rules}.
We denote by $\mathcal{P}_{j}$ the resulting module, which is the direct sum as a vector space of the spaces  $\BW^{b}_{j}$ and $\BW^{u}_{j}$. It is clear that $\mathcal{P}_{j}$ has a proper 
submodule $\BW^{b}_{j}$. Taking then a quotient by this submodule,
the result is isomorphic to $\BW^{b}_{j}$. Note then that the operator $b'$ is nilpotent and thus non-diagonalizable -- it has Jordan cells of rank $2$, as well as the Hamiltonian. 
This means that  $\mathcal{P}_{j}$ is indecomposable and has the structure
%is a submodule and the top subquotient $\mathcal{P}_{j} / \BW_{j} \simeq \BW_{j} $ is isomorphic to the socle.
%Therefore, the projective module $\mathcal{P}_{j}$ has the following indecomposable structure 
\begin{equation}\label{self-ext}
    \xymatrix@R=6pt@C=6pt{
    &&\\
                                         &  \mathcal{P}_{j} \quad =& \\
         &         &
     }
    \xymatrix@R=21pt@C=6pt{
                                          \BW_{j}\ar[d] & \\
             \BW_{j}   &
     }
%\begin{array}{ccc}
%\BW_{j} &&\\
%&\hskip-.2cm\searrow&\\
%&&\hskip-.3cm \BW_{j},
%\end{array} 
\end{equation}
where we denote $\BW_{j}=\BW^b_{j}$, in consistency with notations~\eqref{BWj-def} for the corresponding JS algebra. 
%The Hamiltonian has Jordan cells of rank~$2$ as well.
% the arrow represents the action of the algebra.
Recall then that the JS algebra, as any finite-dimensional associative algebra, is decomposed (as a left module under the multiplication) into a direct sum of projective modules with multiplicities given by  dimensions of their top irreducible subquotients, which are $\BW_j$ in our case. Comparing the dimension of the sum $\oplus_j \mathrm{dim}(\BW_j)\mathcal{P}_j$ with the dimension of the blob algebra, we conclude that  
the $\mathcal{P}_j$ are projective covers of the generically irreducible module $\BW_{j}$.  The decomposition also shows us that all isomorphism classes of simple modules over $\mathcal{B}^b(2N,n,y=0)$ are exhausted by the modules $\BW_j$, with $0\leq j\leq N-1$.

\begin{figure}\centering
%%     \includegraphics[scale=0.5]{GrahamLehrer_q=i_2.eps}
% \begin{equation*}
%   \xymatrix@R=24pt@C=18pt@W=4pt@M=4pt
%   {{}&\BW_{j}^{u}\ar[dr]\ar[dl]&&\\
%     \BW_{j-1}^{b}\ar[d]\ar[drr]
%     &&\BW_{j-1}^{u}\ar[d]\ar[dll]\\
%     \BW_{j-2}^{b}\ar[d]\ar[drr]
%     &&\BW_{j-3}^{u}\ar[d]\ar[dll]\\
%    \BW_{j-4}^{b}\ar[d]\ar[drr]
%     &&\BW_{j-4}^{u}\ar[d]\ar[dll]\\
%%     {\bullet}\ar@{}|{\substack{\StJTL{L-6}{\q^2}}\kern-7pt}[]+<-48pt,15pt>\ar[d]\ar[drr]
%%     &&{\bullet}\ar@{}|{\substack{\StJTL{L-6}{\q^{-2}}}\kern-7pt}[]+<40pt,15pt>\ar[d]\ar[dll]\\
%     {\dots}\ar[d]\ar[drr]
%     &&{\dots}\ar[d]\ar[dll]\\
%     \BW_{2}^{b}\ar[dr]
%     &&\BW_{1}^{u}\ar[dl]\\
%     &\BW_{0}&&
%     } \qquad
%        \xymatrix@R=26pt@C=8pt@W=4pt@M=4pt
%   {{}&\Verma{1,1-2j}\ar[dr]\ar[dl]&&\\
%     \Verma{1,2j-1}\ar[d]\ar[drr]
%     &&\Verma{1,3-2j}\ar[d]\ar[dll]\\
%     \Verma{1,2j-3}\ar[d]\ar[drr]
%     &&\Verma{1,7-2j}\ar[d]\ar[dll]\\
%    \Verma{1,2j-7}\ar[d]\ar[drr]
%     &&\Verma{1,9-2j}^{u}\ar[d]\ar[dll]\\
%%     {\bullet}\ar@{}|{\substack{\StJTL{L-6}{\q^2}}\kern-7pt}[]+<-48pt,15pt>\ar[d]\ar[drr]
%%     &&{\bullet}\ar@{}|{\substack{\StJTL{L-6}{\q^{-2}}}\kern-7pt}[]+<40pt,15pt>\ar[d]\ar[dll]\\
%     {\dots}\ar[d]\ar[drr]
%     &&{\dots}\ar[d]\ar[dll]\\
%     \Verma{1,5}\ar[dr]
%     &&\Verma{1,-1}\ar[dl]\\
%     &\Verma{1,1}&&
%     }
%  \end{equation*}
\includegraphics[scale=0.8]{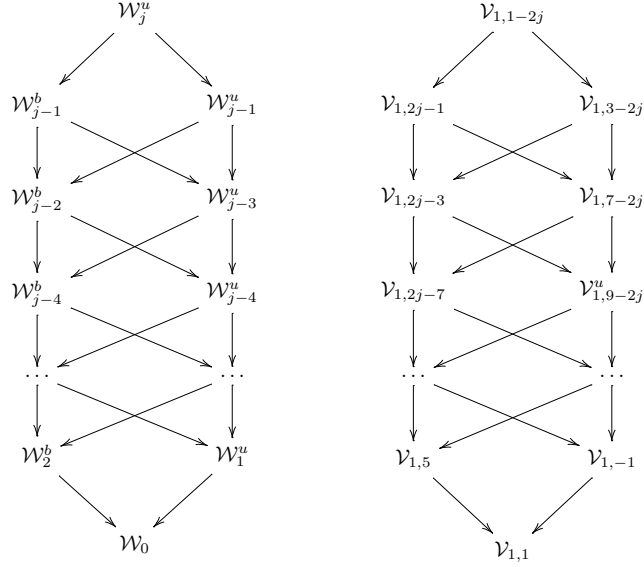}
      \caption{The structure of the projective module
      $\BP_j^{u}$ over the blob algebra $\mathcal{B}(2N,n,y)$ (or $\mathcal{B}^{b}(2N,n,y)$) for $r=1$ and $\q=\mathrm{e}^{i\pi/3}$; we assume here that $j = 1\,\textrm{mod}\,3$. The projective cover $\BP_{j'}^{b/u}$ for each $j'<j$
       is then the submodule generated from
      $\BW_{j'}^{b/u}$, {\it i.e.}, it consists of all nodes and arrows emanating from $\BW_{j'}^{b/u}$. The right diagram describes the structure for the corresponding Virasoro module in terms of Verma module. We will call this Virasoro module projective (see text).}
    \label{fig:proj-thirdroot}
    \end{figure}

    \paragraph{Subquotient structure for $n=y=1$}
We now use the reciprocity property~\eqref{recip} of projective modules  to obtain their structure for the blob algebra in degenerate cases. 
We take again the case  $\q=e^{i\pi/3}$ and consider $r=1$. First, the projective cover for $\BW_0$ is $\BW_0$ itself because there are no other cell modules containing $\BX_0$ as a subquotient. Next we consider  $j = 1\,\textrm{mod}\,3$ for simplicity. By the reciprocity result~\eqref{recip} and using the diagram in Fig.~\ref{figStd_perco_1}, we can write the diagram for the projective cover for $\BX^{u}_j$ shown in Fig.~\ref{fig:proj-thirdroot}.
We note that we also used the fact  from  cellular algebra theory~\cite{GL1} that the filtration of projectives by standard modules should respect/correspond to the filtration in the algebra  --- standard modules with $2j$ through-lines should be above the standard modules with $2j'$ through lines, with $j'<j$. Using this property of projective covers, one can easily write diagrams for the projectives over the blob algebra in terms of standard modules. The arrows in  Fig.~\ref{fig:proj-thirdroot} now correspond to the action of the blob algebra that sends any state from (the subquotient isomorphic to) $\BW_j^{u}$ on the top of the diagram, to a state in $\BW_{j-1}^{b/u}$, {\it etc.} One could say that this projective cover is ``a braid of braids''. More formally, the projective cover $\BP_{j}^{b/u}$ is filtered by the standard modules which have half-number of through lines less than $j$ and that are in the same block (or linkage class) as  $\BW_{j}^{b/u}$. Of course, project
 ive covers are of chain type if the corresponding standard modules they cover are of this type (we will come back to these modules below in section~\ref{sec:class-ind-Vir}). 

Using cellularity of algebras, one can also readily write diagrams for projective covers at other  roots of unity $\q$ than $\q=e^{i\pi/3}$ and integer $r$, probably excepting\footnote{Note that $\mathcal{B}^b(2N,n,0)$ is not a cellular algebra
in contrast to a usual blob algebra. There is no natural way to cut in half
diagrams to obtain reduced states --
the blobbed and unblobbed sectors do not 
make sense here since $b'$ and $1-b'$ are no longer projectors. This is why the reciprocity property cannot be applied in this case in order to get the structure of projective modules.} the case $y=0$. A natural question is then how a projective cover looks in terms of simple subquotients. It is actually very hard to write a general diagram with simple subquotients  for any $L$. We give a few examples below.

\medskip

 Using the dictionary given in Tab.~\ref{tabblobVir}, we can conjecture  that projective\footnote{Of course, we should first fix a category where these Virasoro modules are projective indeed. We expect that it should be a category of Virasoro modules with finite dimensional eigenspaces for $L_0$, locally nilpotent action of the $L_n$'s and with projective objects unique and filtered by Verma modules, and such that the projective object covering the identity module in this category is just the Verma module of weight $0$.} modules over the Virasoro algebra have the braid of braids subquotient structure as in the right-hand side of Fig.~\ref{fig:proj-thirdroot}. This will be discussed further in Sec.~\ref{sec:class-ind-Vir}. We recall that $\Verma{r,s}$ is the Verma module over Virasoro with the weight $h_{r,s}$, and $r$ and $s$ are the usual Kac labels. The arrows in the diagram for the projective modules descend from Verma modules of higher conformal weight to those with a low
 er weight. They thus correspond to an action of positive modes 
 of the stress-energy tensor, {\it i.e.}, the $L_n$'s with $n>0$.

\medskip

Notice finally that it turns out that at degenerate cases all indecomposable but yet reducible projective modules over the (JS) blob algebra are not self-dual and thus not injective. Before giving concrete examples, we first discuss in details the self-duality property. 

\subsection{Dual representations and self-duality}\label{sec:self-dual}

In this section, we discuss the notion of self-duality of a representation of an associative algebra.
Let $\rep$ be a  representation  of an algebra $A$ on a vector-space $V$, {\it i.e.}, $\rep: A\to \Endo(V)$, and assume that $A$ is an algebra with an anti-involution $\cdot^{\dagger}: A\to A$ such that $(ab)^{\dagger} = b^{\dagger}a^{\dagger}$. 
Then we can define a (left) action of $A$ on the dual vector space  $V^*$ -- the vector space of all linear functions $V\to\oC$ -- as
\begin{equation}\label{dual-rep}
\rep^*(a)\bigl(v^*(\cdot)\bigr)
= v^*\bigl(\rep(a^{\dagger})(\cdot)\bigr),\quad \text{with}\quad a\in A,
\; v^*\in V^*,
\end{equation}
where
`$\cdot$' stands for an argument. We also choose a basis in $V^*$ such that $v_i^{*}(v_j)=\delta_{i,j}$ for a basis $\{v_i\}$ in the space $V$.
If an element $a$ is represented by a matrix $(a_{ij})$ on $V$, {\it i.e.}, $\rep(a)(v_i)=\sum_j a_{ij}v_j$, then in the dual representation it is represented by the matrix $\bigl((a^{\dagger})_{ji}\bigr)$. The interchange of indices in the matrix in the dual space means that if the $A$-module $V$ has a subquotient structure like $V_1\to V_2$, where $V_i$ are simple modules, then the dual module $V^*$ has a reversed arrow in its diagram: $V^* = V_2\to V_1$. In this particular example,  flipping the arrow  just means mapping  $V$  to $V^*$ has the kernel isomorphic to $V_2$ and its image is  the submodule $V_1$. In other words, we have a bilinear form  $(\cdot,\cdot)$ on
$V$ such that 
\begin{equation}\label{bil-form}
(\rep(a)(v),w)=(v,\rep(a^{\dagger})(w))
\end{equation}
 and it has a radical $V_2\subset V$. This form can be defined as $(v,w)=v^*(w)$, where the map  $v\mapsto v^*$ is determined using the dual basis $v_i^*$. In general, the dual module has the same subquotient structure as the original module $V$ but where all the arrows representing the action of $A$ are reversed.

Then, a module $V$ is said to be self-dual (or self-conjugate) if there is an isomorphism $V\cong V^*$. It follows from the previous discussion that  the subquotient structure (with simple subquotients)  of the dual of a self-dual module is the same as for the original one $V$. The isomorphism can be explicitly constructed using a symmetric bilinear form on $V$. Indeed, an
isomorphism $\psi$ between $A$-modules $V$ and $V^*$
is given by
\begin{equation}
\psi: V\to V^*, \qquad \psi(v)(\cdot) = (v,\cdot),\qquad v\in V,
\end{equation}
which shows that if there exists a non-degenerate (with zero radical) bilinear form that respects the action of $A$ in the sense of~\eqref{bil-form} then the modules $V$ and $V^*$ are isomorphic.

\medskip
We note finally that the modules encountered in the boundary loop models are not always self-dual.
For the blob algebra, the anti-involution $\cdot^{\dagger}$ is given by the mirror image of the diagram representation of an element from $\blobalg$. The symmetric bilinear form or scalar product on the standard modules was defined above in Sec.~\ref{subsecblobalgeRep}. Then, using these definitions we can in principle study the structure of modules by computing the radical of the bilinear form. This was partially done in~\cite{MartinSaleur} and above at the end of Sec.~\ref{case-2} with the use of Gram matrices. The results are consistent with the diagrams  for standard modules -- the corresponding Gram matrices have zero determinants and their rank gives the dimension of top simple subquotients. The degeneracy of the Gram matrix reflects the fact that (indecomposable and reducible) standard modules are not self-dual. 

The existence of a non degenerate bilinear form is a standard requirement of bulk or boundary  conformal field theory (one has to be more careful with the form involving both bulk and boundary fields, but that is of no interest to us here) \cite{GaberdielRunkel}. It is also expected in physical lattice models, which are defined in terms of local observables such as spins, {\it etc.} 

\subsection{Projective modules: an example}\label{sec:proj-ex}
We now give an example of a projective module on $L=6$ sites for the blob algebra with $\q=e^{i\pi/3}$ and $r=1$. Consider the projective cover $\BP_1^u$ corresponding to $\BX_1^u$. Using Fig.~\ref{fig:proj-thirdroot} and Fig.~\ref{figStd_perco_1}, we can write the following diagram for it (note that the restriction $2j \le L$ makes the diagram finite):
\begin{equation*}
   \xymatrix@R=15pt@C=7pt
   {
   &\\
   &\\
     &\\
     \quad \mathcal{P}^{u}_{1} \quad =   &\\
     &\\
     &
     }
     \xymatrix@R=30pt@C=12pt{
     &\\
              &     {\BW_{1}^{u}} \ar[d]          & \\
                                    & {\BW_{0}} &
     }
   \xymatrix@R=16pt@C=0pt
   {
     &&\\
        &\\
     &\\
     &\quad{=}&\\
     &&\\
     &&
     }\quad
  \xymatrix@R=22pt %% distance between rows
           @C=14pt %% distance between columns
	   @M=3pt  %% a  gap between a node and its arrows
	   @W=3pt 
{
                                    & {\BX_{1}^{u}} \ar[dr] \ar[dl]\ar[d] &\\
     {\BX_{3}^{b}}     \ar[d]\ar[drr]            &    {\BX_0}   \ar[dr]\ar[dl]        &  {\BX_{3}^{u}}\ar[d]\ar[dll]\\
              {\BX_{2}^{b}}      \ar[d]\ar[drr]   
         &               &  {\BX_{1}^{u}}\ar[d]\ar[dll] \\
          {\BX_{3}^{b}}         
         &               &  {\BX_{3}^{u}} 
     }
\end{equation*}

It is a bit harder to control than the structure in terms of simple subquotients for bigger $j$ and $L$ because it is not always obvious which arrows should be placed in the diagram and which ones should not. 
We do not give general pictures (a few more diagrams are given below) but only note that non-self-duality of reducible projective covers for the blob algebra, as can be seen in this example, is a general feature. These projective modules are thus  not injective  (we will show it explicitly later on when we give examples of tilting modules in Sec.~\ref{tilt-mod-hilb}). On the other hand, boundary spin-chains which we consider in the next sections are examples of self-dual representations and this is why projective modules are not the appropriate concept for us.

\subsection{Tilting modules}
\label{tilt-mod}

In the following, we will consider quantum spin-chains rather than statistical loop models.
The spin-chains we consider have a non-degenerate bilinear form given explicitly, for example, in terms of spins. These spin-chains provide a special class of representations with two essential properties: (i) they are filtered by standard modules and (ii) they are self-dual in the sense described above.
Direct summands in such representations are called tilting modules. 

To deal with tilting modules we need the notion of quasi-hereditary algebras. Quasi-hereditarity of an associative algebra   in
particular implies a one-to-one correspondence between (isomorphism classes of) irreducible
modules and the standard modules\footnote{Note that it is not true for any cellular algebra. In particular, the Temperley--Lieb algebra with $n=0$ is not quasi-hereditary.}, and that all projective modules should be filtered by (or composed of) the standard ones. We refer the reader to a complete study of quasi-hereditary algebras in~\cite{DlRin,[Donk]}. It was shown probably for first time in~\cite{MartinSaleur}, see also~\cite{Martin}, that the blob algebra $\blobalg$ is a quasi-hereditary algebra, actually for any choice of parameters except for the case $n=0$ or $\q=i$.

A \textit{tilting} module (over a quasi-hereditary algebra $A$) is a module  that has a filtration by standard modules -- these are $\BW^{u/b}_{j}$ in our case -- and an inverse
filtration by the corresponding duals  -- the so-called costandard  modules $\bigl(\BW^{u/b}_{j}\bigr)^*$ which have reversed arrows; see a general discussion about duals in Sec.~\ref{sec:self-dual}. Recall that a
  filtration of an $A$-module $M$ by $A$-modules $W_i$, with $0\leq
  i\leq n-1$, is a sequence
  of embeddings $0=M_0\subset M_{1}\subset\dots \subset M_i
  \subset\dots \subset M_{n-1}\subset M_n=M$ such that the quotient $M_{i+1}/M_i$ of ``neighbor'' submodules is isomorphic to $W_i$, or in simple words we can say that $M$ is a gluing of $W_i$'s. 
The tilting modules are thus self-dual by definition. The word dual refers here to the natural symmetric, generically non-degenerate,
bilinear form such that $b^\dagger=b$ and $e_i^\dagger = e_i$ (see section~\ref{subsecblobalgeRep}).
Several explicit examples will be given below. 
It is known from the general theory~\cite{DlRin} of quasi-hereditary algebras that  there exists a tilting $\mathcal{B}(2N,n,y)$-module that contains a standard submodule $\BW^{u/b}_{j}$
for any $j$, and that such an indecomposable tilting module is {\it unique}, up to an isomorphism of course. Then, we can introduce the tilting module $\mathcal{T}^{u/b}_{j}$ \textit{generated} from a standard module $\BW^{u/b}_{j}$
as the indecomposable tilting module containing this standard module as a submodule. This property
uniquely defines the tilting module: one should replace each simple subquotient of this standard module by a costandard module 
having this simple module in its socle --- the unique simple subquotient that has only incoming arrows. The result is then
automatically a tilting module, by construction.

We will come back to the tilting modules and their examples in the following after we work out some examples of quantum spin-chains explicitly.

\subsection{The \res~blob algebra projective modules at $\q$ a root of unity}
\label{subsecRestric-proj}

We now turn to the representation theory of the quotient algebra $\mathcal{B}^{\rm res}(2N,n,y) \equiv \mathcal{B}(2N,n,y) / P^{u}_{r} $, with $r$ integer and $\q$ a root of unity. We shall focus on the example of $r=2$ and $\q=\mathrm{e}^{i\pi/3}$
(the case $r=1$ corresponds to the TL algebra for which explicit projective modules were
given in {\it e.g.}~\cite{RS3}). We also recall that $r \in \left( 0, x+1 \right) $ by definition.
The standard modules in the case $r=2$ are  $\BK^{b}_{j} \equiv \BW^{b}_{j}/\BW^{u}_{j+2}$; see also section~\ref{subsecRestric-stand}.
The projective modules of the algebra $\mathcal{B}^{\rm res}(2N,n,y)$, as it is a cellular algebra, can also be built out of the standard modules.
For instance, for $r=2$, we have
\begin{equation*}
\BP^{\rm res}_{j}=~~~~~\left\{\begin{array}{cl}
 \begin{array}{ccc}
\BX^{b}_{0}&&\\
&\hskip-.2cm\searrow&\\
&&\hskip-.3cm \BX^{b}_{1}\end{array}&\hbox{$j=0$,}\nonumber\\
\begin{array}{ccccc}
     &&\hskip-.7cm \BX^{b}_{j}&&\\
     &\hskip-.2cm\swarrow&\searrow&\\
     \BX^{b}_{j-2}&&&\hskip-.3cm \BX^{b}_{j+1}\\
     &\hskip-.2cm\searrow&\swarrow&\\
     &&\hskip-.7cm \BX^{b}_{j}&&
     \end{array}&\hbox{$j\equiv0$ (mod 3) and $j>0$,}\nonumber\\
     &\nonumber\\
 \begin{array}{ccccc}
     &&\hskip-.7cm \BX^{b}_{j}&&\\
     &\hskip-.2cm\swarrow&\searrow&\\
     \BX^{b}_{j-1}&&&\hskip-.3cm\BX^{b}_{j+2}\\
     &\hskip-.2cm\searrow&\swarrow&\\
     &&\hskip-.7cm \BX^{b}_{j}&&
     \end{array}&\hbox{$j\equiv1$ (mod 3),} \nonumber\\
     &\nonumber\\ 
      \BX^{b}_{j}&\hbox{$j\equiv2$ (mod 3).}\nonumber\\
      &\nonumber\\
      \end{array}\right. 
\end{equation*}
Note that some of these projective modules are not self-dual. All those which are self-dual are also tilting modules. This can be checked by a direct construction of the tilting modules following the discussion in the previous subsection~\ref{tilt-mod}.
%All these modules go over to Virasoro smoothly in the scaling limit, except in the case $j=0$,
%this will be discussed in more detail in the next section.

Another important point here is the diamond-type subquotient structure very similar to what is found 
in the case of the Temperley-Lieb algebra. Only rank $2$ Jordan cells in the Hamiltonian
or in the transfer matrix can occur, mapping the top to the socle of the diamonds.
The projective modules for the full algebra $\mathcal{B}(2N,n,y)$ are of course much more complicated, see Fig.~\ref{fig:proj-thirdroot},
and they allow for higher-rank Jordan cells. Some of these higher-rank modules will be described below in section~\ref{sectionFaithfulMirrorSpinChain}. We can thus say that the quotient algebra  provides
a generalization of the TL algebra to the case $r \neq 1$, with the same degree of complexity. 
%, as the most complicated projective indecomposable modules are at worst of the diamond type. 
One could thus expect a spin-chain realization of this quotient blob algebra with a centralizer similar to that of TL -- which is the quantum group $U_{\q} \mathfrak{sl}_2$. Such spin-chains are studied in the next section.

\section{Boundary conditions in XXZ and supersymmetric spin chains, \res\  blob algebra $\mathcal{B}^{\rm res}(2N,n,y)$ and Virasoro staggered modules}\label{secboundaryXXZrestrictedStagg}

Now that we have the basics of the representation theory of the (quotient) blob algebra in hand, 
we turn to the explicit spin-chain realizations of these algebras. 
We first introduce boundary XXZ and supersymmetric spin chains that provide natural representations
of the \res~blob algebra. This corresponds physically to the addition of a spin~$j'\geq1/2$ at the left boundary. While the boundary spin $j'$ does not exceed $x/2$, for $\q=e^{i\pi/(x+1)}$ a root of unity, these spin-chains are self-dual representations of the blob algebra, see definitions in section~\ref{sec:self-dual}. And in these cases, the spin-chains are decomposed onto tilting modules described in section~\ref{subsecRestric-proj} above.
Arguing that the centralizer of the \res~blob algebra is the (Lusztig) quantum group $U_{\q} \mathfrak{sl}_2$, we easily get a decomposition of the spin-chain. The space of states can be organized as a bimodule over
the commuting action of the quantum group and of the \res~blob algebra. As the subquotient structure of the tilting modules does not depend on the number of sites $L$ (this will not be true for the full blob algebra), the scaling limit $L\to\infty$ of the bimodule diagrams can be taken in a rather straightforward way. The result is that the (quotient) blob algebra tilting modules described in section~\ref{subsecRestric-stand} go over to Virasoro staggered modules which are glueings of two Kac modules  introduced in section~\ref{subSecSinglyCritical}. Below we also  compute the associated indecomposability parameters in the staggered modules.  

Note that since the JS blob algebra and the blob algebra are isomorphic except when $y=0$ (see Sec.~\ref{subsecSecJSblob}), 
we will mostly work with the full blob algebra and only mention the differences with the JS version
when relevant.

\subsection{Boundary XXZ spin chain}
\label{subsecBoundaryXXZ}

\subsubsection{XXZ spin chain and the blob}\label{sec:XXZ-def}
 We begin with the well known XXZ spin-$\frac{1}{2}$ chain with $U_{\q} \mathfrak{sl}_2$ symmetry~\cite{PasquierSaleur}
\begin{equation}\label{XXZ_H}
\displaystyle H_{\rm XXZ} = \frac{1}{2} \sum_{i=1}^{2N-1} \left( \sigma^x_i \sigma^x_{i+1} + \sigma^y_i \sigma^y_{i+1}  + \frac{\q + \q^{-1}}{2} \sigma^z_i \sigma^z_{i+1}  \right) + \frac{\q - \q^{-1}}{4} \left( \sigma^z_1 - \sigma^z_{N} \right),
\end{equation}
acting on the Hilbert space $\HXXZ{2N}=(\frac{1}{2})^{\otimes 2N}$, where $(\frac{1}{2}) = \mathbb{C}^2$ is the fundamental representation of $U_{\q} \mathfrak{sl}_2$.
This Hamiltonian can be rewritten, up to an irrelevant additive constant, as $ H = - \sum_{i=1}^{N-1} e_i$ with
\begin{equation}
\displaystyle e_i = \frac{\q + \q^{-1}}{4}  - \frac{1}{2} \left( \sigma^x_i \sigma^x_{i+1} + \sigma^y_i \sigma^y_{i+1}  + \frac{\q + \q^{-1}}{2} \sigma^z_i \sigma^z_{i+1}  \right) - \frac{\q - \q^{-1}}{4} \left( \sigma^z_i - \sigma^z_{i+1} \right),
\end{equation}
where the $e_i$'s provide a representation (usually called 6-vertex or XXZ representation) of the Temperley-Lieb algebra $\mathrm{TL}(N,n)$, with $n=\q+\q^{-1}$.
The operator content of this spin chain is given by the generating function~\cite{PasquierSaleur}
 \begin{equation}
\displaystyle Z_{\rm XXZ} = \mathrm{lim} \ \mathrm{Tr} \ \mathrm{e}^{- \beta(H_{\rm XXZ} - L e_{\infty})}  = \frac{q^{-c/24}}{P(q)} \sum_{j=0}^{\infty} (2j+1) \left( q^{h_{1,1+2j}}-q^{h_{1,-1-2j}}\right)  ,
\end{equation}
where $L=2N$, $ q = \mathrm{e}^{-\pi \beta v_F/L}$, and $v_F=\frac{\pi \sin \gamma}{\gamma}$ is the Fermi velocity, and the groundstate energy density $e_{\infty} = \mathrm{lim}_{L \rightarrow \infty} E_0(L)/L$ is given by~\eqref{eq_einf}.
%, with $E_0(L)$ being the energy of the groundstate of the Hamiltonian $H_{\rm XXZ}$. 
%Using Bethe ansatz, one can show~\cite{PottsBethe} that the mean value of the TL generators in the groundstate is 
%\begin{equation}
%\label{eq_einf}
%\displaystyle e_{\infty} = \sin^2 \gamma \int_{-\infty}^{+\infty} \frac{{\rm d} x}{\cosh \pi x} \ \frac{1}{\cosh 2 \gamma x - 2 \cos \gamma}.
%\end{equation}
The full algebraic study of this spin chain was conducted in Ref.~\cite{RS3} (see also Ref.~\cite{GV}). It was shown that an important 
tool here is to study this spin chain as a representation (bimodule) over $U_{\q} \mathfrak{sl}_2 \otimes \mathrm{TL}(N,n)$. 
This algebraic study has proven to be relevant for understanding the structure of the
underlying LCFTs.
We would like to generalize this analysis to the case of the blob algebra. The first step is of course to find a spin 
chain representation of the blob algebra preserving the $U_{\q} \mathfrak{sl}_2$ symmetry.
A simple way to modify the XXZ spin chain in order to obtain such a representation consists in adding a different spin $j' \in \frac{1}{2} \mathbb{N}^{*}$ at the edge~\cite{SaleurBauer, JS, TheseJerome},
on an additional site that we label with $i=0$.
The resulting Hamiltonian will therefore act on the larger Hilbert space $\HXXZ{2N}[j']=(j')\otimes(\frac{1}{2})^{\otimes 2N}$, with dimension $(2j'+1)2^{L}$, so we have $\HXXZ{2N}[0]=\HXXZ{2N}$.
The model still has a $U_{\q} \mathfrak{sl}_2$ symmetry, but the centralizer (also sometimes called ``commutant'') of this quantum group -- that is the space of endomorphisms $\mathrm{End}_{U_{\q} \mathfrak{sl}_2} \left[ (j') \otimes(\frac{1}{2})^{\otimes L} \right]$ that commute with the $U_{\q} \mathfrak{sl}_2$ action  -- will be different from the Temperley-Lieb algebra.

Let us consider a generic case when $\q$ is not a root of unity and define a new operator, commuting with $U_{\q} \mathfrak{sl}_2$, that projects onto the representation of spin $j'+\frac{1}{2}$ in the first tensor 
product $j' \otimes \frac{1}{2}$:
\begin{equation} \label{eq_b0XXZ}
\displaystyle b = \sum_{m=-j'-1/2}^{j'+1/2} \Ket{j'+1/2,m}\Bra{j'+1/2,m},
\end{equation}
where it is understood that this operator acts non-trivially in the tensor product $j' \otimes \frac{1}{2}$ and as the identity on all other tensorands.
One can easily check that the operator $b$ satisfies $b^2=b$ along with all the other commutation relations required in the blob algebra.
The parameter $y$ of the blob algebra is obtained through
\begin{equation}
\displaystyle e_1 b e_1 = y e_1.
\end{equation}
In particular, acting with this relation on a state $\Ket{m\uparrow\downarrow \dots}$, one obtains
\begin{equation}
y= \q^{-1} \Braket{j',1/2;m \uparrow| j' +1/2,m+1/2}^2 + \q  \Braket{j',1/2;m \downarrow| j' +1/2,m-1/2}^2,
\end{equation}
where the inner products are Clebsch--Gordan coefficients of the form $\Braket{j_0, j_1;m_0,m_1 | j,m}$.
Explicit expression for those can be found in, {\it e.g.}, Ref.~\cite{RefQG}, from which we find that
\begin{subequations} 
\begin{eqnarray}
\Braket{j',1/2;m \uparrow| j' +1/2,m+1/2} &=& \q^{\frac{m-j}{2}} \sqrt{\frac{\left[j+m+1 \right]_{\q}}{\left[2j+1 \right]_{\q}}},\\
 \Braket{j',1/2;m \downarrow| j' +1/2,m-1/2} &=& \q^{\frac{m+j}{2}} \sqrt{\frac{\left[j-m+1 \right]_{\q}}{\left[2j+1 \right]_{\q}}} .
\end{eqnarray}
\end{subequations}
Some algebra using the relation $\left[a \right]_{\q} \q^b+ \left[b \right]_{\q} \q^{-a} = \left[a+b \right]_{\q}$
finally yields 
\begin{equation}
y= \dfrac{ \left[2 j'+2\right]_{\q}}{ \left[2 j'+1\right]_{\q}},
\end{equation}
that is, we find that the relations of the blob algebra are satisfied provided that we set $r=2 j' +1$.
Note that $j'=0$ corresponds to $r=1$, {\it i.e.}, to the usual XXZ case as remarked above.
For now on, we will consider the case of $r$ integer in the fundamental domain, that is,
$0<r<x+1$.

\subsubsection{The Hamiltonian}\label{bnd-ham}
We then introduce a new Hamiltonian
\begin{equation} \label{eq_HbXXZ}
\displaystyle H = - \alpha b + H_{\rm XXZ}
\end{equation}
that we shall refer to as boundary XXZ spin chain in the following (with $\alpha >0$).
At this stage, it might be useful to have explicit expressions for the blob operator $b$~\eqref{eq_b0XXZ}
for a few concrete examples. Let us begin with the case $j'=\frac{1}{2}$, with Hilbert space  $\HXXZ{2N}[\frac{1}{2}]=\frac{1}{2}\otimes(\frac{1}{2})^{\otimes 2N}$.
In the basis $\lbrace \Ket{\uparrow\uparrow}, \Ket{\uparrow\downarrow},\Ket{\downarrow\uparrow},\Ket{\downarrow\downarrow}\rbrace$ for the first two sites $i=0$
and $i=1$, it reads
 \begin{equation}
b= \left( \begin{array}{cccc} 1 & 0 & 0& 0  \\ 0 & \dfrac{\q}{\q+\q^{-1}} &  \dfrac{1}{\q+\q^{-1}}& 0  \\0 &  \dfrac{1}{\q+\q^{-1}} &  \dfrac{\q^{-1}}{\q+\q^{-1}}& 0  \\0 & 0 & 0& 1   \end{array} \right).
\end{equation}
Note that in terms of the Temperley-Lieb generator $e_0$ that would act on the sites $i=0$ and $i=1$
if one had defined the XXZ spin chain on $L+1$ sites $i=0,1,\dots,2L$; we have $b=1-e_0/n$.
In this case, we see that the \res~blob algebra is isomorphic to a TL algebra defined on $L+1$ sites,
with an additional generator $e_0$. This is to be expected as we still couple the system with a spin $\frac{1}{2}$,
only the sign of the coupling differs.
This expression is consistent with the ghost strings construction of~\cite{JS}, since $b = 1-e_0/n$ is
nothing but a Jones-Wenzl projector acting on the first two sites. We also remark that the resulting
Hamiltonian have a ferromagnetic interaction on the first two sites whereas the spin chain is antiferromagnetic
in the bulk. Obviously, this expression is not defined for $n=0$, we shall come back to the meaning of this
in the context of the Virasoro algebra. It is worth noting that the scaling limit of the spin-chain with such a boundary 
ferromagnetic interaction will be different from the usual TL spin-chain, and we obtain new Virasoro modules as a consequence. 
Finally, in the case $j'=1$, one has 
 \begin{equation}
b= \left( \begin{array}{cccccc} 1 & 0 & 0& 0 &0&0 \\ 0 & \dfrac{\q^2}{1+\q^2+\q^{-2}} & \dfrac{\q \sqrt{1+\q^{-2}}}{1+\q^2+\q^{-2}}& 0 &0&0  \\0 & \dfrac{\q \sqrt{1+\q^{-2}}}{1+\q^2+\q^{-2}} & \dfrac{1+\q^{-2}}{1+\q^2+\q^{-2}}& 0 &0&0  \\ 0 & 0 & 0& \dfrac{1+\q^{2}}{1+\q^2+\q^{-2}} &\dfrac{\q^{-1} \sqrt{1+\q^{-2}}}{1+\q^2+\q^{-2}}&0  \\ 0 & 0 & 0& \dfrac{\q^{-1} \sqrt{1+\q^{-2}}}{1+\q^2+\q^{-2}} &\dfrac{\q^{-2}}{1+\q^2+\q^{-2}}&0  \\ 0 & 0 & 0& 0 &0&1   \end{array} \right)
\end{equation}
in the basis $\lbrace \Ket{1,\uparrow}, \Ket{1,\downarrow}, \Ket{0,\uparrow}, \Ket{0,\downarrow}, \Ket{-1,\uparrow}, \Ket{-1,\downarrow}\rbrace$.

\subsubsection{\rres~blob algebra} 
We have found in subsection~\ref{sec:XXZ-def} that the centralizer of $U_{\q} \mathfrak{sl}_2$ in $\HXXZ{2N}[j']=j'\otimes(\frac{1}{2})^{\otimes 2N}$ contains
a representation of the blob algebra. Moreover, we note, by simple calculations for small values of $r=2 j' +1$, that the Jones-Wenzl projector $P^{u}_{r}=P^{u}_{2j'+1}$ defined in eq.~\eqref{eqJonesWenzl}
actually vanishes with the definition of $b$ given by eq.~\eqref{eq_b0XXZ}.
We therefore find that the centralizer contains  a smaller algebra -- actually, the \res~blob algebra (introduced in \eqref{eqDefRestrictedblob})
\begin{equation}\label{eqDefRestrictedblob2}
\mathcal{B}^{\rm res}(2N,n,y)  \equiv \mathcal{B}(2N,n,y) / P^{u}_{r}, 
\end{equation}
with $n=\q+\q^{-1}$ and $y= \frac{ \left[2 j'+2\right]_{\q}}{ \left[2 j'+1\right]_{\q}}$. We will see below that the (full) centralizer of $U_{\q} \mathfrak{sl}_2$ is actually exhausted by $\mathcal{B}^{\rm res}(2N,n,y)$ even at root of unity cases, this result for generic cases was also discussed in~\cite{TheseJerome}. For recent work in the mathematics literature dealing with the blob algebra as a centralizer, see~\cite{HMRam,ZZdaugherty}.

\subsubsection{Generic decomposition}\label{sec:gener-dec} 
Let us first decompose the Hilbert space $\HXXZ{2N}[j']= (j^\prime)\otimes(\frac{1}{2})^{\otimes 2N}$, with   ${\rm dim}\HXXZ{2N}[j'] = (2j'+1) 2^{2N}=r 2^{2N}$, 
with respect to the quantum group. This is easily done in the generic case and we obtain
\begin{multline}
\label{eqDecomGenericQG}
\left. \HXXZ{2N}[j']\right|_{U_{\q} \mathfrak{sl}_2}  = \bigoplus_{j=0}^{N} \left[\binom{2N}{N + j}- \binom{2N}{N + j + r} \right] (j^\prime + j)\\ 
\oplus\bigoplus_{j=1}^{\lfloor j'=\frac{r-1}{2}\rfloor} \left[\binom{2N}{N + j} -  \binom{2N}{N - j + r}\right] (j^\prime - j),
\end{multline}
where we denote by $(j)$ the $j$-th spin representation of $\LQG$, of  dimension $2j+1$.
Using this decomposition over the quantum group, it is then straightforward to obtain the decomposition over its centralizer, with dimensions of irreducibles given by multiplicities. Note that the multiplicities in the decomposition~\eqref{eqDecomGenericQG} are 
given by the dimensions~\eqref{eqCharKacSingle} of the generically irreducible representations (the standard modules) of $\mathcal{B}^{\rm res}(2N,n,y)$. We can thus conclude that the centralizer of $\LQG$ on $\HXXZ{2N}[j']$ is indeed given by  $\mathcal{B}^{\rm res}(2N,n,y)$, and vice versa because of the double-centralizing property.
When decomposed with respect to the \res~blob algebra for $\q$ generic and $N$ integer, the Hilbert space thus reads
\begin{equation}\label{eqHresDecompGeneric}
 \left. \HXXZ{2N}[j'] \right|_{B^{\rm res}}  = \bigoplus_{j=0}^{N} (r+2j) \BK^{b}_{j}  \oplus 
\bigoplus_{j=1}^{\lfloor j'=\frac{r-1}{2}\rfloor} (r-2j) \BK^{u}_{j}  ,
\end{equation}
where we see that the standard modules over $B^{\rm res}$ appear with multiplicities that are now nothing  but the dimension
of the representations of $U_{\q} \mathfrak{sl}_2$ of spin $j_{\rm tot} = j' \pm j$ (recall that $r=2j'+1$).
We will come back to this when we study the decomposition in non-generic cases.
We finally note that the decomposition~\eqref{eqHresDecompGeneric} shows that the boundary XXZ spin-chain is a faithful representation of $\mathcal{B}^{\rm res}(2N,n,y)$.

Using the decomposition~\eqref{eqHresDecompGeneric} and the characters in eq.~\eqref{eq_charBres},
one can readily deduce the universal part of the partition function of the quantum
system defined by the Hamiltonian~\eqref{eq_HbXXZ}
 \begin{equation}\label{eqZbXXZ}
\displaystyle \Zxxz{j'} = \lim_{N\to\infty} \mathrm{Tr} \ \mathrm{e}^{- \beta(H - L e_{\infty})}  = \sum_{j=0}^{\infty} (2j+r) K_{r,r+2j}(q)  + 
\sum_{j=1}^{\lfloor j'\rfloor} (r-2j) K_{r,r-2j}(q)  ,
\end{equation}
where $r=2j'+1$ and  $e_{\infty} $ is still given by~\eqref{eq_einf};
see also Ref.~\cite{JS} for some closely related expressions for $Z$.
Using the total spin $j_{\rm tot} = j' \pm j$ of
the system, where $j_{\rm tot} = j' + j$ corresponds to the first sum in~\eqref{eqZbXXZ} while $j_{\rm tot} = j' - j$ corresponds to the second one, 
we can rewrite $Z$ as (see also~\cite{TheseJerome})
 \begin{equation}
\displaystyle \Zxxz{j'} = \sum_{j_{\rm tot}=j'-{\lfloor j'\rfloor}}^{\infty} (2 j_{\rm tot} +1) K_{1+2 j',1+2 j_{\rm tot}}  .
\end{equation}
Of course when $j'=0$, one recovers the usual partition function $Z_{\rm XXZ}$ of the XXZ spin chain~\cite{RS3}.

\subsection{The Hilbert space decomposition at $\q=\exp(i \pi/3)$ and $r=2$}
\label{subsecHdecompositionXXZ}

%\label{changed}

Let us see how we can apply the previous results to obtain the structure of the  space  of states for the boundary 
XXZ spin chain~\eqref{eq_HbXXZ}. We will focus on the case $\q=\mathrm{e}^{i \pi/3}$ -- that is, a central charge $c=0$ for the corresponding
(L)CFT -- and $r=2$, since $r=1$ corresponds to the Temperley-Lieb case\footnote{Recall that  if $\q=\mathrm{e}^{i\frac{\pi}{x+1}}$ and $x$ is integer, the lattice
construction that we have only makes sense for the ``fundamental domain'' $0 < r < x+1$, with $r$ an integer.}, for which the algebraic properties of the space of states 
in the scaling limit were studied in details in Ref.~\cite{RS3}. We will show that the case $r \neq 1$ for the \res~blob algebra $\mathcal{B}^{\rm res}$
is not more complicated than the TL case, and all the conclusions of Ref.~\cite{RS3} can be generalized without difficulty to this case.
Recall that the partition function of this spin chain reads
\begin{equation}
Z_{\rm XXZ}\bigl[\ffrac{1}{2}\bigr] = \sum_{j=0}^{\infty} (2j+2) K_{2,2+2j}= \sum_{j=0}^{\infty} (2j+2) \frac{q^{h_{2,2+2j}}-q^{h_{2,-2-2j}}}{P(q)}.
\end{equation}
Before discussing the continuum limit, let us first work out the full algebraic structure on the example of $L=8$ sites.

Recall that by construction, our boundary spin-$\frac{1}{2}$ XXZ chain has an additional spin $\frac{1}{2}$ at the left boundary,
with a boundary ferromagnetic interaction respecting the quantum group $U_{\q} \mathfrak{sl}_2$ symmetry of the ordinary XXZ spin chain.
We have argued in section~\ref{subsecBoundaryXXZ} that the centralizer of (a finite image of) $U_{\q} \mathfrak{sl}_2$ at generic~$\q$ in this representation -- that is the space of endomorphisms $\mathrm{End}_{U_{\q} \mathfrak{sl}_2} \left[ (j') \otimes(\frac{1}{2})^{\otimes L} \right]$ that commute with the $U_{\q} \mathfrak{sl}_2$ action  -- was nothing but the \res~blob algebra with $n=\q+\q^{-1}$ and $r=2j'+1$. We have also shown that the boundary XXZ spin chain
provides a faithful representation of the \res~blob algebra in the generic cases, see subsection~\ref{sec:gener-dec}. We show in Appendix~B that these statements are also true in non-semisimple cases. Now, our objective is to obtain a decomposition of the Hilbert space with respect to the \res~blob algebra.
Recall that in generic cases the Hilbert space can be decomposed according to eq.~\eqref{eqHresDecompGeneric} as
\begin{equation}
\left. \HXXZ{2N=8}\bigl[\ffrac{1}{2}\bigr] \right|_{B^{\rm res}} = 2 \BK_0 \oplus 4 \BK_1^b \oplus 6 \BK_2^b \oplus 8 \BK_3^b \oplus 10 \BK_4^b.
\end{equation}
In the case $\q=\mathrm{e}^{i \pi/3}$ and $r=2$, the standard modules $\BK_j^b$ become indecomposable with a structure given in section~\ref{subsecRestric-stand}, and
the standard modules themselves get glued together into indecomposable tilting modules with a diamond shape described in section~\ref{subsecRestric-proj}. Because the representation of the quotient blob algebra on $2N$ sites with the boundary spin $j'=\frac{1}{2}$ is isomorphic to the TL algebra on $2N+1$ sites (see section~\ref{bnd-ham}), we can use the centralizing property with the quantum group $\LQG$. Using then the quantum group results (see Appendix~B), 
we find that the spin-chain is decomposed under the \res~blob algebra for $\q=\exp(i \pi/3)$ and $r=2$ onto tilting modules
\begin{equation}
\label{eqDecompoXXZBres}
\left. \HXXZ{2N=8}\bigl[\ffrac{1}{2}\bigr] \right|_{B^{\rm res}} =
2 \times\!\!
\begin{array}{ccccc}
      &&\hskip-.7cm  \BX^{b}_{1}  &&\\
      &\hskip-.2cm\swarrow&\searrow&\\
      \BX^{b}_{0}  &&&\hskip-.3cm \BX^{b}_{3}  \\
      &\hskip-.2cm\searrow&\swarrow&\\
      &&\hskip-.7cm \BX^{b}_{1}  &&
\end{array}
\hskip-0.3cm
\oplus\,
6 \times
\BX^{b}_{2}\,
\oplus\,
2 \times\!\!
\begin{array}{ccccc}
      &&\hskip-.7cm  \BX^{b}_{3}  &&\\
      &\hskip-.2cm\swarrow&\searrow&\\
      \BX^{b}_{1}  &&&\hskip-.3cm \BX^{b}_{4}  \\
      &\hskip-.2cm\searrow&\swarrow&\\
      &&\hskip-.7cm \BX^{b}_{3}  &&
\end{array}
\hskip-0.3cm
\oplus\,
6 \times\!\!
\begin{array}{ccc}
      &&\hskip-.7cm  \BX^{b}_{4}\\
      &\hskip-.2cm\swarrow\\
      \BX^{b}_{3}  &&  \\
      &\hskip-.2cm\searrow\\
      &&\hskip-.7cm \BX^{b}_{4}
\end{array}
\oplus\,
4 \times \BX^{b}_{4},
\end{equation}
in agreement with general decompositions in~\cite{GV}.
All these tilting modules are also projective except for the singlet $\BX^{b}_{4}$ at the end.
The dimension $d_j^0 =$dim$ \BX^b_{j}$ of the simple modules over the (\res) blob algebra can be easily
worked out from the structure of the standard modules. We find $d_4^0=d_4=1$, $d^0_3 = d_3 - d^0_4 = 7$, $d^0_2=d_2=27$, 
$d^0_1=d_1-d^0_3 = 41$  and $d_0^0 = d_0 - d_1^0 = 1$.
The check of the decomposition for the dimensions is
\begin{equation}
\displaystyle 512 = 2^8 = 2 \times (1+41+41+7) + 6 \times 27 + 2 \times (41+7+7+1) + 6 \times (7+1+1)+ 4\times 1. 
\end{equation}

Recall that the case $r=2$ for percolation corresponds to $y=0$. Therefore,    in order to compare results 
with the Virasoro algebra in the continuum limit,
we should consider the JS blob algebra  $\mathcal{B}^b=b \mathcal{B} b$ and the corresponding spin-chain $b \HXXZ{2N=8}[\frac{1}{2}]$. Note that from the definition of the blob operator $b$ -- it projects on the spin-$(j'+\frac{1}{2})$ subrepresentation in the first two tensorands $(j')\otimes(\frac{1}{2})$ -- it is clear that the image of $b$ is isomorphic to the $\LQG$ representation $b \HXXZ{2N}[j']\cong(j'+\frac{1}{2})\otimes(\frac{1}{2})^{\otimes(2N-1)}$.  
The new boundary XXZ
spin-chain $b \HXXZ{2N=8}[\frac{1}{2}] = (1)\otimes(\frac{1}{2})^{\otimes7}$  provides now a representation of the JS algebra. 
We note that the Hamiltonian corresponding to the JS algebra spin-chain has the form $bH$ in accordance with the discussion in section~\ref{subsecSpec}. 

\begin{figure}
\begin{center}
 \includegraphics[scale=1.0]{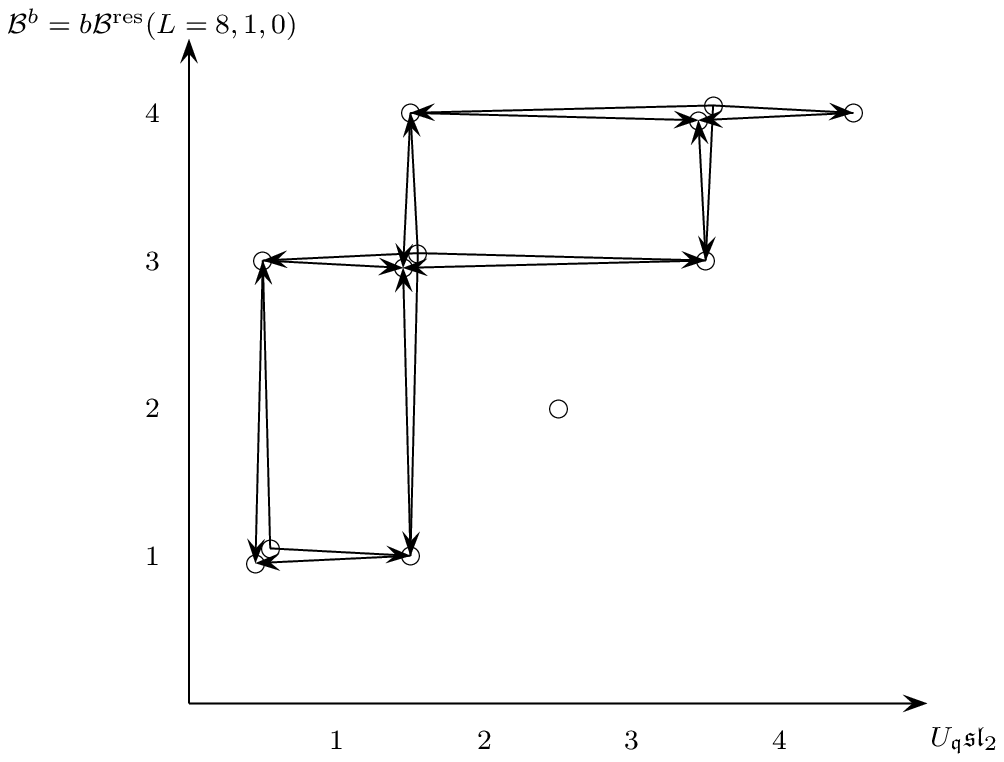}
\end{center}
  \caption{Bimodule diagram for the boundary XXZ spin chain at $\q=\exp(i \pi/3)$ and $r=2$ on $L=8$ sites. It shows the commuting action of the  JS blob algebra (vertical arrows) and of $U_{\q} \mathfrak{sl}_2$ (horizontal arrows).  The same bimodule diagram holds in the case of the supersymmetric $\mathfrak{gl}(m+1|m)$ spin-chain from section~\ref{sec:super} where  $U_{\q} \mathfrak{sl}_2$ should be replaced by a Morita equivalent algebra~\cite{RS2}.}
  \label{figStaircaseJS}
\end{figure}

The decomposition of  $b \HXXZ{2N=8}[\frac{1}{2}]$ over tilting modules is very similar to the \res~blob algebra case in~\eqref{eqDecompoXXZBres}: we first decompose the spin chain $ (1)\otimes(\frac{1}{2})^{\otimes7}$ as a module over $\LQG$ at $\q=\mathrm{e}^{i \pi/3}$, with a similar result to the one in Appendix~B, and then we study the homomorphisms between the direct summands and identify the modules over the JS algebra. The final result is
\begin{equation}
\label{eqDecompoXXZBb}
\left. b\HXXZ{2N=8}\bigl[\ffrac{1}{2}\bigr] \right|_{B^{b}} =
2 \times\!\!
\begin{array}{ccccc}
      &\hskip-.7cm  \hat{\BX}^{b}_{1}  &&\\
      \hskip-.2cm&\searrow&\\
      &&\hskip-.3cm \hat{\BX}^{b}_{3}  \\
      \hskip-.2cm&\swarrow&\\
      &\hskip-.7cm \hat{\BX}^{b}_{1}  &&
\end{array}
\hskip-0.3cm
\oplus\,
6 \times
\hat{\BX}^{b}_{2}\,
\oplus\,
2 \times\!\!
\begin{array}{ccccc}
      &&\hskip-.7cm  \hat{\BX}^{b}_{3}  &&\\
      &\hskip-.2cm\swarrow&\searrow&\\
      \hat{\BX}^{b}_{1}  &&&\hskip-.3cm \hat{\BX}^{b}_{4}  \\
      &\hskip-.2cm\searrow&\swarrow&\\
      &&\hskip-.7cm \hat{\BX}^{b}_{3}  &&
\end{array}
\hskip-0.3cm
\oplus\,
6 \times\!\!
\begin{array}{ccc}
      &&\hskip-.7cm  \hat{\BX}^{b}_{4}\\
      &\hskip-.2cm\swarrow\\
      \hat{\BX}^{b}_{3}  &&  \\
      &\hskip-.2cm\searrow\\
      &&\hskip-.7cm \hat{\BX}^{b}_{4}
\end{array}
\oplus\,
4 \times \hat{\BX}^{b}_{4},
\end{equation}
where we note the first direct summand has a different structure, compare with~\eqref{eqDecompoXXZBres}.
We also find the dimensions $\hat{d}_j$ of the simples $\hat{\BX}^b_j$ with the result $\hat{d}_4^0=1$,  $\hat{d}^0_3 = 6$, $\hat{d}^0_2=21$, and
$\hat{d}^0_1= 28$. 
The check for the dimensions now takes the form
\begin{equation}
 \displaystyle 384 = 3 \times 2^7 = 2 \times (28 + 28 + 6) + 6 \times 21 + 2 \times (28 + 6 + 6 + 1) + 6 \times (6 + 1 + 1) + 4 \times 1 \,.
\end{equation}
Finally, the module structure over the two commuting algebras, $\mathcal{B}^b$ and $\LQG$, can be conveniently represented by the bimodule diagram 
shown in Fig.~\ref{figStaircaseJS}.

The ``tilting'' modules\footnote{Strictly speaking, the first module (for $j=1$) in the list below is not a tilting module (it is self-dual but does not have a filtration by standard modules) but we will nevertheless use the name ``tilting'' for simplicity.} over  $\mathcal{B}^b$ that appear 
in the generalization of the decomposition~\eqref{eqDecompoXXZBb}
to arbitrary size are given by the following diagrams
\begin{equation}
\label{eqTiltingJSQuotient}
~~~~~\left\{\begin{array}{cl}
 \begin{array}{ccc}
\hat{\BX}^{b}_{1}&&\\
&\hskip-.2cm\searrow&\\
&&\hskip-.3cm \hat{\BX}^{b}_{3}\\
&\hskip-.2cm\swarrow&\\
 \hat{\BX}^{b}_{1}&&
\end{array}&\hbox{$j=1$,}\\
\begin{array}{ccccc}
     &&\hskip-.7cm \hat{\BX}^{b}_{j}&&\\
     &\hskip-.2cm\swarrow&\searrow&\\
     \hat{\BX}^{b}_{j-2}&&&\hskip-.3cm \hat{\BX}^{b}_{j+1}\\
     &\hskip-.2cm\searrow&\swarrow&\\
     &&\hskip-.7cm \hat{\BX}^{b}_{j}&&
     \end{array}&\hbox{$j\equiv0$ (mod 3) and $j>0$,}\\
     &\\
 \begin{array}{ccccc}
     &&\hskip-.7cm \hat{\BX}^{b}_{j}&&\\
     &\hskip-.2cm\swarrow&\searrow&\\
     \hat{\BX}^{b}_{j-1}&&&\hskip-.3cm\hat{\BX}^{b}_{j+2}\\
     &\hskip-.2cm\searrow&\swarrow&\\
     &&\hskip-.7cm \hat{\BX}^{b}_{j}&&
     \end{array}&\hbox{$j\equiv1$ (mod 3) and $j>1$,} \\
     &\\ 
      \hat{\BX}^{b}_{j}&\hbox{$j\equiv2$ (mod 3).}\\
      &\\
      \end{array}\right. 
\end{equation}
We note that the module $\hat{\BX}^{b}_{0}$ is also a tilting module but is does not appear in the spin-chain decomposition over the JS algebra.

\subsection{The scaling limit and Virasoro staggered modules}
%and indecomposability parameters $\beta_{r,s}$}
\label{SubSecVirStaggered}

We now briefly discuss how to pass to the continuum limit. 
Physically, the scaling limit consists in focusing on low energy states
and considering the Virasoro modes as given by the generalized 
Koo-Saleur formula~\eqref{eq_KooSaleur}, which we discussed in section~\ref{subsecSpec}. In the scaling limit, 
the boundary XXZ spin chain at $\q=\mathrm{e}^{i\frac{\pi}{x+1}}$ is
described by a CFT with central charge $c$ given by eq.~\eqref{eq-c}. The full Hilbert space in the limit should be described by a semi-infinite version of the  bimodule diagram in Fig.~\ref{figStaircaseJS} over the two commuting algebras, $\mathcal{B}^b$ and $\LQG$.
We then expect some of the JS blob modules given in eq.~\eqref{eqTiltingJSQuotient} to
go over  to staggered Virasoro modules $\VP_{r,r+2j}$~\cite{Rohsiepe,KytolaRidout}, with rank $2$ Jordan cells in $L_0$ mapping the top 
to the socle.  
These modules are composed of two Kac modules, {\it i.e.}, they have a subquotient
structure with a diamond shape.
 Using the subquotient structure~\eqref{eqTiltingJSQuotient}  and  the correspondence $\hat{\BX}^b_{j}\leftrightarrow\VX_{r,r+2j}$ between JS blob algebra simple modules $\hat{\BX}^b_{j}$ and Virasoro simple modules $\VX_{r,r+2j}$ with conformal weight $h_{r,r+2j}$,
we thus have the diagrams, for our $r=2$ example,
\begin{equation}\label{list-stagg}
\VP_{r,r+2j}=~~~~~\left\{\begin{array}{cl}
\begin{array}{ccccc}
     &&\hskip-.7cm h_{2,4} &&\\
     &\hskip-.2cm&\searrow&\\
     &&&\hskip-.3cm h_{2,8}\\
     &\hskip-.2cm&\swarrow&\\
     &&\hskip-.7cm h_{2,4} &&
     \end{array}&\hbox{$j=1$,}\\
\begin{array}{ccccc}
     &&\hskip-.7cm h_{r,r+2j} &&\\
     &\hskip-.2cm\swarrow&\searrow&\\
     h_{r,r+2j-4}&&&\hskip-.3cm h_{r,r+2j+2}\\
     &\hskip-.2cm\searrow&\swarrow&\\
     &&\hskip-.7cm h_{r,r+2j} &&
     \end{array}&\hbox{$j\equiv0$ (mod 3) and $j>0$,}\\
 \begin{array}{ccccc}
     &&\hskip-.7cm h_{r,r+2j}&&\\
     &\hskip-.2cm\swarrow&\searrow&\\
    h_{r,r+2j-2}&&&\hskip-.3cm h_{r,r+2j+4}\\
     &\hskip-.2cm\searrow&\swarrow&\\
     &&\hskip-.7cm h_{r,r+2j} &&
     \end{array}&\hbox{$j\equiv1$ (mod 3) and $j>1$}\end{array}\right.
\end{equation}
where once again we have denoted the simple  Virasoro modules by their conformal weight.
Note that for $j\equiv2$ (mod 3) and $j=0$, we obtain simple Virasoro modules in the scaling limit.
The case $j'$ half-integer is dealt with in a similar way.

We note that the existence of the Virasoro modules given in this list was proven in Ref.~\cite{KytolaRidout}. 
In particular, the first Virasoro module, for $j=1$, is unique up to an isomorphism~\cite{KytolaRidout}. 
It is well-known that the other staggered modules, those for $j>1$, are characterized~\cite{Rohsiepe,KytolaRidout} by indecomposibility parameters and, in general, modules with the same subquotient structure diagram are non-isomorphic if the parameters are different. More details about those numbers can be found in Appendix~A.

\subsection{Boundary supersymmetric spin chains}\label{sec:super}

Recall that other natural representations of the TL$(N,n)$ algebras can be obtained using the Lie superalgebra $\mathfrak{gl}(n+m|m)$~\cite{ReadSaleur01}.
 We consider a model defined on a two-dimensional lattice where each edge carries a $\mathbb{Z}_2$ graded vector space of dimension $n+m|m$. 
We choose these vector spaces to be the fundamental representation~$\square$ of the Lie superalgebra 
$\mathfrak{gl}(n+m|m)$ for $i$ odd, 
and the dual representation $\bar{\square}$ for $i$ even, where $i$ labels the vertical lines. The transfer matrix of the system
(or the Hamiltonian of the associated quantum system in 1+1D) 
then acts on the graded tensor product $\mathcal{H} = (\square \otimes \bar{\square})^{\otimes N}$. 
The TL generators are simply quadratic Casimir invariants,  providing a natural generalization 
of the Heisenberg chain to the $\mathfrak{gl}(n+m|m)$ algebra. We can check that a diagrammatic expansion 
of the transfer matrix yields a dense loop model with a weight $\mathrm{str} \ \mathbb{I} = n + m -m = n$ 
for each closed loop as expected.
This spin chain describes the strong coupling region of a non-linear $\sigma$-model on the complex projective space 
$\mathbb{CP}^{n+m-1|m} = \mathrm{U}(m+n|m) / (\mathrm{U}(1) \times \mathrm{U}(m+n-1|m)) $.

In this case of the supersymmetric spin chains, there are several ways to obtain the blob algebra representations.
One possibility, following Ref.~\cite{Roberto}, is to work with a Hilbert space $\mathcal{H} = \square^{\otimes m} \otimes (\square \otimes \bar{\square})^{\otimes N}$,
and to define the blob operator as the Young symmetrizer on the $m+1$ fundamental representations (note that we have an action of the symmetric group $S_{m+1}$). The continuum limit of these boundary spin-chains was studied in Ref.~\cite{Roberto} in great detail. Meanwhile, one can also
apply the ghost string construction of Ref.~\cite{JS} to the supersymmetric case --- the product of fundamental representations $\square^{\otimes m}$ on the ``left boundary'' should be replaced by the alternating product \;$\;\dots\otimes\bar{\square}\otimes\square\otimes\bar{\square}\otimes (\square \otimes \bar{\square})^{\otimes N}$, the blob operator is then given by the Jones--Wenzl projector --- in order to construct a representation of the blob algebra in a similar way that we did for the XXZ models above. We shall not go into
too much detail here and shall mostly concentrate on the case  $n\ne0$ and $r=2$  that will be useful in the following.
In this case, a simple way to construct a representation of the blob algebra (or more precisely of $\mathcal{B}^{\rm res}(2N,n,y=\left[3\right]_{\q}/\left[2\right]_{\q})$),
is to consider the Hilbert space  $\mathcal{H} = \bar{\square} \otimes (\square \otimes \bar{\square})^{\otimes N}$, and to define $b=1-e_0/n$ where $e_0$
would be the Temperley-Lieb generator acting on the sites $i=0$ and $i=1$. Taking the $\mathfrak{gl}(n+m|m)$ representation of the TL generators,
one has the following Hilbert space decomposition for $L=2N$ even and $n>1$
\begin{equation}
\left. \bar{\square} \otimes (\square \otimes \bar{\square})^{\otimes N} \right|_{\mathcal{B}^{\rm res}} = \bigoplus_{j=0}^{N} \left[2j+2 \right]_{\q'} \BK^{b}_{j},
\end{equation}
where we have introduced $\q'+\q'^{-1}=2m+n$, which is nothing but the trace of the identity operator in the fundamental representation. For $n=1$ or $\mathfrak{gl}(m+1|m)$ spin-chains, the decomposition over the \res~blob algebra is not fully reducible and it is a direct sum over tilting modules like in~\eqref{eqDecompoXXZBres} (for $2N=8$) but with different multiplicities, which can be easily computed as alternating sums of $\left[2j+2 \right]_{\q'}$ numbers. As we consider the case $y=0$ (or $r=2$), before going to the scaling limit we should consider the JS algebra representation on $b\Hilb = \mathrm{ad}\otimes\bar{\square}\otimes (\square \otimes \bar{\square})^{\otimes N-1}$, where $\mathrm{ad}$ is the adjoint representation of $\mathfrak{gl}(m+1|m)$. This spin-chain is decomposed under the JS algebra and its centralizer exactly as in Fig.~\ref{figStaircaseJS} (we note that in this case the centralizer is Morita equivalent to $U_{\q} \mathfrak{sl}_2$~\cite{RS2}).

Finally, in the scaling limit of the  $\mathfrak{gl}(n+m|m)$ boundary spin-chains (critical for $n \leq 2$), the partition function reads 
\begin{equation}
Z_{n|m} = \sum_{j=0}^{\infty} \left[2j+2 \right]_{q'} K_{2,2j+2} .
\end{equation}
We emphasize that only for $n=1$ do we obtain in the limit a logarithmic CFT with a bimodule structure over the $c=0$ Virasoro algebra and its centralizer given by the semi-infinite version of the diagram in Fig.~\ref{figStaircaseJS}.
We also hope that the generalization to other integer values of $r\neq 2$ could be done in a similar fashion using the ghost string construction~\cite{JS}. In this direction, one should first solve the problem of a proper generalization of the Jones--Wenzl projectors (the usual ones are not well-defined in non-semisimple cases) which we leave for a future work.

\section{Faithful spin chain representation of the blob algebra}
\label{sectionFaithfulMirrorSpinChain}

So far we have focused on the \res~blob algebra, which provides a natural generalization
of the Temperley-Lieb algebra, thus allowing one to construct lattice regularizations
of staggered modules involving Kac exponents outside of the first row of the Kac table.
In such models, only rank $2$ Jordan cells could occur and the most complicated 
Virasoro indecomposable modules had a diamond shape. To construct more complicated
Virasoro modules, one may keep the full (JS) blob algebra, and not only the
quotient given by the \res~blob algebra. One would then expect in this case much
more involved indecomposable Virasoro modules, which would be obtained
as the scaling limit of the indecomposable blob modules, with the action of the
Virasoro generators given by the Koo-Saleur formula~\eqref{eq_KooSaleur}.
The convenient tool here turns out to be a faithful and ``full tilting''
spin chain representation~\cite{MartinFaithful0,MartinFaithful1,MartinFaithful2}.
Using our knowledge of the representation theory of the blob algebra,
we will show that one can understand fully the structure of the Hilbert space
of this spin chain even in non-generic cases, and we will
see that Jordan cells of arbitrary rank can occur in the Hamiltonian, provided the number of sites is large enough.
These Jordan cells persist when the scaling limit is taken thus
providing new examples of indecomposable Virasoro modules.
In this section, we will mostly restrict ourselves to the simple example
$\q=\mathrm{e}^{i\pi/3}$ and $r=1$ for which the distinction between 
the JS blob algebra and the blob algebra is irrelevant since the two are
isomorphic. We will therefore work with the full blob algebra keeping in
mind that the results also apply to the more physical JS blob version.

\subsection{Mirror spin chain}
\begin{figure}
\begin{center}
\includegraphics[scale=1]{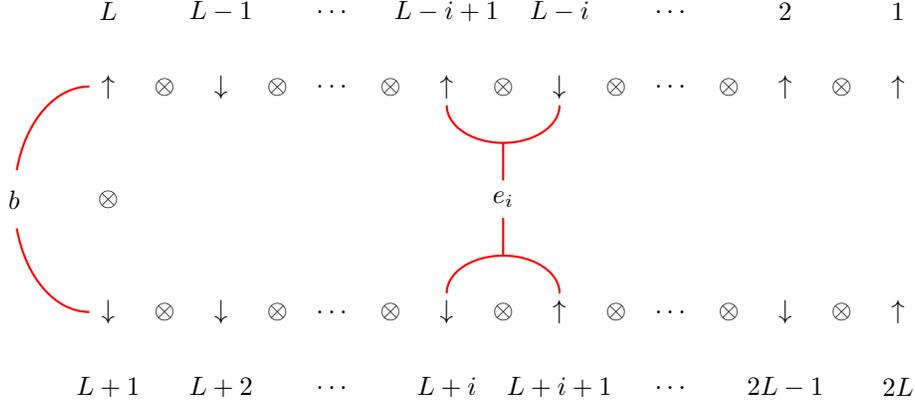}
\end{center}
  \caption{Graphical representation of the action of the operators $e_i$ and $b$ in a mirror spin chain. The red lines describe the corresponding couplings.}
  \label{figMirrorSpinchain}
\end{figure}

Let us introduce a mirror spin chain representation of the blob algebra~\cite{MartinFaithful0}.
The Hilbert space of this spin chain consists of two spin-$\frac{1}{2}$ chains of length $L=2N$ glued together,
so that $\mathcal{H} \cong (\mathbb{C}^2)^{\otimes 2N} \otimes (\mathbb{C}^2)^{\otimes 2N} $.
Let $s=-\mathrm{e}^{i\pi/4} \sqrt{\q}$, $t=\mathrm{e}^{3i\pi/4} \sqrt{\q}$ and $v=i \q^{-r}$.
The action of the TL generators on this mirror spin chain is given by 
\begin{equation}
e_i = \left( \mathbb{I} \otimes \mathbb{I} \otimes \dots \otimes \left( \begin{array}{cccc}
0 & 0 & 0 & 0 \\
0 & s & 1 & 0 \\
0 & 1 & s^{-1} & 0 \\
0 & 0 & 0 & 0 \end{array} \right) \otimes \dots \otimes \mathbb{I} \right) \otimes \left( \mathbb{I} \otimes \mathbb{I} \otimes \dots \otimes \left( \begin{array}{cccc}
0 & 0 & 0 & 0 \\
0 & t & 1 & 0 \\
0 & 1 & t^{-1} & 0 \\
0 & 0 & 0 & 0 \end{array} \right) \otimes \dots \otimes \mathbb{I} \right).
\end{equation}
where the first non trivial tensorand acts on the sites $L-i$ and $L-i+1$, while the second couples the sites $L+i$ and $L+i+1$.
Note that the operator $e_i$ acts in both spin chains simultaneously.
One can check that the $e_i$'s indeed satisfy the Temperley-Lieb relations, in particular $e_i^2=\left[2\right]_s \left[2\right]_t e_i=\left[2\right]_{\q} e_i$.
Meanwhile, the blob operator couples the two spin chains at the edge
\begin{equation}
b = \frac{1}{v+v^{-1}}\mathbb{I} \otimes \mathbb{I} \otimes \dots \otimes \left( \begin{array}{cccc}
0 & 0 & 0 & 0 \\
0 & v & 1 & 0 \\
0 & 1 & v^{-1} & 0 \\
0 & 0 & 0 & 0 \end{array} \right) \otimes \mathbb{I} \otimes  \dots \otimes \mathbb{I},
\end{equation}
acting non-trivially on the sites $L$ and $L+1$.
One checks that the blob operator is a projector ($b=b^2$), as it should be, and a straightforward calculation shows that $e_1 b e_1 = y e_1$, with $y=\frac{\left[r+1\right]_{\q}}{\left[r\right]_{\q}}$.
The action of the operators on the Hilbert space can be represented as in Fig.~\ref{figMirrorSpinchain}.
The mirror spin chain can thus be thought of as two spin chains, folded so that $e_i$ couples the two chains together and $b$ acts at one edge. 
The Hamiltonian has the usual expression
\begin{equation}
H = - \alpha b - \sum_{i=1}^{L-1} e_i,
\end{equation}
with $\alpha>0$. Note that one could also use the Hamiltonian $H = - \sum_{i=1}^{L} b e_i b $ defined on $L+1$ sites thanks 
to the mapping between the JS blob algebra and the blob algebra derived in section~\ref{section1}.
The relations for $b$ and $e_i$ show that this mirror spin chain provides a representation of the blob algebra $\mathcal{B}(2N,n,y)$.
More importantly, one can show~\cite{MartinFaithful2} that this representation is faithful\footnote{To be more precise, it can
be shown that it is also full tilting~\cite{MartinFaithful1,MartinFaithful2}, which is actually a stronger statement.}, so when compared to the XXZ case
studied in section~\ref{secboundaryXXZrestrictedStagg},
we expect much more complicated indecomposable blob modules in non-generic cases.

%
%\begin{table}
%\begin{center}
%\begin{tabular}{|c|c|c|c|c|c|c|c|}
%  \hline
%   & $S_z=-3$  &$S_z=-2$ & $S_z=-1$ & $S_z=0$ &  $S_z=1$ &  $S_z=2$ &  $S_z=2$ \\
%  \hline
%   \vdots & \vdots & \vdots & \vdots & \vdots & \vdots & \vdots & \vdots \\
%  $\BW^u_{2}$ & 7 & 19 & 31 & 37 & 31 & 19 & 7 \\
%  $\BW^b_{2}$ & 1 & 5 & 9 & 11 & 9 & 5 & 1 \\
%  $\BW^u_{1}$ & 0 & 1 & 3 & 3 & 3 & 1 & 0 \\
%  $\BW^b_{1}$ & 0 &0 & 1 & 1 & 1 & 0 & 0 \\
%  $\BW_{0}$ &0  &0 & 0 & 1 & 0 & 0 & 0 \\
%    \hline
%\end{tabular}
%\end{center}
%\caption{Multiplicities of the standard modules $\BW^{b/u}_{j}$ in the decomposition of the Hilbert space $\mathcal{H}_{S_z}$ of the mirror spin chain
%in the sector with spin $S_z$.}
%  \label{tab_faithful}
%\end{table}

The spin $S_z$ is conserved by the Hamiltonian so one can decompose the Hilbert space as 
\begin{equation}
\displaystyle \mathcal{H} = \bigoplus_{S_z=-L}^{L} \mathcal{H}_{S_z}.
\end{equation}
When the representation of the blob algebra is generic (that is, when $r$ is not integer), each sector with fixed $S_z$ can be decomposed onto the standard modules of $\mathcal{B}(2N,n,y)$.
The multiplicities were computed in Ref.~\cite{MartinFaithful1}, and they can be conveniently recast into the generating function $F(x,y) = \sum_{k,S_z} C_{k,S_z} x^k y^{S_z}$, where
$C_{k,S_z}$ is the multiplicity of $\BW^{b}_{j=k/2}$ ({\it resp.} $\BW^{u}_{j=(k-1)/2}$) if $k$ is even ({\it resp.} odd) in the decomposition of $\mathcal{H}_{S_z}$. 
Straightforward algebra yields
\begin{equation}
\displaystyle F(x,y) = \sum_{k,S_z} C_{k,S_z} x^k y^{S_z} = \dfrac{1-x}{x+x^{-1}-(1+y)(1+y^{-1})}.
\end{equation}
The coefficients $C_{k,S_z}$ are then readily obtained by expanding the function $F(x,y)$.
For example, the $S_z=0$ sector on $L=6$ sites can be decomposed as
\begin{equation}
\label{eqHdecompmirror}
\displaystyle \mathcal{H}^{L=6}_{S_z=0} = \BW_{0}  \oplus \BW^b_{1}  \oplus 3  \ \BW^u_{1}  \oplus 11  \ \BW^b_{2}  \oplus 37 \ \BW^u_{2}  \oplus 125 \ \BW^b_{3}  \oplus 431 \ \BW^u_{3}.
\end{equation}
In non generic cases, those standard modules get glued together into bigger, indecomposable, modules of $\mathcal{B}(2N,n,y)$.

The decomposition of the full Hilbert space is actually much simpler~\cite{MartinFaithful1}, and the multiplicity of
the standard modules $\BW^{b}_{j}$ and $\BW^{u}_{j}$ read, respectively, $u_{2j-1}$ and $u_{2j}$;
where $u_n$ satisfies $u_n=4 u_{n-1} - u_{n-2}$ and $u_0=1$, $u_1=3$. The solution of this 
equation can be written as
\begin{equation}
\displaystyle u_n = \sum_{k=0}^{n} 2^k \left( \begin{array}{c} n+k  \\ n-k  \end{array} \right).
\end{equation}
This statement is of course consistent with the identity
\begin{equation}
\displaystyle \mathrm{dim} \mathcal{H} = 4^{2N} = \left(\begin{array}{c}  2N \\ N  \end{array} \right)+  \sum_{j=1}^{N} ( u_{2j-1} + u_{2j}) \left( \begin{array}{c} 2N \\ N-j  \end{array} \right),
\end{equation}
where we have used eq.~\eqref{eqDimStdblob}.
In the continuum limit, one expects the system to be described by a CFT, with central charge $c=1-\frac{6}{x(x+1)}$.
To describe the spectrum, one introduces the generating function of the scaled energy gaps 
\begin{equation}
\displaystyle Z = \mathrm{lim} \ \mathrm{Tr} \ \mathrm{e}^{- \beta(H - L e_{\infty})} = \lim_{L \rightarrow \infty} \sum_{{\rm levels}\ i} q^{\frac{L}{\pi v_F} (E_i- L e_{\infty})} = \mathrm{Tr} \ q^{L_0 - c/24},
\end{equation}
where we recall that $q = \mathrm{e}^{-\pi \beta v_F/L}$, $v_F=\frac{\pi \sin \gamma}{\gamma}$ is the Fermi velocity and $e_{\infty} = \mathrm{lim}_{L \rightarrow \infty} E_0(L)/L$, with $E_0$ the groundstate energy in the case $r=1$.
Thanks to our knowledge of the blob algebra~\cite{JS}, we immediately deduce the operator content of the underlying CFT
\begin{subequations}
\begin{eqnarray}
 Z &=& \sum_{j=1}^{\infty} u_{2j-1} Z_{j}^{b} +\sum_{j=0}^{\infty} u_{2j} Z_{j}^{u},\\
          &=& \frac{q^{-c/24}}{P(q)} \sum_{j=0}^{\infty} \left[ \left(\sum_{k=0}^{2j-1} 2^k \left( \begin{array}{c} 2j-1+k  \\ 2j-1-k  \end{array} \right)\right) q^{h_{r,r+2j}} + \left(\sum_{k=0}^{2j} 2^k \left( \begin{array}{c} 2j+k  \\ 2j-k  \end{array} \right)\right) q^{h_{r,r-2j}}\right]
\end{eqnarray}
\end{subequations}
This formula for the critical exponents and their multiplicities can of course be checked numerically.

\subsection{Hilbert space decomposition and tilting modules}
\label{tilt-mod-hilb}

For our purposes, the main interest of this mirror spin chain  $\Hilb$  is that it 
provides a faithful representation of the (JS) blob algebra~\cite{MartinFaithful2}, 
which we will denote as $\rep$ in what follows. Moreover, it can
be shown that it is also full tilting~\cite{MartinFaithful1,MartinFaithful2}, meaning that it can be fully
decomposed onto a special class of modules called tilting modules introduced in Sec.~\ref{tilt-mod}. We will
focus on non-generic situations for now on, and we shall consider as
an example the case of $n=1$ ({\it i.e.}, $\q=\mathrm{e}^{i \pi/3}$) and $y=1$.
The structure of the standard modules of the blob algebra in this case was
discussed in Sec.~\ref{section3}, and there is no subtlety with the correspondence
with Virasoro. We will therefore  use notations for the blob algebra and not for JS algebra in this section for simplicity. Other non-generic cases can be treated similarly.
Note that we directly jump to the doubly critical case where $r$ is integer
{\it and} $\q$ is a root of unity which is the most complicated situation.
The case $r$ integer and $\q$ generic yields simpler incomposable
modules that will be discussed afterwards (see sec.~\ref{subSecTiltingsMirrorSingly}).

\medskip

We use a non-degenerate bilinear form $(\cdot,\cdot)$ on
$\chVv\times\chVv$ given explicitly in terms of spins and we interpret the complex parameter $\q$ as a formal variable. The generators $b$ and $e_j$ of
$\blobalg$ are self-adjoint with respect to this bilinear form,
{\it i.e.} $\rep(e_j)^{\dagger}=\rep(e_j)$. Together with non-degeneracy of the bilinear form,
this means that the representation $\rep$ is isomorphic do the dual
one on the space $\chVv^*$ of linear functionals. Indeed an
isomorphism $\psi$ between $\blobalg$-modules $\chVv$ and $\chVv^*$
is given by
\begin{equation}
\psi: \chVv\to \chVv^*, \qquad \psi(v)(\cdot) = (v,\cdot),
\end{equation}
where we recall a general discussion in Sec.~\ref{sec:self-dual}; that the $\blobalg$-action on $\chVv^*$ is defined as in~\eqref{dual-rep}; and that `$\cdot$' stands for an argument. The non-degeneracy of the bilinear
 form implies that the kernel of $\psi$ is zero.

Self-duality of the module $\chVv$ implies that the subquotient
structure (with simple subquotients) is not affected by reversing all the arrows representing the
$\blobalg$-action. On the other hand, the faithfulness of the
representation of $\blobalg$ implies that all projective modules
should be present in the spin-chain decomposition. We have seen before in Sec.~\ref{sec:proj-ex}
 that projective $\blobalg$-modules are not self-dual and
therefore are not injective and can in principle be
submodules in some bigger modules. These bigger and self-dual $\blobalg$-modules
indeed exist (the tilting modules actually) and we show that the projectives can be
embedded into them, or generally into a direct sum of tilting modules. 
We finally emphasize that as far as spin-chains are concerned, tilting
modules are more
fundamental than projective modules in the sense that tilting modules
are the building blocks (direct summands)
of the spin-chains.

\begin{figure}
\begin{center}
%\begin{pspicture}(0,0)(12,5)
%\psset{xunit=10mm,yunit=10mm}
%%\psgrid[subgriddiv=1,griddots=10,gridlabels=10pt](0,0)(10,5)
%% \psline[linecolor=black,linewidth=0.5pt,arrowsize=5pt]{->}(0,0)(4.0,0.0) 
%% \psline[linecolor=black,linewidth=0.5pt,arrowsize=5pt]{->}(0,0)(0.0,6.0) 
%
%% \psdots[dotstyle=o,linecolor=black,dotscale= 1.5 1.5](0.0,0.0)
%
%\rput[Bc](0.5,2.5){$\mathcal{T}_{0}$}
%\rput[Bc](1,2.5){$=$}
%\rput[Bc](6.5,2.5){$\simeq$}
%
% \rput[Bc](3.5,0.5){$\BW_{0}$}
%
% \psline[linewidth=0.5pt,arrowsize=5pt]{<-}(3.3,0.8)(2.2,2.1) 
% \psline[linewidth=0.5pt,arrowsize=5pt]{<-}(3.7,0.8)(4.8,2.1) 
%
% \rput[Bc](2,2.5){$\BW^{b}_{2}$}
% \rput[Bc](5,2.5){$\BW^{u}_{1}$}
% \psline[linewidth=0.5pt,arrowsize=5pt]{<-}(2,2.8)(2,4.3) 
% \psline[linewidth=0.5pt,arrowsize=5pt]{<-}(5,2.8)(5,4.3) 
% \psline[linewidth=0.5pt,arrowsize=5pt]{<-}(2.25,2.8)(4.75,4.3) 
% \psline[linewidth=0.5pt,arrowsize=5pt]{<-}(4.75,2.8)(2.25,4.3) 
%
% \rput[Bc](2,4.7){$\BW^{b}_{3}$}
% \rput[Bc](5,4.7){$\BW^{u}_{3}$}
%
%
%
% \rput[Bc](9.5,4.5){$\BW_{0}^{\ast}$}
%
% \psline[linewidth=0.5pt,arrowsize=5pt]{->}(9.3,4.2)(8.2,2.9) 
% \psline[linewidth=0.5pt,arrowsize=5pt]{->}(9.7,4.2)(10.8,2.9) 
%
% \rput[Bc](8,2.5){$\BW^{b\ast}_{2}$}
% \rput[Bc](11,2.5){$\BW^{u\ast}_{1}$}
% \psline[linewidth=0.5pt,arrowsize=5pt]{->}(8,2.2)(8,0.7) 
% \psline[linewidth=0.5pt,arrowsize=5pt]{->}(11,2.2)(11,0.7) 
% \psline[linewidth=0.5pt,arrowsize=5pt]{->}(8.25,2.2)(10.75,0.7) 
% \psline[linewidth=0.5pt,arrowsize=5pt]{->}(10.75,2.2)(8.25,0.7) 
%
% \rput[Bc](8,0.3){$\BW^{u\ast}_{3}$}
% \rput[Bc](11,0.3){$\BW^{b\ast}_{3}$}
%
%
%
%\end{pspicture}
 \includegraphics[scale=0.82]{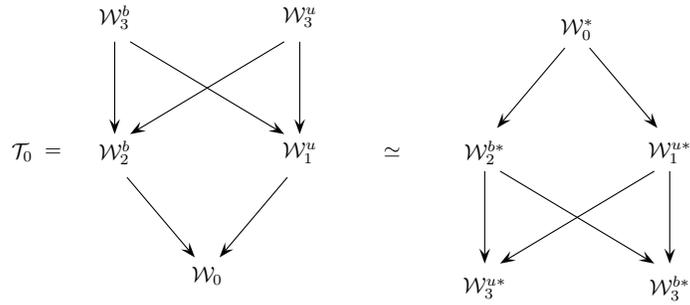}
\end{center}
  \caption{Structure of the tilting module $\mathcal{T}_0$ on $L=6$ sites for the blob algebra at $\q=\mathrm{e}^{i\pi/3}$ and $r=1$
  in terms of standard modules and costandard modules (the costandard modules correspond to the dual of the standard modules, they have the same
  structure as the latter expect that the arrows are reversed). This tilting module is by definition self-dual.}
  \label{FigblobTilting1}
\end{figure}

\begin{figure}
\begin{center}
 \includegraphics[scale=0.82]{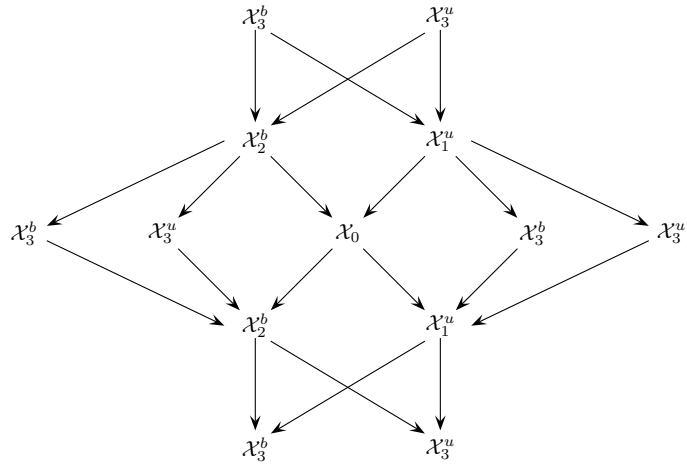}
\end{center}

  \caption{Tilting module $\mathcal{T}_0$ on $L=6$ sites for the blob algebra at $\q=\mathrm{e}^{i\pi/3}$ and $r=1$. }
  \label{figblobTilting2}
\end{figure}

Recall that  there exists a one-to-one correspondence between standard and tilting modules -- a tilting $\mathcal{B}(2N,n,y)$-module that contains a standard submodule $\BW^{u/b}_{j}$, for any $j$, is unique up to an isomorphism. In Sec.~\ref{tilt-mod}, we introduced the tilting module generated from a standard one. Its structure can be deduced  knowing the subquotient structure for each standard module.
As an example of this construction, we will restrict ourselves to the case $L=6$ and we will fully work out the structure of the Hilbert space in the sector $S_z=0$. Using the structure of the standard modules worked out in Sec.~\ref{section3},
it is quite straightforward in this case to deduce the structure of the associated tilting modules. Let us begin by the ``biggest'' tilting module~$\mathcal{T}_0$. Its structure in terms of standard and costandard (duals to standard) modules is given in Fig.~\ref{FigblobTilting1}.
Once again, the structure in terms of costandard modules is readily obtained starting from the structure of the standard module $\BW_0$ and
replacing its simple subquotients by the corresponding costandard modules. The extension of this structure for arbitrary sizes in straightforward.
The resulting structure in terms of simple modules is given in Fig.~\ref{figblobTilting2}. One can check that this module is indeed self-dual -- {\it i.e.},
it has the same structure if one reverses all the arrows. Note that non-self-dual projective covers discussed in Sec.~\ref{sec:proj-gen} are submodules in the tilting module $\mathcal{T}_0$. For example, the projective module $\BP_1^{u}$ described in Sec.~\ref{sec:proj-ex} is embedded in the following way. Its top subquotient $\BX_1^{u}$
is mapped onto the corresponding unique subquotient in $\mathcal{T}_0$, then each $\BX_3^{b/u}$ subquotient is mapped onto a non-trivial linear combination of the two $\BX_3^{b/u}$'s in $\mathcal{T}_0$. The mapping of the remaining subquotients is obviously fixed. We can also identify the projective cover for $\BX_3^{b/u}$ inside $\mathcal{T}_0$. It consists of nodes and arrows `emanating' from the corresponding subquotient in  $\mathcal{T}_0$ in Fig.~\ref{figblobTilting2}. This result might be compared with the general results about projectives in Sec.~\ref{sec:proj-gen} and  Fig.~\ref{fig:proj-thirdroot}.

\begin{figure}
\begin{center}
 \includegraphics[scale=0.82]{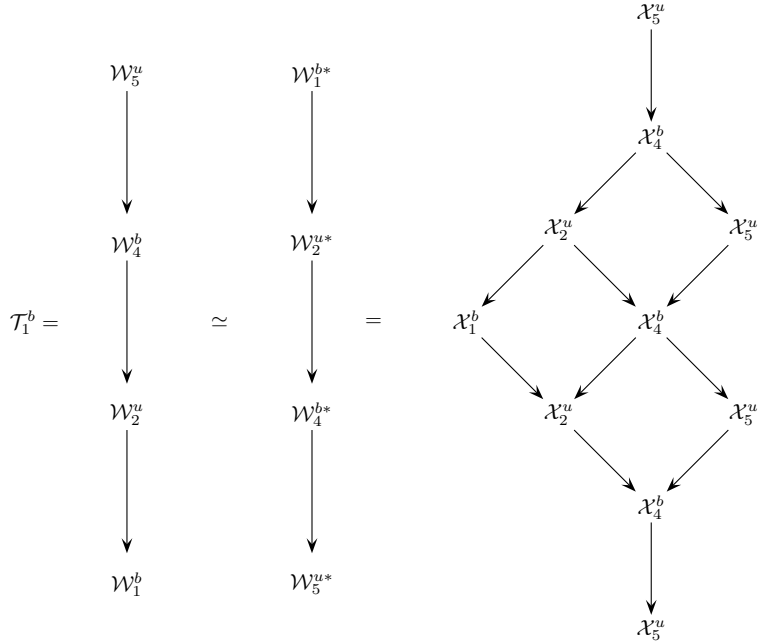}
\end{center}
  \caption{Structure of the tilting module $\mathcal{T}^{b}_{1}$ for the blob algebra at $\q=\mathrm{e}^{i\pi/3}$ and $r=1$, on $L=10$ sites;
  in terms of standard, costandard, and simple modules.}
  \label{figblobTilting3}
\end{figure}

We note that the other tilting modules can be obtained in a similar fashion. We find
%\begin{equation}
%\begin{pspicture}(0,0)(7,3)
%\psset{xunit=8mm,yunit=8mm}
%%\psgrid[subgriddiv=1,griddots=10,gridlabels=10pt](0,0)(7,3)
% \rput[Bc](4,0){$\BX^{u}_{3}$}
% \rput[Bc](4,3){$\BX^{u}_{3}$}
% \rput[Bc](7,0){$\BX^{b}_{3}$}
% \rput[Bc](7,3){$\BX^{b}_{3}$}
% \rput[Bc](5.5,1.5){$\BX^{u}_{1}$}
%
% \rput[Bc](-1,1.5){$\mathcal{T}^{u}_{1}=$}
%  \rput[Bc](3.5,1.5){$=$}
%
%    \psline[linewidth=0.5pt,arrowsize=5pt]{->}(4.25,2.75)(5.25,1.75) 
%    \psline[linewidth=0.5pt,arrowsize=5pt]{->}(5.75,1.25)(6.75,0.25) 
%    \psline[linewidth=0.5pt,arrowsize=5pt]{->}(6.75,2.75)(5.75,1.75) 
%    \psline[linewidth=0.5pt,arrowsize=5pt]{->}(5.25,1.25)(4.25,0.25) 
%    
% \rput[Bc](0,2.15){$\BW^{b}_{3}$}
% \rput[Bc](3,2.15){$\BW^{u}_{3}$}
% \rput[Bc](1.5,0.75){$\BW^{u}_{1}$}
%
%
%    \psline[linewidth=0.5pt,arrowsize=5pt]{->}(0.25,2.0)(1.25,1.0) 
%    \psline[linewidth=0.5pt,arrowsize=5pt]{->}(2.75,2.0)(1.75,1.0) 
%   
%
%\end{pspicture}
%\end{equation}
\begin{equation}
   \xymatrix@R=15pt@C=7pt
   {
     &\\
     \quad \mathcal{T}^{u}_{1} \quad =   &\\
     &\\
     &
     }
     \xymatrix@R=28pt@C=12pt{
     {\BW_{3}^{b}} \ar[dr]         &               &  {\BW_{3}^{u}} \ar[dl]\\
                                    & {\BW_{1}^{u}} &
     }
   \xymatrix@R=18pt@C=0pt
   {
     &&\\
     &\quad{=}&\\
     &&\\
     &&
     }\!\!\!\!\!
  \xymatrix@R=22pt %% distance between rows
           @C=14pt %% distance between columns
	   @M=3pt  %% a  gap between a node and its arrows
	   @W=3pt 
{
     {\BX_{3}^{b}} \ar[dr]         
         &               &  {\BX_{3}^{u}} \ar[dl]\\ 
                                    & {\BX_{1}^{u}} \ar[dr] \ar[dl] &\\
     {\BX_{3}^{b}}                 &               &  {\BX_{3}^{u}}
     }
\end{equation}
The module $\mathcal{T}^{b}_{2}$ has a similar structure with the node $\BX^{u}_{1}$ replaced by $\BX^{b}_{2}$.
Recall that the standard module $\BW^{b}_{1}$ had a single chain structure. The corresponding tilting thus reads
%\begin{equation}
%\begin{pspicture}(0,0)(7,3)
%\psset{xunit=8mm,yunit=8mm}
%%\psgrid[subgriddiv=1,griddots=10,gridlabels=10pt](0,0)(7,3)
% \rput[Bc](5.5,-0.25){$\BX^{u}_{2}$}
% \rput[Bc](5.5,3.25){$\BX^{u}_{2}$}
% \rput[Bc](5.5,1.5){$\BX^{b}_{1}$}
%
% \rput[Bc](-1,1.5){$\mathcal{T}^{b}_{1}=$}
%  \rput[Bc](3.5,1.5){$=$}
%
%    
% \rput[Bc](1.5,2.5){$\BW^{u}_{2}$}
% \rput[Bc](1.5,0.55){$\BW^{b}_{1}$}
%
%
%    \psline[linewidth=0.5pt,arrowsize=5pt]{->}(1.5,2.1)(1.5,0.9) 
%    
%    \psline[linewidth=0.5pt,arrowsize=5pt]{->}(5.5,3)(5.5,1.9) 
%    \psline[linewidth=0.5pt,arrowsize=5pt]{->}(5.5,1.1)(5.5,0) 
%
%   
%
%\end{pspicture}
%\end{equation}
\begin{equation}
   \xymatrix@R=15pt@C=7pt
   {
     &\\
     \quad \mathcal{T}^{b}_{1} \quad =   &\\
     &\\
     &
     }
   \xymatrix@R=28pt@C=5pt{
     & {\BW_{2}^{u}} \ar[d] & \\
     & {\BW_{1}^{b}}        &
     }
   \xymatrix@R=18pt@C=0pt
   {
     &&\\
     &\quad{=}&\\
     &&\\
     &&
     }\!\!\!\!\!
  \xymatrix@R=18pt %% distance between rows
           @C=18pt %% distance between columns
	   @M=3pt  %% a  gap between a node and its arrows
	   @W=3pt  
{
    & {\BX_{2}^{u}} \ar[d]
                                          %% [dd] means two times down
                                          %% while @/^/ or @/_/ is for direction
  & \\
    & {\BX_{1}^{b}} \ar[d]  & \\
    & {\BX_{2}^{u}}         &
     }
\end{equation}
Meanwhile, the tilting modules $\mathcal{T}_2^u=\BW_2^u=\BX_2^u$, $\mathcal{T}_3^b=\BW_3^b=\BX_3^b$, and $\mathcal{T}_3^u=\BW_3^u=\BX_3^u$ are simple.
With this in hands, we find that the generic decomposition~\eqref{eqHdecompmirror} becomes here
\begin{multline}
   \xymatrix@R=15pt@C=7pt
   {
     &\\
     \quad \mathcal{H}_{\rule{0pt}{7.5pt}%
       S_z=0}^{\rule{0pt}{-3.5pt}%
       L=6}\quad =   &\\
     &\\
     &
     }
   \xymatrix@R=24pt@C=12pt{
     {\BW_{3}^{b}} \ar[d]\ar[drr]  &               &  {\BW_{3}^{u}} \ar[d]\ar[dll]\\
     {\BW_{2}^{b}} \ar[dr]         &               &  {\BW_{1}^{u}} \ar[dl]\\
                                    & {\BW_{0}} &
     }
   \xymatrix@R=18pt@C=0pt
   {
     &&\\
     &\quad{\oplus}&\\
     &&\\
     &&
     }
   \xymatrix@R=28pt@C=5pt{
     & {\BW_{2}^{u}} \ar[d] & \\
     & {\BW_{1}^{b}}        &
     }
   \xymatrix@R=18pt@C=0pt
   {
     &&\\
     &\quad{\oplus}\qquad2&\\
     &&\\
     &&
     }\!\!\!\!\!
   \xymatrix@R=28pt@C=12pt{
     {\BW_{3}^{b}} \ar[dr]         &               &  {\BW_{3}^{u}} \ar[dl]\\
                                    & {\BW_{1}^{u}} &
     }
\\
   \xymatrix@R=18pt@C=0pt
   {
     &&\\
     &\quad{\oplus}\quad10&\\
     &&\\
     &&
     }\!\!\!\!\!
   \xymatrix@R=28pt@C=12pt{
     {\BW_{3}^{b}} \ar[dr]         &               &  {\BW_{3}^{u}} \ar[dl]\\
                                    & {\BW_{2}^{b}} &
     }
   \xymatrix@R=18pt@C=0pt
   {
     &&\\
     &\quad\oplus\quad 36\;\BW_{2}^{u}\quad \oplus \quad 112\;\BW_{3}^{b} \quad \oplus \quad 418\;\BW_{3}^{u}, &\\
     &&\\
     &&
     }
\end{multline}
that is,
\begin{equation}
\displaystyle \mathcal{H}_{S_z=0}^{L=6} = \mathcal{T}_0 \oplus  \mathcal{T}^b_{1} \oplus 2 \ \mathcal{T}^u_{1} \oplus 10 \ \mathcal{T}^b_{2} \oplus 36 \ \mathcal{T}^u_{2} \oplus 112 \ \mathcal{T}^b_{3} \oplus 418 \ \mathcal{T}^u_{3}.
\end{equation}
The main interesting feature of this decomposition is that it shows that we should expect rank $3$ Jordan cells  
in the Hamiltonian for eigenvalues corresponding to the simple module $\BX^{u/b}_{3}$. This can
be seen in the structure of the tilting $\mathcal{T}_0$ shown in Fig.~\ref{figblobTilting2}. For example, there are $4$ simple
subquotients $\BX^{b}_{3}$ in $\mathcal{T}_0$, with two of them that are not linked by an arrow (they both lie in the middle of the diagram).
We expect in this case the Hamiltonian of the spin chain to map the top $\BX^{b}_{3}$ subquotient to a {\it linear combination}
of the two middle states, which in turn can be mapped onto the bottom $\BX^{b}_{3}$ (the one in the socle) by acting once again with the Hamiltonian.
As another example, we also expect $\BX^{b}_{4}$ in the tilting $\mathcal{T}^{b}_{1}$ to be mixed into a rank $3$ Jordan cell for $H$ 
for $L \geq 8$ (see Fig.~\ref{figblobTilting3}).

Of course, there are many other rank $2$ Jordan cells, including for example the Jordan cell that  involves states in  the ``diamond'' containing $\BX^{b}_{2}$ 
in Fig.~\ref{figblobTilting2}. Checking the existence of higher rank Jordan cells numerically in this example
is quite tricky as the Hilbert space has already dimension $\mathrm{dim} (\mathcal{H}_{S_z=0}^{L=6}) =924$ on 
$L=6$ sites for which rank $3$ Jordan cells  occur first time. Similar observations were made in the context of periodic spin-chains~\cite{GRSV}.

It is not hard to see that the rank of the Jordan cells is stabilized for a fixed subquotient or a critical exponent, {\it e.g.}, for (all states in) $\BX_3^{b/u}$ the Jordan cell is at most of rank $3$. Meanwhile, the rank is growing for states in subquotients $\BX_j^{b/u}$ having larger~$j$. So, the rank of the biggest Jordan cell will become arbitrarily large  in such a tilting module  when the length of the spin chain is growing. This feature is quite remarkable and shows that much more complicated 
Virasoro modules should be expected in the continuum limit which we discuss below in our classification section~7.

 It is worth noting that when the representation
theory of the blob algebra is fully non-generic ($\q$ root of unity {\it and} $r$ integer), only two kinds
of different tilting modules occur. The first ``class'' of tilting modules is associated with standard
modules with a braid structure, this is the case for $\mathcal{T}_0$ given in Figs.~\ref{FigblobTilting1} and~\ref{figblobTilting2} for $L=6$ and below in Fig.~\ref{fig:tilt-lim} for any even number $L$ (the left diagram).
The other class corresponds to standard modules with a single chain of embeddings. An example of such modules
is provided by  $\mathcal{T}^{b}_{1}$ whose structure is given in Fig.~\ref{figblobTilting3} and in general in Fig.~\ref{fig:tilt-ch-lim} (the left diagram).
Both classes show Jordan cells for the Hamiltonian  of arbitrary rank when the number of sites $L$ tends to infinity.

\section{A classification of indecomposable Virasoro modules}\label{sec:class-ind-Vir}
In this section, we first describe the scaling limit of the tilting modules constructed in the previous section where we studied the faithful `mirror' representation of the blob algebra. The Virasoro modules obtained in this way are of a special type, as they are self-dual and have Jordan cells for the Hamiltonian ($L_0$) of any finite rank. This gives way to our classification of self-dual Virasoro modules relevant for physics given  in section~\ref{subsec:conjecture},  based on the explicit examples of tilting modules obtained in the scaling limit of these boundary spin-chains.  We also describe in section~\ref{sec:non-self-dual} non self-dual modules of the blob algebra that should arise in boundary loop models, together with their connection with the general class of Virasoro staggered modules.

\subsection{The scaling limit of tilting modules and self-dual Virasoro modules}

We have shown in section~\ref{tilt-mod-hilb} that it is possible
to decompose the Hilbert space of the mirror spin chain
onto the so-called tilting modules. Those modules
have a complicated indecomposable structure in non-generic 
cases, far more involved than the usual diamond shape
found in the representation theory of the TL algebra for example.
A direct consequence of this feature is the appearance of arbitrary 
rank Jordan cells in the Hamiltonian. 
We briefly discuss here
the scaling limit of such spin chains.

In the scaling limit, we expect the low-energy physics of the mirror spin chain 
to be described by a (boundary) CFT, which may be logarithmic in non-generic cases.
We have seen before the striking correspondence between the (JS) blob algebra and Virasoro,
so that the spectrum of the transfer matrix (or of the Hamiltonian) in a simple module
is described by the character of the corresponding Virasoro simple module.
We have argued in great detail before how we thus expect the simple modules of the
(JS) blob algebra to go over to Virasoro simple modules in the scaling limit,
with the action of the Virasoro generators given by eq.~\eqref{eq_KooSaleur}. We have also  seen in section~\ref{secboundaryXXZrestrictedStagg} that tilting modules over the \res~blob algebra go over to well-known self-dual indecomposable (staggered) Virasoro modules.
The indecomposable tilting modules for the full blob algebra obtained in the previous section are therefore very good
candidates to provide new indecomposable Virasoro modules relevant for physics.

We  discuss below the modules obtained in the scaling limit of the mirror spin-chains (over the blob algebra)
for $n=1$ ({\it i.e.} $\q=\mathrm{e}^{i \pi/3}$) and $y=1$. As $r=1$, there are no subtleties
between the representation theory of the blob algebra $\mathcal{B}$ and of its JS version $b \mathcal{B} b$.
Therefore, the scaling limit can be taken safely using our results about spin-chain decomposition over the blob algebra\footnote{We should actually consider the new spin chain $b\mathcal{H}$ for the JS algebra as we did previously. However, in the case $r=1$ the structure of modules should not be changed, only multiplicities are different.}.
The resulting objects obtained in the  limit are conjectured to be modules 
over the Virasoro algebra at central charge $c=0$.

%\newpage
%\thispagestyle{empty}
%{\large
% \begin{equation*}
%   \xymatrix@R=25pt@C=16pt@W=4pt@M=4pt
%   {     \dots\ar[d]&\\
%    \BW_{j}^{u}\ar[d]&\\
%     \BW_{j-1}^{b}\ar[d]
%     &\\
%     \BW_{j-3}^{u}\ar[d]
%     &\\
%%    \BW_{j-4}^{b}\ar[d]\ar[drr]
%%     &&\BW_{j-4}^{u}\ar[d]\ar[dll]\\
%%     {\bullet}\ar@{}|{\substack{\StJTL{L-6}{\q^2}}\kern-7pt}[]+<-48pt,15pt>\ar[d]\ar[drr]
%%     &&{\bullet}\ar@{}|{\substack{\StJTL{L-6}{\q^{-2}}}\kern-7pt}[]+<40pt,15pt>\ar[d]\ar[dll]\\
%     {\dots}\ar[d]\\
%     \BW_{1}^{b}&
%     } \qquad
%             \xymatrix@R=30pt@C=1pt@W=4pt@M=4pt
%{&&\\
%&&\\
%\xrightarrow{\mbox{}\quad L\to\infty\quad}&&\\
%&&\\
%&&\\
%}\qquad
%   \xymatrix@R=26pt@C=22pt@W=4pt@M=4pt
%   {     \dots\ar[d]&\\
%    \Verma{1,1-2j}\ar[d]&\\
%     \Verma{1,2j-1}\ar[d]
%     &\\
%     \Verma{1,7-2j}\ar[d]
%     &\\
%%    \BW_{j-4}^{b}\ar[d]\ar[drr]
%%     &&\BW_{j-4}^{u}\ar[d]\ar[dll]\\
%%     {\bullet}\ar@{}|{\substack{\StJTL{L-6}{\q^2}}\kern-7pt}[]+<-48pt,15pt>\ar[d]\ar[drr]
%%     &&{\bullet}\ar@{}|{\substack{\StJTL{L-6}{\q^{-2}}}\kern-7pt}[]+<40pt,15pt>\ar[d]\ar[dll]\\
%     {\dots}\ar[d]\\
%     \Verma{1,3}&
%     } 
%\end{equation*}
%}
%\newpage  

 \begin{figure}\centering
 \includegraphics[scale=0.89]{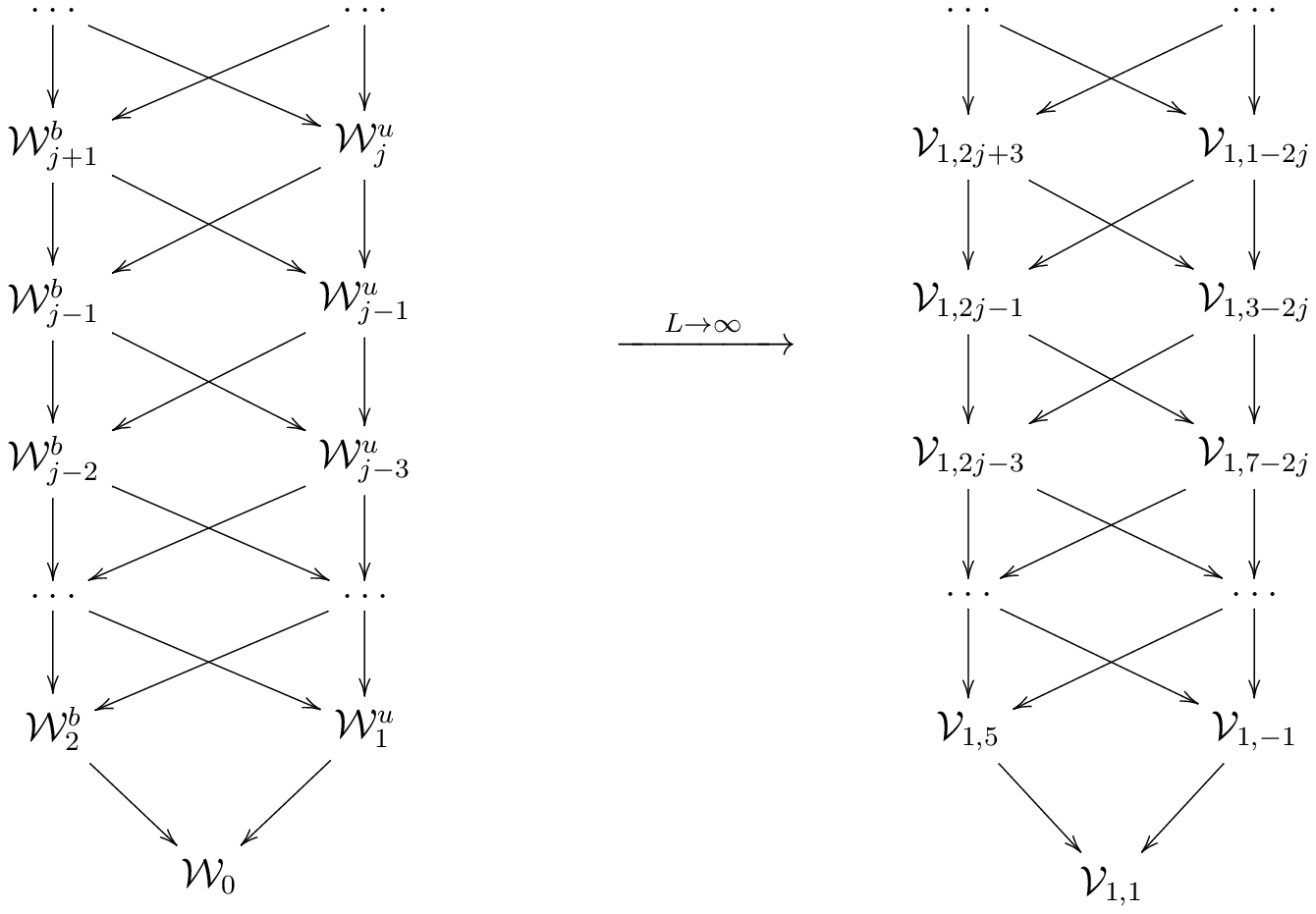}
% \begin{equation*}
%   \xymatrix@R=24pt@C=16pt@W=4pt@M=4pt
%   {     \dots\ar[d]\ar[drr]
%     &&\dots\ar[d]\ar[dll]\\
%    \BW_{j+1}^{b}\ar[d]\ar[drr]
%     &&\BW_{j}^{u}\ar[d]\ar[dll]\\
%     \BW_{j-1}^{b}\ar[d]\ar[drr]
%     &&\BW_{j-1}^{u}\ar[d]\ar[dll]\\
%     \BW_{j-2}^{b}\ar[d]\ar[drr]
%     &&\BW_{j-3}^{u}\ar[d]\ar[dll]\\
%     {\dots}\ar[d]\ar[drr]
%     &&{\dots}\ar[d]\ar[dll]\\
%     \BW_{2}^{b}\ar[dr]
%     &&\BW_{1}^{u}\ar[dl]\\
%     &\BW_{0}&&
%     } \qquad
%             \xymatrix@R=36pt@C=1pt@W=4pt@M=4pt
%{&&\\
%&&\\
%\xrightarrow{\mbox{}\quad L\to\infty\quad}&&\\
%&&\\
%&&\\}
%        \xymatrix@R=26pt@C=8pt@W=4pt@M=4pt
%   {     \dots\ar[d]\ar[drr]
%     &&\dots\ar[d]\ar[dll]\\
%        \Verma{1,2j+3}\ar[d]\ar[drr]
%     &&\Verma{1,1-2j}\ar[d]\ar[dll]\\
%     \Verma{1,2j-1}\ar[d]\ar[drr]
%     &&\Verma{1,3-2j}\ar[d]\ar[dll]\\
%     \Verma{1,2j-3}\ar[d]\ar[drr]
%     &&\Verma{1,7-2j}\ar[d]\ar[dll]\\
%     {\dots}\ar[d]\ar[drr]
%     &&{\dots}\ar[d]\ar[dll]\\
%     \Verma{1,5}\ar[dr]
%     &&\Verma{1,-1}\ar[dl]\\
%     &\Verma{1,1}&&
%     }
%  \end{equation*}
      \caption{The structure of the tilting module
      $\BT_0$ over the blob algebra $\mathcal{B}(2N,n,y)$ (or $\mathcal{B}^{b}(2N,n,y)$) for $r=1$ and $\q=\mathrm{e}^{i\pi/3}$; we assume that $j=1\,\mathrm{mod}\,3$. The right diagram describes its scaling limit in terms of the corresponding Verma modules $\Verma{r,s}$ over the Virasoro algebra with $c=0$ and weight $h_{r,s}$. The resulting Virasoro module is (graded)
 self-dual by construction.}
    \label{fig:tilt-lim}
    \end{figure}

\begin{figure}\centering
 \includegraphics[scale=0.84]{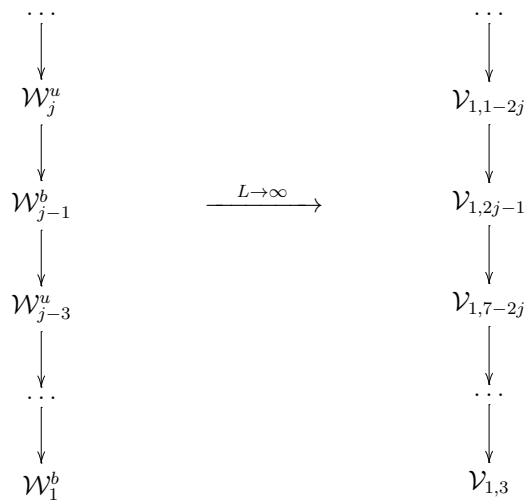}
% \begin{equation*}
%   \xymatrix@R=22pt@C=16pt@W=4pt@M=4pt
%   {     \dots\ar[d]&\\
%    \BW_{j}^{u}\ar[d]&\\
%     \BW_{j-1}^{b}\ar[d]
%     &\\
%     \BW_{j-3}^{u}\ar[d]
%     &\\
%%    \BW_{j-4}^{b}\ar[d]\ar[drr]
%%     &&\BW_{j-4}^{u}\ar[d]\ar[dll]\\
%%     {\bullet}\ar@{}|{\substack{\StJTL{L-6}{\q^2}}\kern-7pt}[]+<-48pt,15pt>\ar[d]\ar[drr]
%%     &&{\bullet}\ar@{}|{\substack{\StJTL{L-6}{\q^{-2}}}\kern-7pt}[]+<40pt,15pt>\ar[d]\ar[dll]\\
%     {\dots}\ar[d]\\
%     \BW_{1}^{b}&
%     } \qquad
%             \xymatrix@R=30pt@C=1pt@W=4pt@M=4pt
%{&&\\
%&&\\
%\xrightarrow{\mbox{}\quad L\to\infty\quad}&&\\
%&&\\
%&&\\}\qquad
%   \xymatrix@R=22pt@C=16pt@W=4pt@M=4pt
%   {     \dots\ar[d]&\\
%    \Verma{1,1-2j}\ar[d]&\\
%     \Verma{1,2j-1}\ar[d]
%     &\\
%     \Verma{1,7-2j}\ar[d]
%     &\\
%%    \BW_{j-4}^{b}\ar[d]\ar[drr]
%%     &&\BW_{j-4}^{u}\ar[d]\ar[dll]\\
%%     {\bullet}\ar@{}|{\substack{\StJTL{L-6}{\q^2}}\kern-7pt}[]+<-48pt,15pt>\ar[d]\ar[drr]
%%     &&{\bullet}\ar@{}|{\substack{\StJTL{L-6}{\q^{-2}}}\kern-7pt}[]+<40pt,15pt>\ar[d]\ar[dll]\\
%     {\dots}\ar[d]\\
%     \Verma{1,3}&
%     } 
%\end{equation*}
      \caption{Structure of the tilting module
      $\BT^b_{1}$ over the blob algebra $\mathcal{B}(2N,n,y)$  for $r=1$ and $\q=\mathrm{e}^{i\pi/3}$; we assume that $j=2\,\mathrm{mod}\,3$. The right diagram describes its scaling limit in terms of the corresponding Verma
      modules $\Verma{r,s}$ over the Virasoro algebra with $c=0$ and weight $h_{r,s}$.
       The resulting Virasoro module is also self-dual by construction.
      }
    \label{fig:tilt-ch-lim}
    \end{figure}

The Virasoro indecomposable module containing the ground state is obtained
as the scaling limit of the tilting module $\mathcal{T}_0$ shown on the left-hand side of Fig.~\ref{fig:tilt-lim} for arbitrarily large $L$. The corresponding Virasoro module obtained in the scaling limit has a structure in terms of Verma modules given by the right-hand side of Fig.~\ref{fig:tilt-lim} 
(recall that standards become Verma modules in the limit).
The diagram in terms
of simple Virasoro modules 
is given in Fig.~\ref{figVirIndec} where we give only a small part centered around the vacuum state.  
It is worth noting that the resulting 
picture in terms of Verma modules is much easier to understand than the full structure in terms of 
simple modules given in Fig.~\ref{figVirIndec}. We see that the resulting indecomposable Virasoro 
module is a ladder of Verma modules, which have themselves a ladder/braid structure
in terms of irreducible Virasoro modules. The arrows now  represent the action of positive Virasoro modes. 
We also note that on odd-length spin chains, the tilting module $\mathcal{T}^{b}_{\frac{1}{2}}$ has a similar ``braid-of-braids'' structure but with different exponents, of course.

It is  interesting to compare the structure of tilting Virasoro modules with the conjectured structure of projective Virasoro modules  in Sec.~\ref{sec:proj-ex} and in Fig.~\ref{fig:proj-thirdroot}. Note that while the towers for projective Virasoro modules are of finite length -- they have finitely many Verma subquotients, and there is a top simple subquotient -- the structure  of the tilting modules is an infinite ladder which has no top subquotient, though the eigenvalues of $L_0$ are of course bounded from below by $0$. The projective modules from Fig.~\ref{fig:proj-thirdroot} which  cover the $\Verma{1,j}$ subquotients are only submodules in the vacuum tilting module. Note also that in the diagram in Fig.~\ref{fig:tilt-lim} there are no simple Virasoro subquotients corresponding to the conformal weight $h_{1,j}$ that would be above the node $\Verma{1,j}$. This is consistent with our conjecture that Virasoro modules in  Fig.~\ref{fig:proj-thirdroot} are projective.

Note that we have seen that the standard modules over the (JS) blob algebra
correspond in the scaling limit to the full Verma modules, without any quotient being taken.
In particular, this means that the resulting module $\BT_0$ containing our vacuum state $\vac$ still contains the state $L_{-1} \vac$ (which corresponds to the lower node marked by $h_{1,-1}$ in Fig.~\ref{figVirIndec}) 
thus breaking the translational part of the $SL(2,\mathbb{C})$ invariance of the vacuum state usually expected in a CFT. Meanwhile, we have $L_1\vac=L_0\vac=0$. Note that our boundary theory is not expected to contain the real, {\it i.e.}, translationally-invariant, vacuum sector because we considered the limit of boundary spin-chains. Our boundary theory is thus defined on a strip with two different boundary conditions, or on a half plane where one non-trivial boundary condition is imposed on the, say, negative real axis, while a free boundary condition is fixed on the positive part of the real line. This obviously breaks the translational invariance.
 Therefore, the structure of the limit of our tilting module is physically reasonable.
 Nevertheless, we will call the resulting indecomposable Virasoro module the vacuum module, as it contains our ground state.
The descendant on level $2$  of the ground state $\vac$ (or of an operator with conformal weight $h_{1,1}=0$)  is  a primary field marked in Fig.~\ref{figVirIndec} by $\tilde{T}(z)$. This primary field has a logarithmic partner $\tilde{t}(z)$ with conformal weight $h_{1,5}=2$. Note that  $\tilde{T}(z)$ is not the stress-energy tensor because it is a descendant of a non-translationally-invariant operator with $h=0$. We leave the physical interpretation of the field $\tilde{T}(z)$ for a future work.
  
Finally, note   that the tilting Virasoro module  in Fig.~\ref{figVirIndec} contains Jordan cells for $L_0$ of arbitrary rank. For example, there is 
a rank-$3$ Jordan cell for the conformal weight $h_{1,7}=5$, a rank-$4$ Jordan cell for the conformal weight $h_{1,11}=15$ and so on.
% TO DO: the pattern is more complicated.
%More generally, there is a rank $k$ Jordan cell in $L_0$ for the conformal weight $h_{1,1+2k}$, when~$k \ge 2$.
  
\medskip

\begin{figure}
\begin{center}
\includegraphics[scale=0.80]{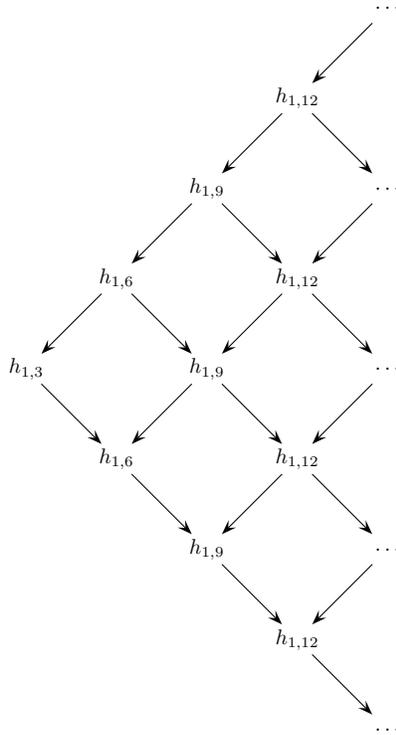}
\end{center}
  \caption{Structure in terms of simple modules of the indecomposable Virasoro module at $c=0$ corresponding to the scaling limit of the tilting module $\mathcal{T}_{1}^{b}$.
All the down-left arrows in  the figure represent the action of positive Virasoro modes, whereas the other arrows correspond
to negative modes. There are once again Jordan cells for $L_0$ of any rank occurring in the spectrum.}
  \label{figVirIndec2}
\end{figure}

All the other tilting modules are also good candidates for 
indecomposable modules arising in a physical LCFT.
A nice example exhibiting a structure different from that of the vacuum module 
is shown in Fig.~\ref{fig:tilt-ch-lim} for the tilting $\mathcal{T}^{b}_{1}$. Its scaling limit in terms of Verma modules is shown on the right part of the diagram --  the corresponding Virasoro module has the structure of a single chain of Verma modules, themselves being single chains
of simple Virasoro modules. We also give in Fig.~\ref{figVirIndec2} its more detailed structure in terms of simples where down-right arrows correspond to the action of negative Virasoro modes while down-left arrows correspond to positive modes action.
The character of such a module is readily computed 
\begin{equation}
\displaystyle \mathrm{Tr}_{\mathcal{T}^{b}_{1}} q^{L_0-c/24} = \frac{1}{P(q)}\sum_{k=0}^{\infty}(1+k) \chi_{1,3+3k} = \frac{1}{P(q)}\sum_{k=0}^{\infty} q^{h_{1,3+3k}}.
\end{equation}
On an odd-length spin chain, the module $\mathcal{T}^{u}_{\frac{1}{2}}$ has a similar ``chain-of-chains'' structure  with different exponents.

The very complicated structure of these new modules imply the existence
of Jordan cells in the operator $L_0$ of arbitrary rank. In the example
of the module shown in Fig.~\ref{figVirIndec2}, there is a rank $k$
Jordan cell in $L_0$ for any conformal weight $h_{1,3+3k}$. These
Jordan cells imply in turn logarithms in correlation functions.
As a simple example, let us consider the case of a rank $3$ Jordan cell
so that $L_0$ expressed in the basis $\bigl(\phi(z),\psi_1(z),\psi_2(z)\bigr)$ reads 
\begin{equation}
L_0 = 
\left( \begin{array}{ccc}
h & 1 & 0 \\
0 & h & 1 \\
0 & 0 & h\end{array} \right).
\end{equation}
One can then show that global conformal invariance enforces
\begin{subequations}
\begin{eqnarray}
\left\langle \phi(z) \phi(0)\right\rangle &=& \left\langle \psi_1(z) \phi(0)\right\rangle = 0 \\
\left\langle \psi_1(z) \psi_1(0)\right\rangle &=& \left\langle \psi_2(z) \phi(0)\right\rangle = \frac{\beta}{z^{2h}} \\
\left\langle \psi_2(z) \psi_1(0)\right\rangle &=& \frac{\theta_1 -2 \beta \log z} {z^{2h}} \\
\left\langle \psi_2(z) \psi_2(0)\right\rangle &=& \frac{\theta_2-2 \theta_1 \log z+2 \beta \log^2 z } {z^{2h}},
\end{eqnarray}
\end{subequations}
where $\theta_1$ and $\theta_2$ are some irrelevant constants.
This can be generalized to Jordan cells of any rank $k$~\cite{Jordancells}, with logarithmic
terms $\log^{k-1} z$ in two-point correlation functions. 
Note that it would be very interesting to 
study those indecomposable modules directly from the Virasoro side.
We expect that the fact that our modules are tilting, self-dual in particular,  might give  restrictions on their possible logarithmic couplings.

\subsubsection{Remarks on the singly critical case: staggered modules for generic central charge}
\label{subSecTiltingsMirrorSingly}

We now comment on the mirror spin chain decomposition for the singly critical case, that is, for $r$ positive integer
and $\q$ generic. As explained in section~\ref{subSecSinglyCritical}, the standard modules have the following subquotient structure
\begin{equation}
\BW^{b/u}_{j}=
\begin{array}{ccc}
\BX^{b/u}_j &&\\
&\hskip-.2cm\searrow&\\
&&\hskip-.3cm \BX^{u}_{r \pm j},
\end{array} 
\end{equation}
while the remaining standard modules $\BW^{u}_{j+r}=\BX^{u}_{j+r}$ are irreducible.
The scaling limit of these objects was related to the Verma modules over the Virasoro algebra.

Following the same line of reasoning as in  sections~\ref{tilt-mod} and~\ref{tilt-mod-hilb}, it is straightforward to see
that the indecomposable tilting modules over the blob algebra that arise in the mirror spin chain decomposition 
must have the form
\begin{equation}
\label{EqstaggGene}
    \xymatrix@R=12pt@C=6pt{
                                         & \BX^{u}_{r \pm j}  \ar[dl] & \\
     \BX^{b/u}_{j} \ar[dr]         &               &  \\
             &      \BX^{u}_{r \pm j}         &  
     }
\end{equation}     
where $0\leq j \leq N$ and $j < r/2$ in the unblobbed `$u$' case. The Hamiltonian on these modules has Jordan cells of rank~$2$.

In the scaling limit, these modules yield indecomposable staggered modules, that exist for {\it any} central charge
\begin{equation} 
    \xymatrix@R=12pt@C=6pt{
                                         & h_{r,-s}  \ar[dl] & \\
     h_{r,s} \ar[dr]         &               &  \\
             &      h_{r,-s},         &  
     }
\end{equation}        
with a non-digonalizable action of $L_0$ mapping the top to the bottom. In the expression of the conformal weights, we used $s=r+2j$.
Note that the existence of these staggered modules for generic central charge was already remarked in Ref.~\cite{Kytola,KytolaRidout}. Just like in
section~\ref{SubSecVirStaggered}, these modules
are characterized by an  indecomposability parameter, or a logarithmic coupling. We provide calculations of these characterizing numbers in Appendix~A.

\subsubsection{Staggered modules as self-glueings of Verma modules}
Recall that for generic $\q={\rm e}^{\frac{i\pi}{x+1}}$ and $y=0$ (or irrational $r=x$) we found in section~\ref{sec:proj-gen} projective modules $\BP_j$ over the JS blob algebra which are self-glueings (or self-extensions) of the standard modules $\BW_j$; see a diagram in eq.~\eqref{self-ext}. The Hamiltonian acts by Jordan cells of rank~$2$ on these modules, mapping the unique top subquotient to the isomorphic submodule in the bottom. In the scaling limit, we thus should obtain Virasoro modules at central charge $c$ given by (\ref{eq-c}) with the structure
\begin{equation*}
    \xymatrix@R=21pt@C=6pt{
                                         & \mathcal{V}_{x,x+2j}  \ar[d] & \\
         &         \mathcal{V}_{x,x+2j}    &
     }
\end{equation*}
where $\mathcal{V}_{x,x+2j}$  are Verma modules with conformal weight $h_{x,x+2j}$, and $j>0$.
We also call these modules staggered modules, since they are glueings of highest-weight modules and $L_0$ should have Jordan cells of rank~$2$ on them. Note that similar self-glueings of highest-weight modules over Virasoro algebra were discussed in Ref.~\cite{KytolaRidout}, and it was proven there that the Virasoro modules of this type exist and  are unique.

\medskip
Finally, we remark that all the results of this section can be extended to other values of $r$ and of the central charge, or integer $x>2$.
The general procedure is very clear and  the structure of the scaling limit of tiltings can be readily obtained by following the same route described in sections~\ref{section3}, and~\ref{section4}, and~\ref{tilt-mod-hilb}  (in particular, see our few general results on the structure of standard modules in App.~C).
% can be implemented in a fashion similar
%to what we have done for the $c=0$, and $r=1, 2$ cases. 
The general structure
of the indecomposable (and self-dual) modules can then  be obtained from that
of the corresponding Verma module. \textit{The possible subquotient structures are
shown in table~\ref{tableTiltGenStruct}.}

\begin{table}
\renewcommand{\arraystretch}{1.5}
\setlength{\tabcolsep}{1cm}
\begin{center}
\begin{tabular}{|@{}p{3cm}@{}|@{\hspace{0.5cm}}c@{\hspace{0.5cm}}|@{\hspace{0.0cm}}c@{\hspace{0.3cm}}|@{\hspace{0.5cm}}c@{\hspace{0.5cm}}|@{\hspace{0.5cm}}p{3cm}@{\hspace{0.5cm}}|}
  \hline
   &Point & Link & Chain & Braid/Ladder \\
   \hline
  \hline
  
 Verma module $\mathcal{V}_{h_0}$ & $ \xymatrix@R=18pt@C=6pt{
                                           \\
                                          {\bullet}\ar@{}|{ \substack{h_0}\kern-7pt}[]+<7pt,7pt>    
     } $
  
   & 
   $ \xymatrix@R=18pt@C=6pt{
                                          {\bullet}\ar@{}|{ \substack{h_0}\kern-7pt}[]+<7pt,7pt> \ar[d]  \\
                                          {\bullet}\ar@{}|{ \substack{h_1}\kern-7pt}[]+<7pt,7pt>    
     } $
    & 
  $ \xymatrix@R=12pt@C=6pt{
                                          {\bullet}\ar@{}|{ \substack{h_0}\kern-7pt}[]+<7pt,7pt> \ar[d]  \\
                                          {\bullet}\ar@{}|{ \substack{h_1}\kern-7pt}[]+<7pt,7pt> \ar[d]  \\
                                          {\bullet}\ar@{}|{ \substack{h_2}\kern-7pt}[]+<7pt,7pt> \ar[d]  \\
                     {\vdots}       
     } $
   &   $ \xymatrix@R=12pt@C=6pt{
                                         & {\bullet}\ar@{}|{ \substack{h_0}\kern-7pt}[]+<7pt,7pt> \ar[dl]\ar[dr] & \\
     {\bullet}\ar@{}|{ \substack{h_1}\kern-7pt}[]+<-22pt,7pt> \ar[d]\ar[drr]  &               &  {\bullet}\ar@{}|{ \substack{h_2}\kern-7pt}[]+<7pt,7pt> \ar[d]\ar[dll]\\
     {\bullet}\ar@{}|{ \substack{h_3}\kern-7pt}[]+<-22pt,7pt> \ar[d]\ar[drr]         &               &  {\bullet}\ar@{}|{ \substack{h_4}\kern-7pt}[]+<7pt,7pt> \ar[d]\ar[dll]\\
     {\vdots}         &               &  {\vdots}
     } $
     \\
  \hline  
   Tilting $\mathcal{T}_{h_0}$ in terms of Verma modules  &  $ \xymatrix@R=18pt@C=6pt{
                                           \\
                                          {\mathcal{V}_{h_0}}    
     } $

   & 
   $ \xymatrix@R=18pt@C=6pt{
                                          {\mathcal{V}_{h_1}} \ar[d]  \\
                                          {\mathcal{V}_{h_0}}    
     } $   
   & 
   $ \xymatrix@R=12pt@C=6pt{
                                            {\vdots} \ar[d]  \\
                                         {\mathcal{V}_{h_2}} \ar[d]  \\
                                           {\mathcal{V}_{h_1}} \ar[d]  \\
                                          {\mathcal{V}_{h_0}} 
     } $
     &   $ \xymatrix@R=12pt@C=6pt{
     {\vdots} \ar[d]\ar[drr]  &               &  {\vdots} \ar[d]\ar[dll]\\
     {\mathcal{V}_{h_3}} \ar[d]\ar[drr]  &               &  {\mathcal{V}_{h_4}} \ar[d]\ar[dll]\\
     {\mathcal{V}_{h_1}} \ar[dr]         &               &  {\mathcal{V}_{h_2}} \ar[dl]\\
                                    & {\mathcal{V}_{h_0}} &
     } $
     \\
     \hline  
   Tilting $\mathcal{T}_{h_0}$ in terms of simple modules & 
   Point

   &
   Half-Diamond
  
   & 
  ``Chain of chains''
     &  ``Braid of braids'' 
     
     \\
    & $ \xymatrix@R=18pt@C=6pt{
                                           \\
                                          {\bullet}\ar@{}|{ \substack{h_0}\kern-7pt}[]+<7pt,7pt>  
     } $

   &
    $ \xymatrix@R=12pt@C=6pt{
                                         & {\bullet}\ar@{}|{ \substack{h_1}\kern-7pt}[]+<7pt,7pt>  \ar[dl] & \\
     {\bullet}\ar@{}|{ \substack{h_0}\kern-7pt}[]+<-22pt,7pt> \ar[dr]         &               &  \\
             &      {\bullet}\ar@{}|{ \substack{h_1}\kern-7pt}[]+<-22pt,7pt>         &  
     } $
  
   & 
       (See Fig.~\ref{figVirIndec2})
     &   (See Fig.~\ref{figVirIndec})     
     \\
  \hline

\end{tabular}
\end{center}
  \caption{General schematic structure of the indecomposable Virasoro modules $\mathcal{T}_{h_0}$ obtained as scaling limit of blob tilting modules. We show the different subquotient structures obtained depending on the structure of the corresponding Verma module $\mathcal{V}_{h_0}$,which corresponds to a standard module in the blob language. Note that we restricted ourselves to the case $c \leq 1$.}
   \label{tableTiltGenStruct}
\end{table}

\subsubsection{Indecomposable Virasoro modules in physical LCFTs: a conjecture}\label{subsec:conjecture}

 The representation theory of the Virasoro algebra is known to be wild~\cite{wild,wild1,Germoni}, which makes classification issues a priori extremely difficult. However, physicists are not just interested in ``all'' Virasoro modules. Rather, they are interested in modules which may appear in physical LCFTs --  CFTs which are  fixed points of  interacting, non unitary, field theories with well defined {\sl local} actions, such as the superprojective sigma models at topological angle $\theta=\pi$, {\it etc}.  If such LCFTs exist, it is reasonable to expect that they must also admit some lattice regularizations, that is, that their properties can be studied by considering models defined on large, but finite, lattices, and exploring the thermodynamic and scaling limits. Such lattice models involving local Hamiltonians and local degrees of freedom are typically  quantum spin-chains with local interactions, {\it e.g.}, those studied in this paper. The point is now that the representation theory of algebras occurring for such finite lattice models -- such as the Temperley Lieb algebra or the blob algebra -- is well under control. We might therefore hope to classify all
  physically relevant indecomposable Virasoro modules by simply defining them as scaling limit of spin-chains modules. Of course, there is a large choice of spin chains, determined not only by the degrees of freedom but also by type of interaction. Nevertheless, experience with unitary  models (say, the $O(3)$ sigma model at $\theta=\pi$) has shown that, if the continuum limit admits only the Virasoro algebra as a chiral algebra, it can  be fully understood  by using lattice models  with the simplest  possible algebra compatible with the symmetries (in the case of the $O(3)$ sigma model, the spin $1/2$ XXX spin chain)\footnote{Bigger lattice algebras would lead similarly to classifications of indecomposable modules for bigger chiral algebras such as the $\mathcal{N}=1$ super Virasoro algebra, {\it etc.}}. It is thus reasonable to expect the Temperley Lieb algebra and its blob extension to describe a very large class of models. Based on the results discussed in the foregoing sections, we finally  give the following conjecture. 
\begin{center}
\fbox{
\begin{minipage}{0.7\textwidth}
 {\bf Conjecture:}  \textit{All 
indecomposable Virasoro modules  appearing in physical LCFTs (for which Virasoro is the maximal chiral algebra) should be obtained
as scaling limit of lattice tilting modules over the blob algebra or self-dual (sub)quotients thereof. The corresponding self-dual Virasoro modules are classified  in table~\ref{tableTiltGenStruct}.}

\end{minipage}
}
\end{center}

A few related remarks are in order here:
\begin{itemize}

\item   It is  important to note that any consistent boundary CFT with two not necessarily different boundary conditions $\alpha$ and $\beta$ (our boundary spin-chains correspond usually to  different $\alpha$ and $\beta$) should have non-degenerate two-point correlation functions of the boundary fields~\cite{GabRunW} living in the  corresponding spaces $\Hilb_{\alpha\alpha}$ and $\Hilb_{\beta\beta}$. Algebraically, the two-point functions determine a Virasoro-symmetric bilinear form on each of these spaces  of boundary fields. Because CFT is determined by a consistent set of correlators, if a boundary field, say, for a condition $\alpha$ has  zero two-point correlators  with all boundary fields from $\Hilb_{\alpha\alpha}$  this means the CFT is actually described by a quotient of $\Hilb_{\alpha\alpha}$  by the radical of the bilinear form. Of course, the resulting two-point correlator will be then non-degenerate. 
The non-degeneracy condition implies that each Virasoro module describing the space $\Hilb_{\alpha\alpha}$ (or $\Hilb_{\beta\beta}$) of boundary fields is a self-dual module (compare with the discussion in section~\ref{sec:self-dual}). This statement about self-duality partially support our conjecture. The situation is more complicated for the space  $\Hilb_{\alpha\beta}$ of boundary condition changing fields (fields sitting at a point joining the two  boundaries $\alpha$ and $\beta$). Actually one only requires  the bilinear form on $\Hilb_{\alpha\beta}\times\Hilb_{\beta\alpha}$ to be non-degenerate (note that we expect $\Hilb_{\beta\alpha}\cong\Hilb_{\alpha\beta}^*$). Therefore we do not have any obvious reasons to require self-duality  for the spaces $\Hilb_{\alpha\beta}$. Nevertheless, all logarithmic models studied in the literature~\cite{GabRunW}  (as well as here) have only {\sl self-dual} spaces $\Hilb_{\alpha\beta}$ of boundary conditions changing fields,
and we believe this should happens for any physical boundary LCFT. 
%Our conjecture thus describes an important class of the spaces $\Hilb_{\alpha\beta}$ which are self-dual as Virasoro modules. 

\item The restriction that the maximal local chiral algebra be Virasoro is crucial. We note indeed that there exists~\cite{[FGST3],GabRunW} a consistent (although the corresponding action or sigma model is not known) LCFT at $c=0$ with so-called triplet chiral algebra $\mathcal{W}_{2,3}$ that has Virasoro modules which are not in our classification ---  those having rank-$3$ Jordan cells for the $h=0$ states. But in this LCFT the  maximal local chiral algebra is $\mathcal{W}_{2,3}$, which is bigger than Virasoro.
Therefore,  we do not pretend to classify all indecomposable Virasoro modules that may appear in {\it any} LCFT. Meanwhile, we note that the blob algebra modules reproduce also all known Virasoro modules from the logarithmic $(1,p)$ models~\cite{GKfus,JR,[BGT]} including the chiral symplectic fermions theory (for $p=2$) where the maximum local chiral algebra -- the triplet W-algebra~\cite{[K-first]} -- contains the Virasoro as a proper subalgebra.

\item  One could also wonder what would come out of the algebraic study of the two-boundary TL
algebra with a blob operator on both sides of the strip~\cite{DubailTwoBd}. 
We expect this algebra to be related to the fusion process of two Virasoro modules. Indeed, using the ideas of Refs~\cite{RS2,RS3}
that considered fusion as an induction process, the fusion of two blob modules
would naturally yield modules over the two-boundary TL algebra. The study of such a fusion
product for the blob algebra would be very interesting indeed and should help understanding the results
of Refs.~\cite{PRZ,RP} from a more algebraic perspective.

\end{itemize}

\medskip

\textit{We now discuss  shortly below 
%the other part of our classification
a very different class of Virasoro modules. The main difference is that they are non self-dual.} We stress once again that this class of modules is not supposed to appear as a space of boundary fields in a physical LCFT.  Nevertheless,  non-self dual modules can be relevant in the context of statistical loop models. Moreover, we believe that it is interesting to explore how far the representation theory of the blob algebra can go in describing Virasoro modules just from an abstract point of view.

\subsection{Non self-dual staggered modules and  loop models}\label{sec:non-self-dual}

 Of course, the relationship between the blob and Virasoro algebras goes beyond our classification of self-dual modules, and taking the scaling limit of non self-dual indecomposable blob modules  provides a quick way to obtain candidates for indecomposable Virasoro modules, which may be of interest on their own, or in relation with loop models, for instance. We note here that loop models, which  provide in general  non self-dual modules, do not lead to  physical field theories  in the sense we defined previously. It is indeed known that their continuum limit does not coincide with any local field theory, a fact that should not be so surprising, since they are themselves non local. Of course loop models can be described using local field theories, but these must contain more degrees of freedom than those in the loop models, the latter appearing then only as a sort of subsector. The additional degrees of freedom do restore self-duality, as one can observe  for instance when going from the usual Temperley Lieb loop models to the XXZ spin chain \cite{RS3}. 
 
To proceed,  we describe now a more general class of indecomposable modules over the (JS) blob algebra and compare them (when it is possible) with  known results on the Virasoro side. This general class of modules is obtained by taking quotients of projective modules. Projective modules over the blob algebra were studied in section~\ref{sec:proj-gen}, and their structure in terms of standard modules was described  in Fig.~\ref{fig:proj-thirdroot}. We also discussed the scaling limit of these modules, which we dubbed projective modules over Virasoro. These modules are {\sl not} self-dual just because  they have a top, which is a simple subquotient, and have no socles (see definitions in section~\ref{sec:proj-gen}). As can be easily seen, these non-self dual Virasoro modules can be realized as submodules of tilting modules and  have therefore  a very similar classification -- ``chain-of-chains'' or ``braid-of-braids'' at critical values of the central charge.
In the example given in the right-hand side of  Fig.~\ref{fig:proj-thirdroot}, the projective Virasoro modules can be seen as glueings of two or more Verma modules.  Recall that a staggered Virasoro module~\cite{KytolaRidout} is essentially defined as a glueing (an extension) of {\sl two} Verma modules such that $L_0$ has Jordan cells of rank~$2$. 
  Our new modules are glueings of three and more Verma modules and they presumably have Jordan cells of rank $2$ and more. So we  consider them as proper generalizations of the staggered Virasoro modules. We can also take  quotients of these  modules
  -- these may yield a  family of generalized staggered modules with higher rank Jordan cells.
 
Let us now consider some interesting quotients. For the example of $r=1$ and $j=1\,\textrm{mod}\,3$ used in  Fig.~\ref{fig:proj-thirdroot}, one can take the quotient corresponding to one of the two  diagrams $\BW_{j}^{u} \to \BW_{j-1}^{b/u}$. The structure of the resulting module in terms of simple subquotients can be schematically described by the diagram (``a glueing of two towers")
%\begin{equation*}
%   \xymatrix@C=5pt@R=20pt@M=2pt@W=2pt{%
%   &&\\
%    &\BW_{j}^{u}\ar[dl]&\\
%     \BW_{j-1}^{b}&&\\
%     &&
% }   \quad 
%    \xymatrix@C=5pt@R=20pt@M=2pt@W=2pt{%
%   &&\\
%      &&\\
%     &=&\\
%        &&\\
%     &&
% } \quad
%  \xymatrix@C=15pt@R=15pt@M=2pt{%
%    &&{\stackrel{\BX_j^u}{\bullet}}\ar[dl]\ar[dr]\ar[drr]&&&\\
%    &{\stackrel{\BX_{j-1}^b}{\bullet}}\ar[dr]\ar[ddr]&
%    &{\stackrel{\BX_{j+1}^{u}}{\bullet}}\ar[ddl]\ar[dr]\ar[drr]\ar[dl]&{\stackrel{\BX_{j+2}^{b}}{\times}}\ar[d]\ar[dr]\ar[ddll]\ar[dll]&\\
%    &&{\stackrel{\BX_{j}^{b}}{\times}}\ar[ddrr]\ar[ddr]&&\ddots&\ddots\\
%    &&  {\stackrel{\BX_{j}^{u}}{\bullet}}\ar[dr]\ar[drr] &&&\\
%    && &\ddots&\ddots&
% } 
%\end{equation*}
\begin{equation*}
   \xymatrix@C=5pt@R=20pt@M=2pt@W=2pt{%
   &&\\
    &\BW_{j}^{u}\ar[dl]&\\
     \BW_{j-1}^{b}&&\\
     &&
 }   \quad 
    \xymatrix@C=5pt@R=20pt@M=2pt@W=2pt{%
   &&\\
      &&\\
     &=&\\
        &&\\
     &&
 } \quad
  \xymatrix@C=15pt@R=15pt@M=2pt{%
    &&{\stackrel{\BX_j^u}{\bullet}}\ar[dl]\ar[dr]\ar[drr]&&&\\
    &{\stackrel{\BX_{j-1}^b}{\bullet}}\ar[dr]\ar[ddr]&
    &{\stackrel{\BX_{j+1}^{u}}{\bullet}}\ar[ddl]\ar[dr]\ar[drr]&{\stackrel{\BX_{j+2}^{b}}{\times}}\ar[d]\ar[dr]\ar[dll]&\\
    &&{\stackrel{\BX_{j}^{b}}{\times}}\ar[ddrr]\ar[ddr]&&\ddots&\ddots\\
    &&  {\stackrel{\BX_{j}^{u}}{\bullet}}\ar[dr]\ar[drr] &&&\\
    && &\ddots&\ddots&
 } 
\end{equation*}
We can take then further quotients, for example by the  subquotients marked by $\times$. The result is then nothing but a familiar diamond-shape module, whose scaling limit corresponds to a Virasoro staggered module discussed in section~\ref{secboundaryXXZrestrictedStagg}. We could also consider quotients generated by states deeper in  the two towers, with as a result non-self dual modules. It is interesting to remark that these modules also exist on the Virasoro side (probably not uniquely). Indeed, recall that in general a staggered Virasoro module is a glueing  of two
highest-weight modules with a non-diagonalizable action of $L_0$ and they might be also obtained as quotients of glueings of two Verma modules; see more details in~\cite{KytolaRidout}. Such quotients are characterized by indecomposability parameters and might exist for continuous values of these parameters. It is important to note that the non-self dual blob modules are unique and their scaling limit has thus fixed indecomposability parameters (logarithmic couplings). The lattice models and the scaling limit we use in this paper hence pick up specific values of the indecomposability parameters. 

%It is possible  to introduce an even larger family of indecomposable modules,  the ``projective'' Virasoro modules  described by the right diagram in Fig.~\ref{fig:proj-thirdroot}, whose existence was conjectured based on   the correspondence in Tab.~\ref{tabblobVir} between the JS blob algebra and Virasoro.

%There are of course continuously many non-self-dual staggered modules which are not reproduced by our results but these modules  can  appear in physical models only for specific logarithmic couplings. We believe that (JS) blob algebra modules reproduce these specific staggered modules in the scaling limit, as was discussed in section~\ref{sec:non-self-dual}. 

Another less trivial example  corresponds to the quotient of the projective module $\BP^u_j$  (at $r=1$ and $j=2\,\mathrm{mod}\,3$) by its submodule $\BP^b_{j-4}$. This quotient is given by the left diagram in terms of standard modules:
 \begin{equation*}
{\scriptsize
   \xymatrix@R=18pt@C=8pt@W=2pt@M=2pt
   { &&&\\
    &&\BW_{j}^{u}\ar[dl]&\\
     &\BW_{j-1}^{b}\ar[dl]& &\\
     \BW_{j-3}^{u} &&&
     } \quad 
%   \xymatrix@R=18pt@C=8pt@W=2pt@M=2pt
%   { 
%    &&\BX_{j}^{u}\ar[dl]\ar[dr]&&&\\
%     &\BX_{j-1}^{b}\ar[dl]\ar[dr]& & \BX_{j+2}^{b}\ar[dl]\ar[dr]&&\\
%     \BX_{j-3}^{u}\ar[dr] &&\BX_{j}^{u}\ar[dl]\ar[dr]&&\BX_{j+3}^{u}\ar[dl]\ar[dr]&\\
%          &\BX_{j-1}^{b}\ar[dr]& & \BX_{j+2}^{b}\ar[dl]\ar[dr]&&\;\ddots\\
%               &&\BX^u_j\ar[dr]& & \;\ddots&\\
%                              &&&\;\ddots &&\\
%     }
      \xymatrix@R=22pt@C=1pt@W=4pt@M=4pt
{&&\\
&&\\
\xrightarrow{\mbox{}\quad L\to\infty\quad}&&\\
&&\\
&&}\quad
   \xymatrix@R=18pt@C=1pt@W=2pt@M=2pt
   { 
    &&h_{1,1-2j}\ar[dl]\ar[dr]&&&\\
     &h_{1,2j-1}\ar[dl]\ar[dr]& &h_{1,5+2j}\ar[dl]\ar[dr]&&\\
     h_{1,7-2j}\ar[dr] &&h_{1,1-2j}\ar[dl]\ar[dr]&&h_{1,-5-2j}\ar[dl]\ar[dr]&\\
          &h_{1,2j-1}\ar[dr]& &h_{1,5+2j} \ar[dl]\ar[dr]&&\quad\ddots\quad\\
               &&h_{1,1-2j}\ar[dr]& & \quad\ddots\quad&\\
                              &&&\quad\ddots\quad &&\\
     }
     }
 \end{equation*}
where we also describe the scaling limit of this quotient in terms of irreducible Virasoro (at $c=0$) representations labeled by $h_{r,s}$. Note that this  Virasoro module should be seen as  a generalized staggered module having Jordan cells of rank~$2$ and~$3$. It would be very interesting of course to construct this module (and its further quotients) using the techniques of~\cite{KytolaRidout}.

% We review here the theory~\cite{KytolaRidout} of the so-called \textit{staggered} modules over
%$\Vir(p',p)$. 
% Using Prop.~$4.6$
%and Cor.~$4.7$ from~\cite{KytolaRidout}, the study of any
%staggered module can actually be reduced to the simpler analysis of ``Verma-type'' staggered modules which are gluing of two Verma modules
%\begin{equation}\label{Verma-stag-seq}
%0\longrightarrow \Verma_{h_l}\longrightarrow \cVP(h_l,h_r) \longrightarrow \Verma_{h_r}\longrightarrow 0,
%\end{equation}
%where $\Verma_{h}$ denotes the Verma module generated from the highest-weight
%vector with conformal weight~$h$. Let $\Hmod_{h}$ be a quotient of the Verma module $\Verma_{h}$, then any staggered module $\VP$
%defined by a non-splitted exact sequence
%\begin{equation*}
%0\longrightarrow \Hmod_{h_l}\longrightarrow \VP(h_l,h_r) \longrightarrow \Hmod_{h_r}\longrightarrow 0,
%\end{equation*}
% \textit{is} a quotient of the ``Verma-type'' staggered module $\cVP$ defined in~\eqref{Verma-stag-seq} above.

Finally  consider the loop models  discussed in section~\ref{subsecblobalgStatmech}. Suppose   the evolution  starts in a sector with, say,  the maximum number of through-lines and ends up in the sector with zero through-lines. The Hilbert space of this loop model is then in general a glueing of the (JS) blob algebra standard modules. Such a glueing by construction is not a tilting module (it is not self-dual) and is very close to the modules (quotients of projectives) we just described. It would be interesting to identify the corresponding indecomposability parameters (following {\it e.g.}~\cite{VJS}).

\medskip

\section{Conclusion}

Our conjecture in section~\ref{subsec:conjecture} (see also Tab.~\ref{tableTiltGenStruct}) about the classification of physically relevant Virasoro indecomposable modules is obviously the central result of this paper. If true, it does provide some elements to tame the wilderness of the Virasoro algebra in the non semi-simple case, even though the type of modules one can encounter still remains dauntingly varied and complicated. 

Admittedly, our understanding of the precise relationship between lattice algebras and their continuum limit counterparts remains in its infancy. We believe that the case of the (JS) blob algebra and the Virasoro algebra is particularly tantalizing, and deserves more thorough and rigorous scrutiny. From a more practical perspective, it would be most useful to have a Coulomb gas formalism adapted to the (JS) blob algebra, where general indecomposable Virasoro modules could be built, in the spirit of Ref.~\cite{[FFHST]}.

An obvious generalization of this work would be to explore lattice algebras related with higher value of the $U_{\q} \mathfrak{sl}_2$ spin -- the simplest case would be the Birman-Wenzl-Murakami algebra, which appears in the spin-one case, and should be related to the non semi-simple $\mathcal{N}=1$ super Virasoro algebra. More important maybe is the extension to the case of bulk LCFTs, which are related to lattice models with periodic boundary conditions, and thus yet other lattice algebras, such as the Jones--Temperley--Lieb algebra \cite{RS2,GRS1,GRS3}. According to \cite{MartinSaleur}, this problem bears some interesting relations with the blob algebra case, and we will return to this soon. 

\bigskip
\noindent {\bf Acknowledgments}: We thank I.Yu.~Tipunin and N.~Read for many discussions and collaborations on related matters.
AMG thanks P.P. Martin and K. Kyt\"ol\"a for stimulating discussions, and is also grateful to K. Kyt\"ol\"a for his kind hospitality in Helsinki University during 2012. 
RV also wishes to thank A. Lazarescu for insightful discussions. This work was supported in part by the ANR Project 2010 Blanc SIMI 4: DIME. The work of AMG was supported in 
part by Marie-Curie IIF fellowship and the RFBR grant 10-01-00408.

\section*{Appendix A:  Staggered modules
and indecomposability parameters~$\beta_{r,s}$}\label{sec:stag-beta}
\renewcommand\thesection{A}
\renewcommand{\theequation}{A\arabic{equation}}
\setcounter{equation}{0}

In this section, we gather a few results concerning the so-called indecomposability parameters (or logarithmic couplings)
that characterize the staggered modules~\cite{KytolaRidout} encountered in this paper.

\subsection{Doubly critical case and diamond staggered modules}

Indecomposability parameters for the modules in Sec.~\ref{SubSecVirStaggered} can be computed using the general formula of Ref.~\cite{VJS}.
Let us consider the family of staggered modules simply depicted as
\begin{equation}
\VP = \begin{array}{ccccc}
      &&\hskip-.7cm h_\psi  &&\\ 
      &\hskip-.2cm\swarrow&\searrow&\\
      h_\xi &&&\hskip-.3cm h_\rho \\
      &\hskip-.2cm\searrow&\swarrow&\\
      &&\hskip-.7cm h_\phi &&
\end{array},
\end{equation}
for the general LCFT with central charge $c=1-\frac{6}{x_0 (x_0+1)}$. This family includes the staggered modules $\VP_{r,r+2j}$ from eq.~\eqref{list-stagg} obtained in  the scaling limit of our XXZ-like boundary spin-chains.
We denote the generating vectors in $\VP_{r,r+2j}$  in terms of the fields $\primt_{j}(z)$, $\primr_{j}(z)$, $\priml_{j}(z)$, and $\primb_{j}(z)$.
%where we set $\primt_{j}=\lim_{z\to0}\primt_{j}(z)\vacr$, {\it etc}.
 The state $\priml_{j}=\lim_{z\to0}\priml_{j}(z)\vac$ has the lowest conformal weight in $\VP_{r,r+2j}$.
The whole module can be generated by acting with Virasoro generators on $\primt_{j}$; in particular, the dilatation
operator $L_0$ mixes the fields $\primt_{j}(z)$ and $\primb_{j}(z)$ into a rank-$2$ Jordan cell.
The state $\primb_{j}$ is singular and it is a
descendant of $\priml_{j}$:
\begin{equation}
\displaystyle \primb_{j} = A_j \priml_{j},
\end{equation}
where the operator $A_j$ belongs to the universal enveloping algebra of Virasoro. The singular-vector 
condition $L_1 \primb_{j}= L_2 \primb_{j}=0$ fixes uniquely $A_j$ once a normalization has been properly chosen.
The module $\VP_{r,r+2j}$ is uniquely characterized by a number called logarithmic coupling (or  indecomposability parameter) $\beta_{r,r+2j}$~\cite{MathieuRidout1, KytolaRidout}. It is defined through the equation
\begin{equation}
\label{eqDefBeta}
\displaystyle A_j^\dag \primt_{j} = \beta_{r,r+2j} \priml_{j}.
\end{equation}
Note that it can also be simply expressed using the usual Virasoro bilinear form as $\beta_{r,r+2j} = \langle \primb_{j} \left. \right| \primt_{j} \rangle$,
with the normalization $\langle \priml_{j} \left. \right| \priml_{j} \rangle = 1$.
These logarithmic couplings are closely related to the coefficients that appear in front of logarithmic terms in correlation functions and OPEs~\cite{MathieuRidout1,GV}. 

The indecomposability parameters $\beta_{r,r+2j}$ for the family of staggered modules can be computed from 
small deformations around this theory $x=x_0+ \epsilon$ as \cite{VJS}
\begin{equation}
\label{b_formula}
\displaystyle \beta = - \lim_{\epsilon \rightarrow 0} \frac{ \Braket{\phi | \phi}}{h_{\psi}-h_{\xi}-n} = -  \frac{\left. \frac{\rm d}{\mathrm{d} \epsilon} \Braket{\phi | \phi}\right|_{\epsilon=0}  }{\left. \frac{\rm d}{\mathrm{d} \epsilon} (h_{\psi}-h_{\xi}) \right|_{\epsilon=0}}   ,
\end{equation}
where $n=(\left. h_{\psi}-h_{\xi})\right|_{\epsilon=0}$. In this expression, $\Braket{\phi | \phi}$ is the norm-squared of the state $\Ket{\phi}$,
which is a descendant of $\Ket{\xi}$ that becomes singular at $\epsilon=0$ (we refer the interested reader to~\cite{VJS} for 
more detail about this formula).
Using the same normalization\footnote{That is, we chose a normalization such that $\Ket{\phi} = (L_{-n}+\dots) \Ket{\xi}$.} as in Ref.~\cite{VJS},
we computed some of the indecomposability parameters associated with these staggered modules.
The results are gathered in Tab.~\ref{general_b_figure_c0}.
We also show the results for the ``dilute'' part of the Kac label ($s \leq r$)
that could be reached from the lattice by a blob version of dilute $\mathcal{O}(n)$ models~\cite{DubailDilute}.

\renewcommand{\arraystretch}{2.5}

\begin{table}
\begin{center}%\small
%\vspace{-2cm}
%\begin{turn}{90}
\begin{tabular}{c||c|c|c|c|c|c|c|c|c|c}
   \cline{1-4}
  \multicolumn{1}{|c||}{$r=5$} & $\dfrac{67375}{676}$ & $\dfrac{175}{9}$ & $\dfrac{105}{4}$  &  \multicolumn{7}{c}{}  \\
  \cline{1-4}
  \multicolumn{1}{|c||}{$r=4$} & $\bullet$ & $\bullet$ & $\bullet$ &  \multicolumn{7}{c}{}    \\
   \cline{1-4}
  \multicolumn{1}{|c||}{$r=3$} & $\dfrac{5}{6}$ & $\dfrac{1}{3}$ & $\bullet$  &  \multicolumn{7}{c}{}       \\
  \hline
  \multicolumn{1}{|c||}{$r=2$} & $\bullet$ & $\bullet$ & $\bullet$  & $\bullet$ & $\bullet$ & $\bullet$   & $-\dfrac{20}{9}$ & $-\dfrac{2800}{81}$ & $\bullet$ &  \multicolumn{1}{|c|}{$-\dfrac{61600}{2601}$ }   \\
  \hline
  \multicolumn{1}{|c||}{$r=1$} & $\bullet$ & $\bullet$ & $\bullet$  & $-\dfrac{1}{2}$ & $-\dfrac{5}{8}$ & $\bullet$   & $-\dfrac{35}{3}$ & $-\dfrac{13475}{216}$ & $\bullet$ &  \multicolumn{1}{|c|}{$-\dfrac{17875}{324}$ } \\

  \hline
  \hline
 \multicolumn{1}{|c||}{ $\beta_{r,s}$} & $s=1$ & $s=2$ & $s=3$  & $s=4$ & $s=5$ & $s=6$   & $s=7$ & $s=8$ & $s=9$ & \multicolumn{1}{|c|}{$s=10$}  \\    
     \hline

\end{tabular}

%\end{turn}
\end{center}
\caption{
Indecomposability parameters $\beta_{r,s}$ (logarithmic couplings) for the boundary percolation problem ($x=2$ and $c=0$).
The symbols  $\bullet$ correspond to the cases where no interesting $\beta_{r,s}$ coefficient can be defined. It can either mean that
the corresponding module is irreducible, or reducible with a structure that does not allow for a logarithmic coupling.
}
\label{general_b_figure_c0}
\end{table}

\renewcommand{\arraystretch}{1.0}

%The general subquotient structure of the Virasoro staggered module with conformal weight $h_{r,r+2j}=h_{r,1+2j'}$ 
%is not much different than the case $r=1$. We focus on ``minimal'' theories $\mathcal{LM}(p,p+1)$ theories with $x=p$. 
%For such theories, the staggered modules have the following structure  
%\begin{equation}
%\VP_{r,r+2j}=~~~~~\left\{\begin{array}{cl}
%\begin{array}{ccccc}
%     &&\hskip-.7cm \VX_{j'}&&\\
%     &\hskip-.2cm\swarrow&\searrow&\\
%     \VX_{j'-1}&&&\hskip-.3cm \VX_{j'+p}\\
%     &\hskip-.2cm\searrow&\swarrow&\\
%     &&\hskip-.7cm \VX_{j'}&&
%     \end{array}&\hbox{$j'=0$ (mod $\frac{p+1}{2}$) and $j'>0$,}\nonumber\\
%     &\nonumber\\
%\begin{array}{ccccc}
%     &&\hskip-.7cm \VX_{j'}&&\\
%     &\hskip-.2cm\swarrow&\searrow&\\
%     \VX_{j'-2}&&&\hskip-.3cm \VX_{j'+p-1}\\
%     &\hskip-.2cm\searrow&\swarrow&\\
%     &&\hskip-.7cm \VX_{j'}&&
%     \end{array}&\hbox{$j'=\frac{1}{2}$ (mod $\frac{p+1}{2}$) and $j'>1$,}\nonumber\\
%     &\nonumber\\
%\vdots \\
%\begin{array}{ccccc}
%     &&\hskip-.7cm \VX_{j'}&&\\
%     &\hskip-.2cm\swarrow&\searrow&\\
%     \VX_{j'-p}&&&\hskip-.3cm \VX_{j'+1}\\
%     &\hskip-.2cm\searrow&\swarrow&\\
%     &&\hskip-.7cm \VX_{j'}&&
%     \end{array}&\hbox{$j'=\frac{p-1}{2}$ (mod $\frac{p+1}{2}$) and $j'>p-1$,}\nonumber\\
%     &\nonumber\\
%\VX_{j'} &\hbox{$j'\equiv\frac{p}{2}$ (mod $\frac{p+1}{2}$).}\nonumber\\
%     &\nonumber\\
%\end{array}\right.
%\end{equation}
%where we recall that $\VX_{j'}=\VX_{r,r+2j}$ with $1+2j'=r+2j$.

All these remarks can be extended to other LCFTs with $x \geq 3$ without 
much more work. Some of these results concerning the new indecomposability parameters
could even be tested numerically in the boundary XXZ spin chain introduced in section~\ref{subsecBoundaryXXZ}, following the lines of Ref.~\cite{VJS}.

% Let us finish by a concrete example than could be
%verified numerically along the lines of Ref.~\cite{VJS} in the boundary XXZ spin chain
%at $q=\mathrm{e}^{i\pi/3}$ and $r=2$ (to be discussed in the next subsection).
%Let us consider a spin chain with length $L$ odd. 
%The first Jordan cell occurring in the spectrum has $j=\frac{5}{2}$ and corresponds
%to a conformal weight $h=\frac{5}{8}$. The staggered module associated with this
%Jordan cell reads
%\begin{equation}
%\VP_{\frac{5}{2}} = \begin{array}{ccccc}
%      &&\hskip-.7cm \psi_{\frac{5}{2}}  &&\\
%      &\hskip-.2cm\swarrow&\searrow&\\
%      \xi_{\frac{5}{2}} &&&\hskip-.3cm \rho_{\frac{5}{2}} \\
%      &\hskip-.2cm\searrow&\swarrow&\\
%      &&\hskip-.7cm \phi_{\frac{5}{2}} &&
%\end{array}= \begin{array}{ccccc}
%      &&\hskip-.7cm h_{2,7}  &&\\
%      &\hskip-.2cm\swarrow&\searrow&\\
%       h_{2,5} &&&\hskip-.3cm  h_{2,11} \\
%      &\hskip-.2cm\searrow&\swarrow&\\
%      &&\hskip-.7cm  h_{2,7} &&
%\end{array}.
%\end{equation}
%We have the relation
%\begin{equation}
%\phi_{\frac{5}{2}} = \left( L_{-2} - \frac{2}{3} L_{-1}^2\right) \xi_{\frac{5}{2}}.
%\end{equation}
%This Jordan cell is characterized by the indecomposability parameter $\beta_{2,7} = -\frac{20}{9}$
%that does not occur in the first row of the Kac table (in which case it could be reached with the Temperley-Lieb algebra).
%Using eq.~\eqref{eq_KooSaleur} and the numerical methods of Ref.~\cite{VJS}, this result
%could be verified numerically in concrete lattice models.

\subsection{The singly critical case}
We mentioned in section~\ref{subSecTiltingsMirrorSingly} that the staggered Virasoro modules for generic central charges $c$
are also (as for critical $c$) characterized by an  indecomposability parameter that we shall denote $\beta_{r,s}$.
Some of these numbers were computed  as a function of the central charge using SLE related methods~\cite{Kytola}. 
We  compute the coefficients $\beta_{r,s}$
using eq.~\eqref{b_formula} and the methods of~\cite{VJS}, considering now $r$ as a deformation parameter.  
To compare our results with Ref.~\cite{Kytola}, we will use the SLE parametrization 
\begin{subequations}
\begin{eqnarray}
c &=&1 - \dfrac{6 (\kappa-4)^2}{4 \kappa}, \\
h_{r,s} &=& \dfrac{\kappa^2 (r^2-1) - 8 \kappa (r s-1) + 16 ( s^2-1)}{16 \kappa}.
\end{eqnarray}
\end{subequations}
We will consider the blob algebra for generic $\q$ (so generic $\kappa$), and $r=r_0+\epsilon$,
with $r_0 \in \mathbb{N}^*$. For non-zero small $\epsilon$, the representation theory of the blob
algebra is generic, so that the standard modules $\BW^{b/u}_{j}$ (corresponding to the conformal weight $h_{r,r \pm 2j}=h_{r_0+\epsilon,r_0 + \epsilon \pm j}$)
and $\BW^{u}_{r_0 \pm j}$ (corresponding to $h_{r,r - 2 (r_0 \pm j)} = h_{r_0+\epsilon,-r_0 + \epsilon \mp j}$) are irreducible.
However, when $\epsilon \rightarrow 0$, these standard modules can be glued together to form the indecomposable module~\eqref{EqstaggGene}.

Let $s_0=r_0 \pm 2j$. We will denote $\Ket{\phi} = (L_{-1}^{s_0 r_0} + \dots) \Ket{\xi_{h_{r_0+\epsilon,s_0+\epsilon}}}$ the descendant
of the primary $\Ket{\xi_{h_{r_0+\epsilon,s_0+\epsilon}}}$, with conformal weight $h_{r_0+\epsilon,s_0+\epsilon} + s_0 r_0$, that becomes
singular at $\epsilon = 0$. Using the main result of~\cite{VJS}, here reproduced as (\ref{b_formula}), the indecomposability parameter $\beta_{r_0,s_0}$ then reads
\begin{equation}
\displaystyle \beta_{r_0,s_0} (\kappa) = - \lim_{\epsilon \rightarrow 0} \frac{ \Braket{\phi | \phi}}{h_{r_0+\epsilon, -s_0+\epsilon}-h_{r_0+\epsilon, s_0+\epsilon}-s_0 r_0}.
\end{equation}
This formula allows to compute $\beta_{r_0,s_0} (\kappa)$ in a very efficient way. As an example, one can readily recover
$\beta_{1,1} = 1 - \frac{\kappa}{4}$ and $\beta_{2,1} = -\frac{1}{16} (\kappa-4)(\kappa+4)(\kappa+2)$, in agreement with~\cite{Kytola}.
In addition, this equation allows us to make more general conjectures for the form of these coefficients, 
for example, we found very strong evidence that tends to show that (see also~\cite{MathieuRidout1,RidoutBulk})
\begin{equation}
\displaystyle \beta_{1,s} = (-1)^s \left( \frac{(s-1)!}{2}\right)^2 \dfrac{\kappa-4s}{\kappa^{2(s-1)}} \prod_{i=1}^{s-1} \left( \kappa-4 i\right)\left( \kappa+4 i\right).
\end{equation}
This formula was verified for $s \leq 12$.

\section*{Appendix B: Quantum Group decomposition of the Boundary XXZ spin chain and Bimodule}\label{sec:QGboundaryXXZ}
\renewcommand\thesection{B}
\renewcommand{\theequation}{B\arabic{equation}}
\setcounter{equation}{0}

\begin{figure}
\begin{center}
\includegraphics[scale=1.0]{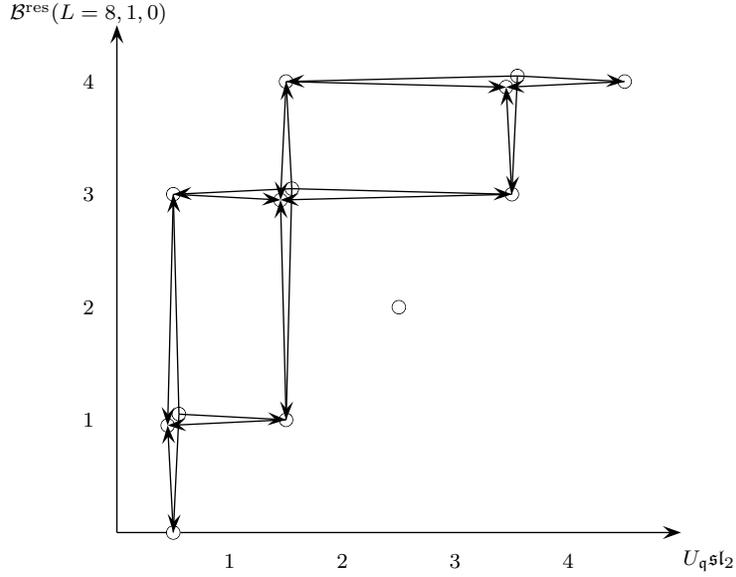}
\end{center}
  \caption{Bimodule diagram for the space of states of the boundary XXZ spin chain at $\q=\exp(i \pi/3)$ and $r=2$ on $L=2N=8$ sites. It shows the commuting action of the \res~blob algebra (vertical arrows) and of $U_{\q} \mathfrak{sl}_2$ (horizontal arrows).}
  \label{figStaircase}
\end{figure}

In this section, we explain how we obtain the Hilbert space decomposition of section~\ref{subsecHdecompositionXXZ}
using quantum group results. Recall that the example we considered had $L=8$, $\q=\mathrm{e}^{i \pi/3}$, and $r=2$ 
so the Hilbert space in terms of representations of $U_{\q} \mathfrak{sl}_2$ reads  $\mathcal{H}=(\frac{1}{2}) \otimes(\frac{1}{2})^{\otimes 8}$.

The generically irreducible representations of $U_{\q} \mathfrak{sl}_2$ are labeled by their spin $j$
and denoted by $(j)$. These $\q$-deformation of usual SU$(2)$ representations 
are called Weyl or standard modules, and they have dimension $2j+1$.
When $\q=\exp(i \pi/3)$, some of these standard modules become reducible with a link structure.
That is, the maximum proper submodule is irreducible, and the quotient
by this submodule is irreducible as well. The modules $(j)$ with $j = 1$ (mod $3/2$) remain irreducible.
The modules $(j)$ with $j = 2$ (mod $3/2$) are reducible with a proper submodule isomorphic
to the top simple subquotient of $(j+1)$, the latter being itself reducible with a proper submodule
isomorphic to the top of $(j+3)$. The non trivial indecomposable modules arising as a direct summand 
in the decomposition $(\frac{1}{2}) \otimes(\frac{1}{2})^{\otimes 8}$ -- the tilting modules -- have the
following subquotient structure~\cite{PasquierSaleur, Martin1, Martin2}

\begin{equation}
T_{\frac{6p+1}{2}}=
\begin{array}{ccccc}
      &&\hskip-.7cm  (\frac{6p+1}{2})_0  &&\\
      &\hskip-.2cm\swarrow&\searrow&\\
      (\frac{6p-3}{2})_0  &&&\hskip-.3cm (\frac{6p+3}{2})_0  \\
      &\hskip-.2cm\searrow&\swarrow&\\
      &&\hskip-.7cm (\frac{6p+1}{2})_0  &&
\end{array}
,\ \ \
T_{\frac{6p+3}{2}}=
\begin{array}{ccccc}
      &&\hskip-.7cm  (\frac{6p+3}{2})_0  &&\\
      &\hskip-.2cm\swarrow&\searrow&\\
      (\frac{6p+1}{2})_0  &&&\hskip-.3cm (\frac{6p+7}{2})_0  \\
      &\hskip-.2cm\searrow&\swarrow&\\
      &&\hskip-.7cm (\frac{6p+3}{2})_0  &&
\end{array},
\end{equation}
where we have denoted $(j)_0$ the simple modules of $U_{\q} \mathfrak{sl}_2$.
Modules with negative spin $j<0$ are understood to be zero and can be omitted.
The module $(\frac{1}{2})=(\frac{1}{2})_0$ is a tilting module. We thus
expect the tensor product $(\frac{1}{2}) \otimes(\frac{1}{2})^{\otimes L}$ 
to be decomposed onto tilting modules. The decomposition for $L=8$ reads
\begin{multline}
\label{eqDecompoXXZuqsl2}
\left. \mathcal{H}_{\rm XXZ}^{2N=8} \right|_{U_{\q}(\mathfrak{sl}_2) } =
1 \times
\begin{array}{ccccc}
      &&\hskip-.7cm  (\frac{7}{2})_0  &&\\
      &\hskip-.2cm\swarrow&\searrow&\\
      (\frac{3}{2} )_0 &&&\hskip-.3cm (\frac{9}{2})_0  \\
      &\hskip-.2cm\searrow&\swarrow&\\
      &&\hskip-.7cm (\frac{7}{2})_0  &&
\end{array}
\hskip-0.3cm
\oplus
7 \times
\begin{array}{ccccc}
      &&\hskip-.7cm  (\frac{3}{2})_0  &&\\
      &\hskip-.2cm\swarrow&\searrow&\\
      (\frac{1}{2})_0  &&&\hskip-.3cm (\frac{7}{2})_0  \\
      &\hskip-.2cm\searrow&\swarrow&\\
      &&\hskip-.7cm (\frac{3}{2})_0  &&
\end{array}
\hskip-0.3cm
\\\oplus
27 \times (\frac{5}{2})_0
\oplus
\hskip0.5cm
41 \times \ \ \ \
\begin{array}{ccc}
        \hskip-.7cm (\frac{1}{2})_0  &&\\
       &\hskip-0.7cm\searrow&\\
      &&\hskip-.58cm (\frac{3}{2})_0  \\
     &\hskip-0.7cm\swarrow&\\
     \hskip-.7cm (\frac{1}{2})_0  &&
\end{array}
\oplus
1 \times
 (\frac{1}{2})_0,
\end{multline}
in agreement with general results in~\cite{GV}.
The check on the dimensions reads
\begin{equation}
\displaystyle 512 = 1 \times (6+6+4+2) +7 \times (2+2+2+6) + 27 \times 6 + 41 \times (2+2+2) + 1 \times 2. 
\end{equation}

As discussed in section.~\ref{subsecBoundaryXXZ}, for $r=2$,
the \res~blob algebra is isomorphic to a TL algebra defined on $2N+1$ sites, so
we know that the centralizer of the $\LQG$ in this example is $\mathcal{B}^{\rm res}(2N,1,0)$.
The decomposition under $\mathcal{B}^{\rm res}$ can be thus inferred from the
structure in terms of $U_{\q}(\mathfrak{sl}_2)$ representations, and {\it vice versa}. This is done
by coming back to the definition of the centralizer (commutant) as the space of endomorphisms that
commute with a given algebra. This was explained in great detail in Ref.~\cite{RS3}.
We will simply remark here that the decompositions~\eqref{eqDecompoXXZBres} and~\eqref{eqDecompoXXZuqsl2}
are fully consistent, and that the multiplicities of the tilting modules in the decomposition
over one algebra correspond to the dimensions of the simple modules in the decomposition with respect to its centralizer.
Studying then all possible homomorphisms between direct summands over $\LQG$ -- these are presented in the bimodule diagram in
Fig.~\ref{figStaircase} by vertical arrows -- gives finally the structure of modules over the centralizer of $\LQG$. We see that the dimensions of simple modules and the structure of the tilting modules obtained at this step exactly coincide with results obtained in section~\ref{subsecRestric-proj}. This analysis confirms that $\mathcal{B}^{\rm res}(2N,1,0)$ is the centralizer of $\LQG$.
% actually the decomposition we obtain is still closely related to the $r=1$ (TL) case.
 Using this strategy, other roots of unity and higher spins at the boundary
can be treated in the very same way, that is, decomposing $ (j') \otimes(\frac{1}{2})^{\otimes L}$
with respect to the quantum group in order to obtain the decomposition over the centralizer of $\LQG$ (note that  general decompositions for $\LQG$ and the TL algebra were obtained in~\cite{GV})
 and identifying then the direct summands with general results for tilting modules over $\mathcal{B}^{\rm res}(2N,n,y)$. This analysis also show that the boundary spin-chains are faithful representations of the \res~blob algebras.

\section*{Appendix C: Dimensions of few simple modules over the (JS) blob algebra}\label{sec:dim-simple}
\renewcommand\thesection{C}
\renewcommand{\theequation}{C\arabic{equation}}
\setcounter{equation}{0}

\newcommand{\dimm}{\mathrm{dim}}
We can easily compute dimensions of simple modules over the blob algebra and its JS version using the subquotient structure of the corresponding standard modules. In this section, we consider the blob algebra with $\q=e^{i\pi/p}$, $p=x+1$, and $y=[r+1]_{\q}/[r]_{\q}$, for $1\leq r\leq p-1$, with both $p$ and $r$ integers, and provide a computation of $\dimm \BX_0$ and $\dimm \hat{\BX}_0$. Having in mind the weight diagram~\cite{Martin} like the one described in Sec.~\ref{sec:st-mod-r1}, we obtain first the subquotient structure for $\BW_0$:
\begin{equation}
{\small
    \xymatrix@R=10pt@C=22pt{
    &\BX_{r}^u\ar[r]\ar[ddr]&\BX_{p}^u\ar[r]\ar[ddr]&\BX_{p+r}^u\ar[r]\ar[ddr]&\BX_{2p}^u\ar[r]\ar[ddr]&\ldots\\
    \BX_0\ar[ur]\ar[dr]&&&&&&\\
    &\BX_{p-r}^b\ar[r]\ar[uur]&\BX_{p}^b\ar[r]\ar[uur]&\BX_{2p-r}^b\ar[r]\ar[uur]&\BX_{2p}^b\ar[r]\ar[uur]&\ldots
    }
}
\end{equation}
where as usual we assume that $\BX^{u/b}_{j>N}\equiv0$.
From this diagram, we also recover the subquotient structure of the standard modules $\BW^u_{r+np}$, $\BW^b_{p-r+np}$, and $\BW^{u/b}_{np}$, which are submodules in $\BW_0$. 
Then, using the dimensions of these submodules we obtain $\dimm \BX_0$ as an alternating sum:
\begin{multline}
\dimm \BX_0 = \dimm \BW_0 -  \sum_{n\geq0}\bigl(\dimm\BW^u_{r+np} + \dimm\BW^b_{p-r+np}\bigr) +   \sum_{n\geq1}\bigl(\dimm\BW^u_{np} + \dimm\BW^b_{np}\bigr) \\
   = 
   \sum_{n\geq0} \left\{2 \binom{2N}{N - np} - \binom{2N}{N - np - r} - \binom{2N}{N - (n+1)p  + r} \right\} - \binom{2N}{N}.
\end{multline}
Similarly, we obtain
\begin{equation*}
\dimm \BX_{r+kp}^u = \dimm \BW_{r+kp}^u -  \sum_{n\geq k+1}\bigl(\dimm\BW^u_{np} + \dimm\BW^b_{np}\bigr) + \sum_{n\geq k+1}\bigl(\dimm\BW^u_{r+np} + \dimm\BW^b_{p-r+np}\bigr), \\
\end{equation*}
and so on.

For the JS blob algebra $\mathcal{B}^{b}(2N,n,y)$ and for $1\leq r\leq p-2$,  using the isomorphism~\eqref{JS-alg-iso} we have
\begin{multline*}
\dimm \hat{\BX}_0 = \dimm \hat{\BW}_0 -  \sum_{n\geq0}\bigl(\dimm\hat{\BW}^u_{r+np} + \dimm\hat{\BW}^b_{p-r+np}\bigr) +   \sum_{n\geq1}\bigl(\dimm\hat{\BW}^u_{np} + \dimm\hat{\BW}^b_{np}\bigr) \\
%   =   \binom{2N-1}{N-1} +
%   \sum_{n\geq1} ( \binom{2N-1}{N - nx - 1} +  \binom{2N-1}{N - nx} )- \sum_{n\geq0}\left\{\binom{2N-1}{N - nx - r - 1} + \binom{2N-1}{N - (n+1)x  + r} \right\}\\
     =   
    \sum_{n\geq0}\left\{\binom{2N-1}{N - np - 1} +  \binom{2N-1}{N - np} - \binom{2N-1}{N - np - r - 1} - \binom{2N-1}{N - (n+1)p  + r} \right\} - \binom{2N-1}{N},
\end{multline*}
Finally, for $r=p-1$ (or $y=0$) we now have a full chain of embeddings for JS algebra standard modules as
\begin{equation}
\hat{\BW}_{0} \hookleftarrow\hat{\BW}^u_{p-1}  \hookleftarrow\hat{\BW}^u_{p}  \hookleftarrow\hat{\BW}^u_{2p-1} \hookleftarrow \dots
\end{equation}
which means that each term in this chain has the subquotient structure of the chain type and not the braid one. We thus obtain the dimension of $\hat{\BX}_0$ in this case  by a single subtraction
\begin{equation*}
\dimm\hat{\BX}_0 = \dimm \hat{\BW}_0 -\dimm \hat{\BW}^u_{p-1} =  \binom{2N-1}{N-1} -  \binom{2N-1}{N - p},\qquad \text{for}\; r=p-1.
\end{equation*}

\end{document}